\documentclass[a4paper, onecolumn]{article}

\usepackage{setspace}
\usepackage{textcomp}
\usepackage{diagbox}
\usepackage{titlesec}
\usepackage[titletoc,toc,page,header]{appendix}
\usepackage{tabularx}
\usepackage{authblk}
\usepackage{booktabs}
\usepackage{graphicx}
\usepackage{amsmath}
\usepackage{geometry}
\usepackage{amssymb}
\usepackage{float}
\usepackage{makecell}
\usepackage{multirow}
\usepackage{tikz,pgf}
\usepackage{cite}
\usepackage{gensymb}
\usepackage{amstext}
\usepackage{bm}
\usepackage{indentfirst}
\usepackage{latexsym}
\usepackage{color}
\usepackage{subfigure}
\usepackage{slashed}
\usepackage{multicol}
\usepackage{lipsum}
\usepackage{cuted}
\usepackage{flushend}
\usepackage{threeparttable}
\usepackage{braket}
\allowdisplaybreaks[4]
\usetikzlibrary{trees}
\usetikzlibrary{decorations.pathmorphing}
\usetikzlibrary{decorations.markings}
\usetikzlibrary{patterns}
\usetikzlibrary{quotes,angles}
\usetikzlibrary{snakes}
\usepackage{mathrsfs}
\usepackage{amsbsy}
\usepackage[normalem]{ulem}
\usepackage{hyperref}
\usepackage[misc]{ifsym}
\usepackage[T1]{fontenc}
\usepackage[utf8]{inputenc}
\usepackage{authblk}

\usetikzlibrary{calc}
\usetikzlibrary{intersections}
\usetikzlibrary{trees}
\usetikzlibrary{decorations.pathmorphing}
\usetikzlibrary{decorations.markings}
\usetikzlibrary{patterns}
\tikzset{
   global scale/.style={
      scale=#1,
      every node/.append style={scale=#1}},
   photon/.style={decorate, decoration={snake}, draw=red},
   nucleon/.style={draw=black, postaction={decorate},
      decoration={markings,mark=at position .55 with{\arrow[draw=black]{>}}}},
   pion/.style={draw=blue, postaction={decorate},
      decoration={markings,mark=at position .55 with{\arrow[draw=blue]{}}}},
    }

\newcommand{\bk}{\boldsymbol{k}}
\newcommand{\bp}{\boldsymbol{p}}

\newcommand{\bq}{\boldsymbol{q}}

\newcommand{\bb}{\boldsymbol{b}}
\newcommand{\bep}{\boldsymbol{\epsilon}}
\newcommand{\bsi}{\boldsymbol{\sigma}}

\allowdisplaybreaks{}

\begin{document}
\title{\Large\textbf{Dispersive Analysis of Low Energy \(\gamma^* N\rightarrow\pi N\) Process}}
\author{Xiong-Hui Cao\(^1\)}
\author{Yao Ma\(^1\)~\thanks{aaron\underline{\ }ma@pku.edu.cn}}
\author{Han-Qing Zheng\(^{1,2}\)~\footnote{Present Address: College of Physics, Sichuan University, Chengdu, Sichuan 610065, Peoples Republic of China}}
\affil{\(^1\)Department of Physics and State Key Laboratory of Nuclear Physics and Technology, \\Peking University, Beijing 100871, Peoples Republic of China}
\affil{\(^2\)Collaborative Innovation Center of Quantum Matter, Beijing 100871, Peoples Republic of China}

\maketitle

\begin{abstract}
We use a dispersion representation based on unitarity and analyticity to study the low energy \(\gamma^* N\rightarrow \pi N\) process in the $S_{11}$ channel.
Final state interactions among the $\pi N$ system are critical to this analysis.
The left-hand part of the partial wave amplitude is imported from $\mathcal{O}(p^2)$ chiral perturbation theory result.
On the right-hand part, the final state interaction is calculated through Omn\`es formula in $S$ wave.
It is found that a good numerical fit can be achieved with only one subtraction parameter, and the eletroproduction experimental data of multipole amplitudes \(E_{0+},\ S_{0+}\) in the energy region below \(\Delta(1232)\) are well described when the photon virtuality $Q^2 \leq 0.1 \mathrm{GeV}^2$.
\end{abstract}

\section{Introduction}

The electromagnetic interactions of nucleon have long been recognized as an important source of information for understanding strong interaction physics~\cite{Chew:1957tf, Adler:1968tw, Amaldi:1979vh, Drechsel:1992pn, Pascalutsa:2006up, Aznauryan:2011qj, Ronchen:2014cna}.
The investigation of pion photoproduction started in the 1950s with the seminal work of Chew
\emph{et al.} (CGLN) \cite{Chew:1957tf}, where the formalism for pion photoproduction on a nucleon target was developed, and fixed-$t$ dispersion relations (DRs) were used as a tool for the analyses of the reaction data. Postulates underlying the DR approach are analyticity, unitarity, and crossing symmetry of a $S$ matrix.
The CGLN formalism was later extended to pion electroproduction~\cite{Fubini:1958zz, Berends:1967vi}, and DR was used in the analyses of the experimental data~\cite{Ball:1961zza, Berends:1967vi, Devenish:1973ta, Crawford:1983yi}.
Based on the recent low energy experiments, partial wave analyses have been performed to study the underlying structure of the reaction amplitudes and describing the properties of the nucleon resonances~\cite{Drechsel:2007if, Doring:2009uc, Gasparyan2010, Ronchen:2014cna}.

Since the 1980s, it has been successful to explore the electroproduction and relevant processes using chiral perturbation theory ($\chi$PT) at low energies~\cite{Bernard:1993bq, Hilt:2013fda, Yao:2018pzc, Yao:2019avf, Navarro:2020zqn}.
For the calculation of loop diagrams, there are several renormalization schemes, which are,  e.g., the heavy-baryon approach in Ref.~\cite{Bernard:1993bq} and the EOMS scheme adopted in Refs.~\cite{Hilt:2013fda, Navarro:2020zqn}, to solve the power-counting breaking problems.
However, $\chi$PT only works well near the threshold and fails at slightly higher energies.
So the unitary method is necessarily adopted in order to suppress the contributions from large energy and recast unitarity of the amplitude.

Some unitarity methods have already been explored (for a recent review, see Ref.~\cite{Yao:2020bxx}).
The couple channel $N/D$ method was used to unitarize $\chi$PT amplitudes in Ref.~\cite{Gasparyan2010}, and the J{\"u}lich model was adopted to study photoproduction and the relevant process in Ref.~\cite{Ronchen:2014cna}.
In this paper, our \(\gamma^* N\rightarrow \pi N\) amplitudes are obtained through the dispersive analysis~\cite{Babelon:1976kv}, in the case we set up with chiral $\mathrm{O}(p^2)$ \(\gamma^* N\rightarrow \pi N\) amplitudes and a \(\pi N\) final state interaction estimated by the Omn\`{e}s solution \cite{Omnes:1958hv} in the single channel approximation.
In order to achieve such a dispersive analysis, efforts have been made in understanding the complicated analytic structure of the amplitudes. 

Based on our dispersion representation,  the multipole amplitudes ($S_{11}E_{0+}$ and $S_{11}S_{0+}$) data from Refs.~\cite{Drechsel:1998hk,Drechsel:2007if,Kamalov:1999hs, Kamalov:2000en, Kamalov:2001qg, Pascalutsa:2006up} below the \(\Delta(1232)\) peak have been fitted.
This work extends our previous analyses on pion photoproduction~\cite{Ma:2020hpe} to the virtual-photon process with a photon virtuality $Q^2$ up to $0.2 \mathrm{GeV}^2$, and finds a good description of the data when $Q^2\leq 0.1\mathrm{GeV}^2$, with only one parameter.
Besides, the comparison between the $\mathcal{O}(p^2)$ calculation in this paper and the one up to $\mathcal{O}(p^4)$ from Ref.~\cite{Hilt:2013fda} is performed and a discrepancy between the two results are noticed in the higher $Q^2$ region.

This paper is organized as follows.
In Sec.~\ref{sec:lhcchpt}, a brief introduction to pion electroproduction is given.
In Sec.~\ref{sec:pwa}, we set up the dispersive formalism for \(\gamma^* N \to \pi N\) process and make an analysis about the singularities which appear in this process.
In Sec.~\ref{sec:fit}, numerical results of multipoles are carried out.
Finally we give our conclusions in Sec.~\ref{sec:Clu}.

\section{Pion electroproduction}\label{sec:lhcchpt}

\subsection{Basics of single pion electroproduction off the nucleon}

In this section we provide a short introduction to the notations describing the electroproduction of pions.
Single pion electroproduction off the nucleon is the process described by
\begin{align}
    e\left(l_{1}\right)+N\left(p_{1}\right) \rightarrow e\left(l_{2}\right)+N\left(p_{2}\right)+\pi^a(q)\ ,
\end{align}
where \(a\) is the isospin index of the pion and $l_1(l_2),\ p_1(p_2),\ \ q$ are incoming (outgoing) electron, incoming (outgoing) nucleon and pion momentum, respectively. 

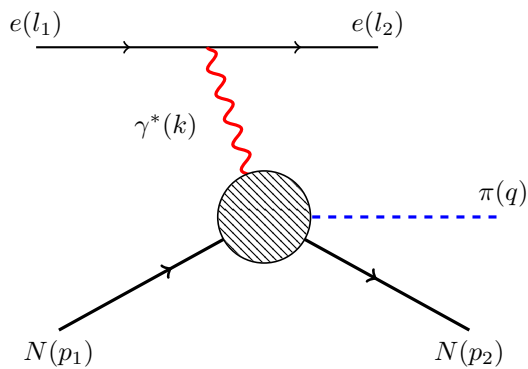
\begin{figure}[H]\centering
        \begin{tikzpicture}[scale = 1.5]
            \draw[nucleon, thick](-2,0)node[above]{$e(l_1)$}to(-0.5,0);
            \draw[nucleon, thick](-0.5,0)to(1,0)node[above]{$e(l_2)$};
            \draw[photon, very thick](-0.5,0)to(0,-1.5);
            \node[left] at (-0.5,-0.7){$\gamma^*(k)$}; 
            \draw[nucleon, very thick](-1.8,-2.5)node[below]{$N(p_1)$}to(0,-1.5);
            \draw[nucleon, very thick](0,-1.5)to(1.8,-2.5)node[below]{$N(p_2)$};
            \draw[pion, dashed, very thick](0,-1.5)to(2.1,-1.5)node[above]{$\pi(q)$};
            \draw[very thick](0,-1.5)circle(0.4);
            
            \fill[white](0,-1.5) circle(0.4);
            \fill[pattern color=black, pattern=north west lines](0,-1.5)circle(0.4);
        \end{tikzpicture}
        \caption{Pion electroproduction in the one-photon-exchange approximation.  $k=l_1-l_2$ represents the momentum of the single exchanged virtual photon. The shaded circle represents the full hadronic vertex.}\label{fig1}
\end{figure}

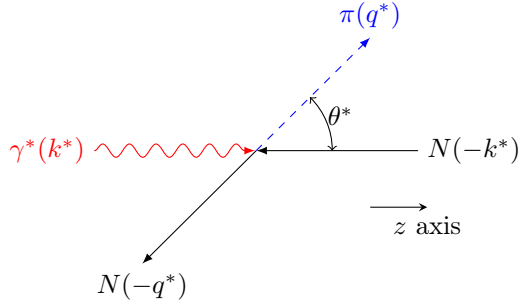
\begin{figure}[H]\centering
        \begin{tikzpicture}[scale = 1.5]
            \draw[-stealth] (1,-0.5) -- (1.5,-0.5)  node[below] {$z$ axis};
            
            \draw[-latex] (1.42,0) coordinate (a) node[right] {$N(-k^*)$}
              -- (0,0) coordinate (b);
            \draw[-latex]  (0,0)
              -- (-1,-1) coordinate (c) node[below] {$N(-q^*)$};
            \draw[-latex, dashed, blue]  (0,0)
              -- (1,1) coordinate (d) node[above] {$\pi(q^*)$};
            \draw[photon,-latex] (-1.42,0 )coordinate (e) node[left] {{\color{red} $\gamma^*(k^*)$}}
              -- (-0.01,0);
            \draw  pic["$\theta^*$", draw=black, <->, angle eccentricity=1.2, angle radius=1cm]
              {angle=a--b--d};
        \end{tikzpicture}
        \caption{Scattering angle $\theta^*$ in the cm frame.}\label{cm}
\end{figure}

Because the interaction between electron and nucleon is pure electromagnetic, for every additional virtual photon exchange, there will be one more fine structure constant $\alpha=e^{2} /(4 \pi) \approx 1 / 137$ suppression factor.
Hence, we can only consider the lowest contribution or the so-called one-photon-exchange approximation; see Fig.~\ref{fig1}.
In this approximation, the invariant amplitude $\mathcal{M}$ is interpreted as the product of the polarization vector $\epsilon_\mu$ of the virtual photon and the hadronic transition current matrix element $\mathcal{M}^\mu$,
\begin{align}\label{eps mu}
\mathcal{M}=\epsilon_{\mu} \mathcal{M}^{\mu}=e \frac{\bar{u}\left(l_{1}\right) \gamma_{\mu} u\left(l_{2}\right)}{k^{2}} \mathcal{M}^{\mu}\ ,
\end{align}
where
\begin{align}
\mathcal{M}^{\mu}=-i e\left\langle N\left(p_{2}\right), \pi(q)\left|J^{\mu}(0)\right| N\left(p_{1}\right)\right\rangle\ ,
\end{align}
with $J^\mu$ the electromagnetic current operator.
Since $k^\mu \epsilon_\mu =0$ in both photoproduction and eletroproduciton, it is possible to separate the pure electromagnetic part of the process from the hadronic part, which is the process,
\begin{align}
\gamma^{*}(k)+N\left(p_{1}\right) \rightarrow N\left(p_{2}\right)+\pi(q)\ ,
\end{align}
where $\gamma^*$ refers to a (spacelike) virtual photon, so we can define $k^2=-Q^2<0$, and $Q^2$ called photon virtuality.
Mandelstam variables $s,\ t, \text{ and } u$ are defined as
\begin{align}
s=\left(p_{1}+k\right)^{2},\quad t=\left(p_{1}-p_{2}\right)^{2},\quad u=\left(p_{1}-q\right)^{2}\ ,
\end{align}
and satisfy $s+t+u=2 m_{N}^{2}+m_{\pi}^{2}-Q^{2}$, where $m_N$ and $m_\pi$ denote the nucleon mass and the pion mass, respectively.
In the center-of-mass (cm) frame, $\pi N$ final state system\footnote{In this section, the superscript $*$ refers to the physical quantity in the cm frame.}, the energies of the photon, $k_0^*$, the pion, $E^*_\pi$ and incoming (outgoing) nucleon, $E_1^*\ (E_2^*)$ are given by
\begin{align}
\begin{aligned}
k_{0}^{*} &=\frac{W^{2}-Q^{2}-m_{N}^{2}}{2 W} ,\quad E_\pi^*=\frac{W^{2}+m_{\pi}^{2}-m_{N}^{2}}{2 W}\ , \\
E_1^* &=\frac{W^{2}+m_{N}^{2}+Q^{2}}{2 W} ,\quad E_2^*=\frac{W^{2}+m_{N}^{2}-m_{\pi}^{2}}{2 W}\ ,
\end{aligned}
\end{align}
where $W=\sqrt{s}$ is the cm total energy. 
The values of the initial and final state momentum in the cm frame are
\begin{align}
    \begin{aligned}
    \left|\bk^{*}\right| &=\sqrt{\left(\frac{W^{2}-m_{N}^{2}-Q^{2}}{2 W}\right)^{2}+Q^{2}}\ , \\
    \left|\bq^{*}\right| &=\sqrt{\left(\frac{W^{2}-m_{N}^{2}+m_{\pi}^{2}}{2 W}\right)^{2}-m_{\pi}^{2}}\ ,
    \end{aligned}
\end{align}
The real photon equivalent energy in laboratory frame $k^\mathrm{lab}$ is given by
\begin{align}
k^\mathrm{lab}=\frac{W^{2}-m_{N}^{2}}{2 m_{N}}\ .
\end{align}
and $k^{\mathrm{cm}}=(m_N/W)k^{\mathrm{lab}}$.
The cm scattering angle $\theta^*$ between the pion three momentum and the $z$ axis, defined by the incoming photon direction, is depicted in Fig.~\ref{cm}

The scattering amplitude of pion electroproduction can be parametrized in terms of the Ball amplitudes \cite{Ball:1961zza}, which are defined in Lorentz-covariant form,
\begin{align}
-i e\left\langle N^{\prime} \pi\left|J^{\mu}(0)\right| N\right\rangle=\bar{u}\left(p_{2}\right)\left(\sum_{i=1}^{8} B_{i} V_{i}^{\mu}\right) u\left(p_{1}\right)\ ,
\end{align}
where $u(p_1$) and $\overline{u}(p_2)$ are the Dirac spinors of the nucleon in the initial and final states, respectively.
Here we use the notation of~\cite{Dennery:1961zz,Berends:1967vi,Bernard:1993bq}, but it is slightly different from \cite{Adler:1968tw,Hilt:2013fda}:
\begin{align}
\begin{aligned} \label{b8}
    V_1^\mu&=\gamma_5\gamma^\mu \slashed{k}\ ,\quad V_2^\mu=2\gamma_5 P^\mu\ , \quad
    V_3^\mu=2\gamma_5q^\mu\ ,\quad V_4^\mu=2\gamma_5 k^\mu\ ,  \\
    V_5^\mu&=\gamma_5\gamma^\mu\ ,\quad V_6^\mu=\gamma_5 P^\mu\slashed{k}\ , \quad
    V_7^\mu=\gamma_5k^\mu \slashed{k}\ ,\quad V_8^\mu=\gamma_5 q^\mu\slashed{k}\ ,
\end{aligned}
\end{align}
where $P=(p_1+p_2)/2$ and $\slashed{k}=\gamma^\mu k_\mu$. Using the electromagnetic current conservation $k_\mu \mathcal{M}^\mu=0$, only six independent amplitudes are required for the description of pion electroproduction. Furthermore, in pion photoproduction ($Q^2 = 0$), only four independent amplitudes survive.

The parameterization of Ref.~\cite{Bernard:1993bq} takes care of current conservation already from the beginning, which contains only six independent amplitudes $A_i$,
\begin{align}\label{A_i}
\mathcal{M}^{\mu}=\bar{u}\left(p_{2}\right)\left(\sum_{i=1}^{6} A_{i} M_{i}^{\mu}\right) u\left(p_{1}\right)
\end{align}
with
\begin{align}
\begin{aligned}
    M_{1}^{\mu} &=\frac{1}{2}\gamma_{5}\left(\gamma^{\mu} \slashed{k}-\slashed{k} \gamma^{\mu}\right)\ , \\ 
    M_{2}^{\mu} &=2\gamma_{5}\left(P^{\mu} k \cdot\left(q-\frac{1}{2} k\right)-\left(q-\frac{1}{2} k\right)^{\mu} k \cdot P\right)\ , \\ 
    M_{3}^{\mu} &=\gamma_{5}\left(\gamma^{\mu} k \cdot q-\slashed{k} q^{\mu}\right)\ , \\ 
    M_{4}^{\mu} &=2\gamma_{5}\left(\gamma^{\mu} k \cdot P-\slashed{k} P^{\mu}\right)-2 m_{N} M_{1}^{\mu}\ , \\ 
    M_{5}^{\mu} &=\gamma_{5}\left(k^{\mu} k \cdot q-k^{2} q^{\mu}\right)\ , \\ 
    M_{6}^{\mu} &=\gamma_{5}\left(k^{\mu}\slashed{k}-k^{2} \gamma^{\mu}\right)\ .
\end{aligned}
\end{align}
Each of them individually satisfies gauge invariance $k_{\mu} M_{i}^{\mu}=0$. 
The scalar functions $A_i$ and $B_i$ can be linked through
\begin{align}
\begin{aligned}\label{AB}
    A_{1} &=B_{1}-m_N B_{6}\ ,\quad A_{2} =\frac{2}{m_{\pi}^{2}-t} B_{2}\ , \quad
    A_{3} =-B_{8}\ ,\quad A_{4} =-\frac{1}{2} B_{6}\ ,\quad A_{6} =B_{7}\ , \\
    A_{5} &=\frac{2}{s+u-2m_N^{2}}\left(B_{1}-\frac{s-u}{2\left(m_{\pi}^{2}-t\right)}     
B_{2}+2 B_{4}\right)=\frac{1}{k^{2}}\left(\frac{s-u}{t-m_{\pi}^{2}} B_{2}-2 B_{3}\right)\ .
\end{aligned}
\end{align}

The CGLN amplitudes $\mathcal{F}_i$ are another common parameterization \cite{Chew:1957tf, Dennery:1961zz}, which plays an important role in experiments and partial wave analyses. These amplitudes are defined in the cm frame via
\begin{align}
\epsilon_{\mu} \bar{u}\left(p_{2}\right)\left(\sum_{i=1}^{6} A_{i} M_{i}^{\mu}\right) u\left(p_{1}\right)=\frac{4 \pi W}{m_{N}} \chi_{2}^{\dagger} \mathbf{F} \chi_{1}\ ,
\end{align}
where $\chi_1$ and $\chi_2$ denote initial and final Pauli spinors, respectively.
Electromagnetic current conservation allows us to work in the gauge where the polarization vector of virtual photon has a vanishing longitudinal component.
In terms of the polarization vector of Eq.~(\ref{eps mu}) this is achieved by introducing the vector \cite{Bernard:1993bq,Davidson:1995jm,Borasoy:2007ku},
\begin{align}
b_\mu=\epsilon_\mu-\frac{\bep\cdot\hat{\bk}}{|\bk|}k_\mu\ ,
\end{align}
where $b_0 \neq 0$, but $\bb\cdot\hat{\bk}=0$ ($\hat{\bk}=\bk/|\bk|$). 
$\mathcal{F}$ may be written as [$\bsi=(\sigma_1,\sigma_2,\sigma_3)$],
\begin{align}
    \mathbf{F}=&i\bsi\cdot\bb \mathcal{F}_1+\bsi\cdot\hat{\bq}\bsi\cdot(\hat{\bk}\times\bb)\mathcal{F}_2+i\bsi\cdot\hat{\bk}\hat{\bq}\cdot\bb \mathcal{F}_3+i\bsi\cdot\hat{\bq}\hat{\bq}\cdot\bb \mathcal{F}_4  \nonumber \\
    &-i\bsi\cdot\hat{\bq}b_0 \mathcal{F}_7-i\bsi\cdot\hat{\bk}b_0 \mathcal{F}_8\ .
\end{align}
We can connect $A_i$ and $\mathcal{F}_i$ through algebraic calculations, and the results can be found in Appendix~\ref{CGLN}.

The CGLN amplitudes can be expanded into multipole amplitudes\cite{Dennery:1961zz},
\begin{align}\label{F8}
\begin{aligned}
\mathcal{F}_{1} &=\sum_{l=0}^{\infty}\left\{\left[l M_{l+}+E_{l+}\right] P_{l+1}^{\prime}(x)+\left[(l+1) M_{l-}+E_{l-}\right] P_{l-1}^{\prime}(x)\right\}\ , \\
\mathcal{F}_{2} &=\sum_{l=1}^{\infty}\left\{(l+1) M_{l+}+l M_{l-}\right\} P_{l}^{\prime}(x)\ , \\
\mathcal{F}_{3} &=\sum_{l=1}^{\infty}\left\{\left[E_{l+}-M_{l+}\right] P_{l+1}^{\prime \prime}(x)+\left[E_{l-}+M_{l-}\right] P_{l-1}^{\prime \prime}(x)\right\}\ , \\
\mathcal{F}_{4} &=\sum_{l=2}^{\infty}\left\{M_{l+}-E_{l+}-M_{l-}-E_{l-}\right\} P_{l}^{\prime \prime}(x)\ , \\
\mathcal{F}_{7} &=\sum_{l=1}^{\infty}\left[l S_{l-}-(l+1) S_{l+}\right] P_{l}^{\prime}(x)=\frac{\left|\bk^{*}\right|}{k_{0}^{*}} \mathcal{F}_{6}\ , \\
\mathcal{F}_{8} &=\sum_{l=0}^{\infty}\left[(l+1) S_{l+} P_{l+1}^{\prime}(x)-l S_{l-} P_{l-1}^{\prime}(x)\right]=\frac{\left|\bk^{*}\right|}{k_{0}^{*}} \mathcal{F}_{5}\ ,
\end{aligned}
\end{align}
with $x=\cos\theta=\hat{\bq} \cdot \hat{\bk}$, $P_l(x)$ the Legendre polynomial of degree $l$, $P^\prime_l=\mathrm{d}P_l/\mathrm{d}x$ and so on. 
Subscript $l$ denotes the orbital angular momentum of the pion-nucleon system in the final state.
The multipoles $E_{l \pm}, M_{l \pm}, \text{ and } S_{l \pm}$ are functions of the cm total energy $W$ and the photon virtuality $Q^2$, and refer to transversal electric, magnetic transitions, and scalar transitions, \footnote{Sometimes the longitudinal multipoles are used instead of the scalar multipoles, they satisfies a relationship $L_{l\pm}=(k_0/|\bk|)S_{l\pm}$} respectively.
The subscript $l_\pm$ denotes the total angular momentum $j = l \pm 1/2$ in the final state.
By inverting the above equations, the angular dependence can be completely figured out\cite{Berends:1967vi},
\begin{align}
\begin{aligned}
    E_{l+} &= \int_{-1}^{1} \frac{d x}{2(l+1)}\left[P_{l} \mathcal{F}_{1}-P_{l+1} \mathcal{F}_{2}+\frac{l}{2 l+1}\left(P_{l-1}-P_{l+1}\right) \mathcal{F}_{3}+\frac{l+1}{2 l+3}\left(P_{l}-P_{l+2}\right) \mathcal{F}_{4}\right]\ ,  \\
    E_{l-} &= \int_{-1}^{1} \frac{d x}{2 l}\left[P_{l} \mathcal{F}_{1}-P_{l-1} \mathcal{F}_{2}-\frac{l+1}{2 l+1}\left(P_{l-1}-P_{l+1}\right) \mathcal{F}_{3}+\frac{l}{2 l-1}\left(P_{l}-P_{l-2}\right) \mathcal{F}_{4}\right]\ ,  \\
    M_{l+} &=\int_{-1}^{1} \frac{d x}{2(l+1)}\left[P_{l} \mathcal{F}_{1}-P_{l+1} \mathcal{F}_{2}-\frac{1}{2 l+1}\left(P_{l-1}-P_{l+1}\right) \mathcal{F}_{3}\right]\ , \\
    M_{l-} &=\int_{-1}^{1} \frac{d x}{2 l}\left[-P_{l} \mathcal{F}_{1}+P_{l-1} \mathcal{F}_{2}+\frac{1}{2 l+1}\left(P_{l-1}-P_{l+1}\right) \mathcal{F}_{3}\right]\ , \\
    S_{l+} &=\int_{-1}^{1} \frac{d x}{2(l+1)}\left[P_{l+1} \mathcal{F}_{7}+P_{l} \mathcal{F}_{8}\right]\ , \\
    S_{l-} &=\int_{-1}^{1} \frac{d x}{2 l}\left[P_{l-1} \mathcal{F}_{7}+P_{l} \mathcal{F}_{8}\right]\ .
\end{aligned}
\end{align}
Please refer to Appendix~\ref{ap:Helicity} for the connections between 
multipoles and partial wave helicity amplitudes.

The isospin structure of the scattering amplitude can be written as
\begin{align}
    A(\gamma^*+N \to \pi^a+N^\prime)=\chi_2^\dagger \bigg \{ \delta^{a3} A^{(+)} +i \epsilon^{a 3 b} \tau^b A^{(-)} +\tau^a A^{(0)}\bigg \} \chi_1 \ ,
\end{align}
where \(\tau_a\) (\(a=1,2,3\)) are Pauli matrices. 
We can define the isospin transition amplitudes by $A^{I, I_3}(A^{\frac{3}{2}, \pm \frac{1}{2}},A^{\frac{1}{2}, \pm \frac{1}{2}})$, where $\{I, I_3\}$ denote  isospin of the final $\pi N$ system.
In the notation $\ket{I, I_3}$, the isospin part of the state vectors for the nucleon and the pion is written as
\begin{gather}
|p\rangle=\left|\frac{1}{2},+\frac{1}{2}\right\rangle\ , \quad|n\rangle=\left|\frac{1}{2},-\frac{1}{2}\right\rangle\ , \\
\left|\pi^{+}\right\rangle=-|1,+1\rangle\ , \quad\left|\pi^{0}\right\rangle=|1,0\rangle\ , \quad\left|\pi^{-}\right\rangle=|1,-1\rangle\ .
\end{gather}
So isospin transition amplitudes can be obtained from \(A^{(\pm)}\) and \(A^{(0)}\) via
\begin{align}
A^{\frac{3}{2}, \frac{1}{2}} &=A^{\frac{3}{2},-\frac{1}{2}}=\sqrt{\frac{2}{3}}\left(A^{(+)}-A^{(-)}\right)\ , \\
A^{\frac{1}{2} , \frac{1}{2}} &=-\sqrt{\frac{1}{3}}\left(A^{(+)}+2 A^{(-)}+3 A^{(0)}\right)\ , \\
A^{\frac{1}{2},-\frac{1}{2}} &=\sqrt{\frac{1}{3}}\left(A^{(+)}+2 A^{(-)}-3 A^{(0)}\right)\ .
\end{align}

In the one-photon-exchange approximation, the differential cross section can be factorized as\cite{Amaldi:1979vh,Drechsel:1992pn}
\begin{align}
\frac{\mathrm{d} \sigma}{\mathrm{d} \mathcal{E}_2 \mathrm{d} \Omega_{l} \mathrm{d} \Omega_{\pi}^{*}}=\frac{\alpha}{2 \pi^{2}} \frac{\mathcal{E}_2}{\mathcal{E}_1}\frac{1}{Q^{2}} \frac{k^{\mathrm{lab}}}{1-\epsilon} \frac{\mathrm{d} \sigma_{v}}{\mathrm{d} \Omega_{\pi}^{*}} \equiv \Gamma \frac{\mathrm{d} \sigma_{v}}{\mathrm{d} \Omega_{\pi}^{*}}\ ,
\end{align}
where $\Gamma$ is the flux of the virtual photon, $\mathcal{E}_{1,2}$ denote the energy of the initial and final electrons in the laboratory frame, respectively.
The parameter $\epsilon$ expresses the transverse polarization of the virtual photon in the laboratory frame, and it is an invariant under collinear transformations.
In terms of laboratory electron variables, it is given by \cite{Amaldi:1979vh}
\begin{align}
\epsilon= \left(1+2 \frac{\bk^{2}}{Q^{2}} \tan ^{2}\left(\frac{\theta_{l}}{2}\right)\right)^{-1}\ ,
\end{align}
where $\theta_l$ is the scattering angle of the electron in the laboratory frame.
The virtual photon differential cross section, $\mathrm{d} \sigma_{v} / \mathrm{d} \Omega_{\pi}^*$, for an unpolarized target without recoil polarization can be written in the form \cite{Drechsel:1992pn},
\begin{align}
\begin{aligned}\label{sigpi}
\frac{\mathrm{d} \sigma_{v}}{\mathrm{d} \Omega_{\pi}^*}=& \frac{\mathrm{d} \sigma_{T}}{\mathrm{d} \Omega_{\pi}^*}+\epsilon \frac{\mathrm{d} \sigma_{L}}{\mathrm{d} \Omega_{\pi}^*}+\sqrt{2 \epsilon(1+\epsilon)} \frac{\mathrm{d} \sigma_{L T}}{\mathrm{d} \Omega_{\pi}^*} \cos \phi_{\pi}^*+\epsilon \frac{\mathrm{d} \sigma_{T T}}{\mathrm{d} \Omega_{\pi}^*} \cos 2 \phi_{\pi}^* \\
&+h \sqrt{2 \epsilon(1-\epsilon)} \frac{\mathrm{d} \sigma_{L T^{\prime}}}{\mathrm{d} \Omega_{\pi}^*} \sin \phi_{\pi}^*+h \sqrt{1-\epsilon^{2}} \frac{\mathrm{d} \sigma_{T T^{\prime}}}{\mathrm{d} \Omega_{\pi}^{*}}\ ,
\end{aligned}
\end{align}
in which $\phi_\pi^*$ is the azimuthal angle of pion and $h$ is the helicity of the incoming electron.
For further details about Eq.~(\ref{sigpi}), especially concerning polarization observables, we refer to Ref.~\cite{Drechsel:1992pn}.
If we integrate the dependence of azimuthal angle, at the end, we will get
\begin{align}\label{signoomig}
\sigma_v=\sigma_T+\epsilon \sigma_L\ .
\end{align}

In the following chapter, we will introduce $\chi$PT as an effective field theory which  allow us to calculate pion production.
The upper limit for the cm total energy $W$, restricted by the fact that we only consider pion and nucleon degrees of freedom, is below the $\Delta (1232)$ resonance peak.
Furthermore, through the experience gained by studying EM form factors\cite{Kubis:2000aa, Kubis:2000zd}, the estimate of the upper limit of momentum transfers is $Q^2\simeq 0.1 \mathrm{GeV}^2$ in $\chi$PT \cite{Hilt:2013fda, Tiator:2016btt}.

\section{Partial wave amplitudes}\label{sec:pwa}

\subsection{$\chi$PT amplitudes and unitarity method}\label{chpt}

We recalculated the pion electroproduction process close to the threshold using $\chi$PT up to \(\mathcal{O}(p^2)\) and confirm the results of~\cite{Bernard:1993bq}.
The invariant scalar functions can be extracted from full amplitudes.
The results are listed in the Appendix~\ref{ap:amp}.
for higher order $\mathcal{O}(p^3)$ contributions and the influence of $\Delta(1232)$ resonance, readers can refer to Ref.~\cite{Navarro:2020zqn}.

In the following part, superscripts and subscripts $I,~J$ (isospin, total angular momentum) are ignored for brevity.
Considering the final-state theorem\cite{Watson:1954uc} and using the dispersion relation, the unitarized $S$ wave amplitude can be written as \cite{Babelon:1976kv, Mao2009, GarciaMartin:2010cw, Moussallam:2013una, Danilkin:2012ua, Danilkin:2018qfn, Danilkin:2019opj, Hoferichter:2019nlq, Ma:2020hpe}
\begin{align}\label{eq:disRep}
    \mathcal{M}(s)=\mathcal{M}_L(s)+\Omega(s)\left(-\frac{s}{\pi}\int_{(m_\pi +m_N)^2}^{\infty}\frac{\big({\rm Im}\ \Omega(s^\prime)^{-1}\big)\mathcal{M}_L(s^\prime)}{s'(s'-s)}{\rm d} s'+\mathcal{P}(s)\right)\ ,
\end{align}
where \(\mathcal{P}(s)\) is subtraction polynomial.
The amplitude \(\mathcal{M}_L\) only contains left-hand cut singularity.
Thus, the pion electroproduction amplitude \(\mathcal{M}(s)\) is determined up to a polynomial.
\(\Omega(s)\) is the so-called Omn\`es function~\cite{Omnes:1958hv},
\begin{align}\label{eq:omnes}
    \Omega(s)=\tilde{\mathcal{P}}(s)\exp\bigg[\frac{s}{\pi}\int_{(m_\pi +m_N)^2}^\infty\frac{\delta(s^\prime)}{s^\prime(s^\prime-s)}{\rm d}s^\prime\bigg]
\end{align}
with \(\tilde{\mathcal{P}}\) representing a polynomial, reflecting the zeros of $\Omega(s)$ in the complex plane and \(\delta^I_{J}(s)\) being the elastic \(\pi N\) partial wave phase shift.

For our calculation, we use the $\chi$PT result to estimate $\mathcal{M}_L$, so as long as function $\Omega(s)$ is known, we can get the amplitude with correct unitarity and analyticity property.

\subsection{Singularity structure of partial wave amplitudes}

Applicability of the Omn\`es method to the amplitudes of interest relies on the ability
to separate the amplitude into a piece having only a left-hand cut and a piece having only a right-hand one. This, \emph{a priori}, is not the case if the left-hand cuts overlapped with the unitary cut.

So we review the analytic structures arising in our calculation and find that the singularities in this virtual process are rather more complicated than real photoproduction. 
There will be some additional cuts in the complex $s$ plane, compared with the photoproduction one.
We follow Ref.~\cite{Kennedy:1962sing} which relies on the Mandelstam double spectral representation to illustrate the analytic structure of the partial wave amplitudes.
According to crossing symmetry, one amplitude can simultaneously describe the three channels of $s,\ t,\ u$,
\begin{align}
    \begin{aligned}
    &s:\qquad \gamma^*+N \to \pi+N^\prime\ , \quad \sigma_1=M^2,\quad\rho_1=(m+M)^2\ ;  \\
    &t:\qquad \gamma^*+\pi \to \overline{N}+N^\prime\ , \quad \sigma_2=m^2,\quad\rho_2=4m^2\ ;  \\
    &u:\qquad \gamma^*+\overline{N}^\prime \to \pi+\overline{N}\ , \quad \sigma_3=M^2,\quad\rho_3=(m+M)^2\ .
    \end{aligned}
\end{align}
Here, for brevity, we define $m=m_\pi,\ M=m_N$, $\sigma_i$ represent the the mass squares of strongly interacting intermediate bound states, and the continuous spectra will begin at $\rho_i$ which is the threshold of two particle intermediate states.
Note here that the Mandelstam variable \(t\) defined in the $s$ plane is related to $z_s=\cos\theta$ via
 \begin{equation}\label{eq_rhot}
    \begin{aligned}
    t=& -Q^{2}+m^{2}-\frac{\left(s-Q^{2}-M^{2}\right)\left(s+m^{2}-M^{2}\right)}{2 s} \\
    &+\left\{\left[s^{2}+2\left(Q^{2}-M^{2}\right) s+\left(Q^{2}+M^{2}\right)^{2}\right]\left(s-s_L\right)\left(s-s_R\right)\right\}^{\frac{1}{2}} \frac{z_{s}}{2 s}\ .
    \end{aligned}
\end{equation}
where $s_L=(m-M)^2,s_R=(m+M)^2$, and we can define $\nu=(s-u)/(4m_N)$ as 
the crossing symmetric variable. The physical $s,u$-channel region is shown in 
the following for $Q^2=0.1\mathrm{GeV}^2$.
The threshold for $\pi$ electroproduction lies at
\begin{align}
  \begin{aligned}
    \nu_{\mathrm{thr}} &=\frac{m_{\pi}\left[\left(2 m_{N}+m_{\pi}\right)^{2}+Q^{2}\right]}{4 m_{N}\left(m_{N}+m_{\pi}\right)}\ , \\
    t_{\mathrm{thr}} &=-\frac{m_{N}\left(m_{\pi}^{2}+Q^{2}\right)}{m_{N}+m_{\pi}}\ .
  \end{aligned}
\end{align}

\begin{figure}[H]\centering
     \begin{tikzpicture}[>=stealth,scale=1,line width=0.8pt]
    \pgfmathsetmacro{\ticker}{0.125} 
    \coordinate (A) at (0,0);
    \coordinate (B) at (0,6);
    \coordinate (C) at (4,6);
    \coordinate (D) at (4,0);
    \draw(A)--(B)--(C)--(D)--cycle;
    \coordinate [label=left:\rotatebox{90}{$t(\mathrm{GeV}^2)$}](E) at ($(B)+(-0.4,-0.2)$);
    \coordinate [label=below:$\nu(\mathrm{GeV})$](F) at ($(D)+(-0.2,-0.4)$);
    \foreach \i  in {1,2,3,4}
    {
    \draw (1*\i,0) --(1*\i,\ticker);
    }
    \foreach \j  in {2,4,6}
    {
    \draw (0,1*\j) --(\ticker,1*\j);
    }
    \coordinate [label=below:$-1$](a) at ($(0,0)$);
    \coordinate [label=below:$-0.5$](b) at ($(1,0)$);
    \coordinate [label=below:$0$](c) at ($(2,0)$);
    \coordinate [label=below:$0.5$](d) at ($(3,0)$);
    \coordinate [label=below:$1$](e) at ($(4,0)$);
    
    \coordinate [label=left:$-0.2$](f) at ($(0,0)$);
    \coordinate [label=left:$-0.1$](g) at ($(0,2)$);
    \coordinate [label=left:$0$](h) at ($(0,4)$);
    \coordinate [label=left:$0.1$](i) at ($(0,6)$);
    \foreach \i  in {0.2,0.4,0.6,0.8,1.2,1.4,1.6,1.8,2.2,2.4,2.6,2.8,3.2,3.4,3.6,3.8} 
    {
    \draw (1*\i,0) --(1*\i,0.8*\ticker);
    }
    \foreach \j  in {0.4,0.8,1.2,1.6,2.4,2.8,3.2,3.6,4.4,4.8,5.2,5.6} 
    {
    \draw (0,1*\j) --(0.8*\ticker,1*\j);
    }
    \draw[line width=1.5pt,red,name path = pathcurve1] ($(4,3.8)$)..controls ($(2,3.5)$)and ($(2,2.5)$)..($(2.4,0)$);
    \draw[line width=1.5pt,blue,name path = pathcurve2] ($(0,3.8)$)..controls ($(2,3.5)$)and ($(2,2.5)$)..($(1.6,0)$);
    \coordinate (G) at (4,2);
    \coordinate (F) at (0,2);
    \path[name path = pathGLine](F)--(G);
    \path[draw,fill,name intersections={of = pathGLine and pathcurve1,by=H}](H)circle(2pt);
    
    \draw ($(0,4)$)--($(4,4)$);
    \draw[green,dashed, very thick] ($(0,4.3)$)--($(4,4.3)$);
    \draw[green,dashed, very thick] ($(1.85,0)$)--($(2.15,6)$);
    \draw[green,dashed, very thick] ($(2.15,0)$)--($(1.85,6)$);
    \draw ($(2,0)$)--($(2,6)$);
    
    \coordinate [label=above:${t=t_{\mathrm{thr}}}$](I) at ($(3,2)$);
    \coordinate [label=above:${s=m_N^2}$](L) at ($(2.7,5)$);
    \coordinate [label=above:${u=m_N^2}$](L) at ($(1.3,5)$);
    \coordinate [label=above:${t=m_\pi^2}$](L) at ($(3,4.3)$);
    \end{tikzpicture}
    \caption{ The Mandelstam plane for $\pi$ electroproduction off the nucleon: The red line shows the 
    boundary of $s$ channel physical region for $Q^2=0.1\mathrm{GeV}^2$. The blue line 
    corresponds to the physical region boundary of $u$ channel process.
    The nucleon and $\pi$ pole positions are indicated by the dotted green lines 
    $s=m_N^2,u=m_N^2$, and $t=m_\pi^2$. The threshold of $\pi$ electroproduction is 
    represented by solid black circle.}\label{Mandelstam}
\end{figure}
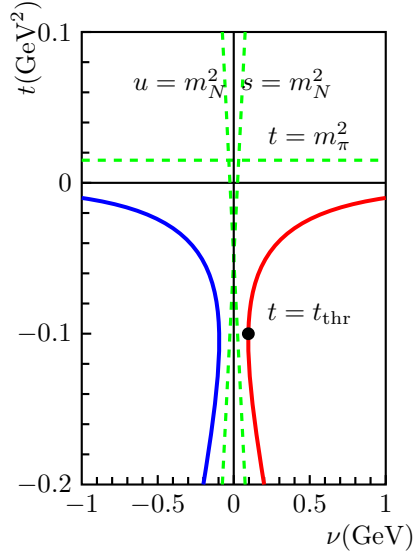

With regard to dynamical cut positions of the partial wave $T$ matrix, we first take the $t$ channel as an illustration.
The full amplitude can be written as a dispersion integral,
\begin{align}
T(s, t)=\int_{\sigma_{2}}^{\infty} \frac{\mathcal{F}\left(s, t^{\prime}\right)}{t^{\prime}-t} d t^{\prime}\ ,
\end{align}
where $\mathcal{F}$ is a spectral function. 
The partial wave amplitude is the projection of the full amplitude onto a rotation function $d^J$,
\begin{align}
T^{J}(s) &=\int_{-1}^{1} \mathrm{d} z_{s} d^{J}\left(z_{s}\right) \int_{\sigma_{2}}^{\infty} \mathrm{d} t^{\prime}\frac{\mathcal{F}\left(s, t^{\prime}\right)}{t^{\prime}-t\left(s, z_{s}\right)} \nonumber\\
&=\int_{\sigma_{2}}^{\infty} \mathrm{d} t^{\prime} \mathcal{F}\left(s, t^{\prime}\right) \int_{-1}^{1} \mathrm{d} z_{s} \frac{d^{J}\left(z_{s}\right)}{\alpha\left(t^{\prime}, s\right)-\beta(s) z_{s}}\ ,
\end{align}
where the integration $\int_{\sigma_{2}}^{\infty}$ denotes the sum of the value at pole $t^\prime=\sigma_2$ and $\int_{\rho_{2}}^{\infty}$.
It can be proven that the final singularity only comes from the form in a logarithmic function,
\begin{align}
\ln (\alpha+\beta)-\ln (\alpha-\beta)\ .
\end{align}

We classify all cuts as follows:
\begin{itemize}
\item {unitarity cut:} \(s\in[s_R,\infty)\) on account of the \(s\)-channel continuous spectrum;
\item \(t\)-channel cut: 1. the arc, on the left of $s=s_c$, stems from \(t\)-channel continuous spectrum for \(4m^2\leq t \leq 4M^2\); 2. $s\in(-\infty,0]$, corresponding to $t$-channel continuous spectrum for $t\geq 4M^2$;
\item \(u\)-channel cut:  \(s\in(-\infty,s_u]\) with \(s_u=\frac{M^{3}-m^{2} M-m\left(M^{2}+Q^{2}\right)}{m+M}\) due to the \(u\)-channel continuous spectrum for \({u\geq(m+M)}^2\);
\item $t$-channel cut from pion pole: due to $t$ channel single pion exchange,  and the branch points locate at $0, C_t, C_t^\dagger$;
\item $u$-channel cut from nucleon pole: due to $t$ channel single nucleon exchange,  and the branch points located at $0, C_u, C_u^\dagger$,
\end{itemize}
where the branch points in the complex plane are (the other three cases are symmetric about the real axis)
    \begin{align}
    C_t&=M^2-\frac{Q^2}{2}+i\sqrt{4M^2Q^2-m^2Q^2-\frac{Q^4}{4}+\frac{M^2}{m^2}Q^4} \ ,  \\
    C_u&=M^2-\frac{1}{2}\frac{m^2}{M^2}Q^2+i\sqrt{4m^2Q^2-\frac{m^2}{M^2}m^2Q^2+\frac{m^2}{M^2}Q^4-\frac{1}{4}\left(\frac{m^2}{M^2}\right)^2Q^4}\ .
    \end{align}
The singularities caused by the pole exchanges of $t, u$ channels are complicated but definitely separated from the unitarity cut.

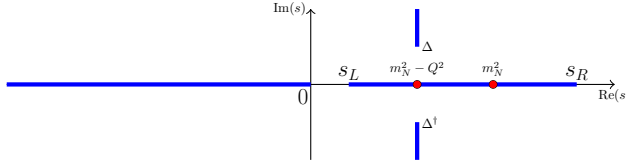
\begin{figure}[H]\centering
    \begin{tikzpicture}[global scale = 0.5]
        \draw[->](-8,0)to(8,0);
        \draw[->](0,-2)to(0,2);

        \draw[ultra thick, blue](-8,0)to (0,0);        

        \draw[ultra thick, blue](1,0)to (7,0);
        \node[above] at (7,0){\LARGE\(s_R\)};     
        \node[above] at (1,0){\LARGE\(s_L\)};        
        
        \draw (4.8,0) circle (0.1);
        \fill[red](4.8,0) circle (0.1);
        \node[above] at (4.8,0.1){\(m_N^2\)};
        
        \draw (2.8,0) circle (0.1);
        \fill[red](2.8,0) circle (0.1);
        \node[above] at (2.8,0.1){\(m_N^2-Q^2\)};
        
        \draw[ultra thick, blue](2.8,1)to (2.8,2);
        \draw[ultra thick, blue](2.8,-1)to (2.8,-2);
        \node[right] at (2.8,1){\(\Delta\)};
        \node[right] at (2.8,-1){\(\Delta^\dagger\)};
        \node[below] at (8,0){\(\mathrm{Re}(s)\)};
        \node[left] at (0,2){\(\mathrm{Im}(s)\)};
        
        \node[below] at (-0.2,0){\LARGE\(0\)};
    \end{tikzpicture}
    \caption{Kinematical singularities, where $\Delta=2MQ$. The red dot represents the nucleon pole, and the two vertical solid rays represent the kinematic cuts from $\sqrt{s^{2}+2\left(Q^{2}-M^{2}\right) s+\left(Q^{2}+M^{2}\right)^{2}}$.}\label{kinsing}
\end{figure}

Aside from the above dynamical singularities, there exist additional kinematical singularities from relativistic kinematics and polarization spinor of fermions, especially in an inelastic scattering process.
These inelastic ones will naturally introduce some square-root functions in the partial wave amplitudes (or multipole amplitudes) which will cause the kinematical singularities.
They provide some of the most obvious characteristics in the case of relativistic theory.
Here kinematical cuts are introduced when the arguments of the square-root functions from Eq.~(\ref{eq_rhot}) are negative.
All the involved arguments together with their corresponding domains with their variable less than zero are listed in Table~\ref{sinre}. 
    \begin{table}[H]\small
        \centering
        \caption{Arguments causing singularities}\label{sinre}
        \begin{threeparttable}   
            \begin{tabular}{@{}ll}
                \toprule
                Arguments&Domain\\
                \midrule
                \(s\)&\(\left(-\infty,0\right)\) \\
                \midrule
                \(s-s_R\)&\(\left(-\infty,s_R\right)\) \\
                \midrule
                \(s-s_L\)&\(\left(-\infty,s_L\right)\) \\
                \midrule
                \(s^{2}+2\left(Q^{2}-M^{2}\right) s+\left(Q^{2}+M^{2}\right)^{2}\)&\((M^2-Q^2 \pm 2iMQ, M^2-Q^2 \pm i\infty)\) \\
                \bottomrule
            \end{tabular}
        \end{threeparttable}
    \end{table}
There is some arbitrariness when fixing the cut position~\cite{Doring:2009yv, Ceci:2011ae}.
For example, compare $\sqrt{\left(s-s_L\right)\left(s-s_R\right)}$ and $\sqrt{s-s_L} \sqrt{s-s_R}$; they may correspond to different cut structure.
The former will have an extra cut, which is perpendicular to the real axis and passes the midpoint of $s_L$ and $s_R$.
So we choose the latter one to make sure that left cuts are lying on the real axis.
In addition, there is a pole like singularity at \(M^2\) that comes from the fact that Eq.~(\ref{F8}) has the $1/(s-M^2)$ term (See Appendix~\ref{CGLN}), which will appear in partial wave amplitudes.
Finally, there may be a pole derived from the gauge invariant amplitudes. 
Relations (\ref{AB}) have introduced the $1/(t-m^2)$ pole singularity, if we consider the partial wave integral, e.g.,
\begin{align}
\int\mathrm{d}z\frac{z}{\left(t(z)-m^2\right)\left(u(z)-M^2\right)} \propto \int\mathrm{d}z\frac{z}{(a+bz)(c-bz)} \propto \frac{1}{a+c} \propto \frac{1}{s-M^2+Q^2}\ ,
\end{align}
where $a=-m^2M^2+M^4-m^2Q^2+M^2Q^2+m^2s-2M^2s+Q^2s+s^2,~c=m^2M^2-M^4+m^2Q^2-M^2 Q^2-m^2s+Q^2s+s^2,~b=\sqrt{s-s_L} \sqrt{s-s_R} \sqrt{s^{2}+2\left(Q^{2}-M^{2}\right) s+\left(Q^{2}+M^{2}\right)^{2}}$.
So the possible additional singularities in our partial wave analysis are displayed in Fig.~\ref{kinsing}.

For a certain channel we are considering, these singularities may not all appear due to the cancellation from linear combinations. Therefore, it must be analyzed in detail when it is used.

\section{Numerical analyses}\label{sec:fit}

We are now in the position to compare the unitary representation of the virtual photoproduction amplitude given in Eq.~(\ref{eq:disRep}) with experimental multipole amplitude data in the \(S_{11}\) channel.
Here we use MAID2007~\cite{Drechsel:1998hk,Drechsel:2007if} and DMT2001~\cite{Kamalov:1999hs, Kamalov:2000en, Kamalov:2001qg, Pascalutsa:2006up} results for the fitting.
These models provide a good description to multipole amplitudes, differential cross sections as well as polarization observables.
They can be used as the basis for the prediction and the analysis of meson photo- and electroproduction data on proton and neutron targets.

\subsection{Fitting procedure}

In the fit, unknown parameters include the low energy constants (LECs) that appeared in \(\mathcal{M}_{\chi PT}(s)\), the subtraction constants in the auxiliary function \(\Omega(s)\) and the ones in the subtraction polynomial \(\mathcal{P}(s)\).
However, the parameters in chiral lagrangian appearing in \(\mathcal{M}_{\chi PT}(s)\) up to $\mathcal{O}(p^2)$ are well fixed. They are \(m_N=0.9383~{\rm GeV}\), \(m_{\pi}=0.1396~{\rm GeV}\), \(e=\sqrt{4\pi\alpha}=0.303\), \(g_A=1.267\), \(F_{\pi}=0.0924~{\rm GeV}\), \(c_6={3.706}/{(4m_N)}\), and \(c_7={-0.12}/{(2m_N)}\)~\cite{PDG2018}\footnote{Neglecting $\chi$PT correction beyond tree level, the two LECs \(c_6\) and \(c_7\) can be related to the anomalous magnetic moments of the nucleon via $c_6=(k_p+k_n)/2m_N,\quad c_7=(k_p-k_n)/4m_N$, with \(k_p\) and \(k_n\) being anomalous magnetic moments of proton and neutron, respectively. Since $k_p$ and $k_n$ are precisely determined by experiments, one can infer the uncertainties of $c_6$ and $c_7$ must be negligible and shall hardly change our results.}.
Hence, \(\mathcal{M}_{\chi PT}(s)\) is parameter free.
Further, we set \(\tilde{\mathcal{P}}(s)=1\) and compute \(\Omega(s)\) by using the partial wave phase shift extracted from the \(\pi N\) \(S\) matrix given in Ref.~\cite{Hoferichter:2015hva}.
Note that it should be a good approximation for a single channel case that the integrations in Eqs. (\ref{eq:omnes}) and (\ref{eq:disRep}) are performed up to \(2.1\rm GeV^2\) (below the $\eta N$ threshold).
Lastly, the subtraction polynomial $\mathcal{P}$ is taken to be a constant, \(\mathcal{P}(s)=a(Q^2)\), i.e., here we only consider once subtraction. \footnote{ The influence of twice subtractions is also examined to test the fit result. It is found that the major physical outputs are almost inert.}
We further assume $a$ to be independent of $Q^2$, since there is no nearby resonance involved.\footnote{Discussions on the similar issue in the mesonic sector can be found in Ref~\cite{Moussallam:2013una}.}
As a reasonable assumption, we do not take into account $Q^2$ dependent subtraction constants in the following.
The above fit method is simultaneously performed on the data from the MAID and DMT models.

\begin{figure}[H]
\centering
\subfigure{
    \begin{minipage}[b]{0.985\linewidth}
    
    \begin{minipage}[b]{0.245\linewidth}
    \includegraphics[width=3.63cm]{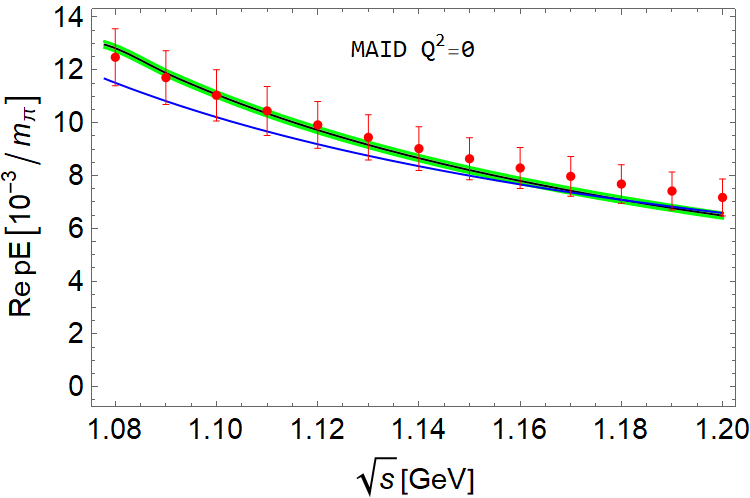}
    \vspace{0.02cm}
    \hspace{0.02cm}
    \end{minipage}
    \begin{minipage}[b]{0.245\linewidth}
    \includegraphics[width=3.63cm]{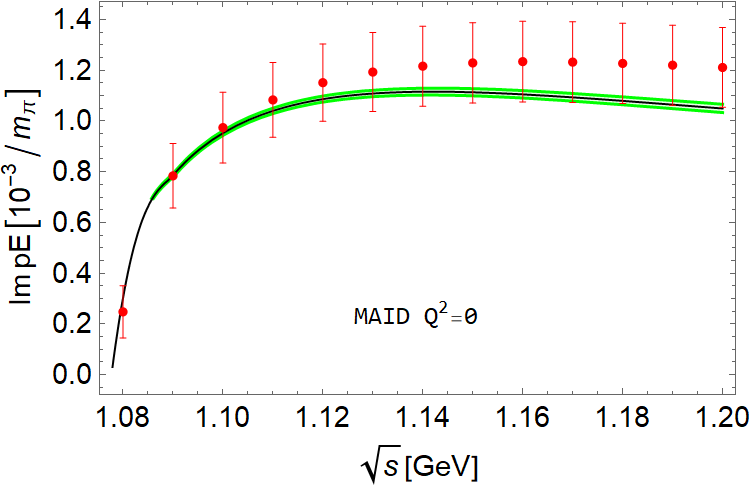}
    \vspace{0.02cm}
    \hspace{0.02cm}
    \end{minipage}
    \begin{minipage}[b]{0.245\linewidth}
    \includegraphics[width=3.63cm]{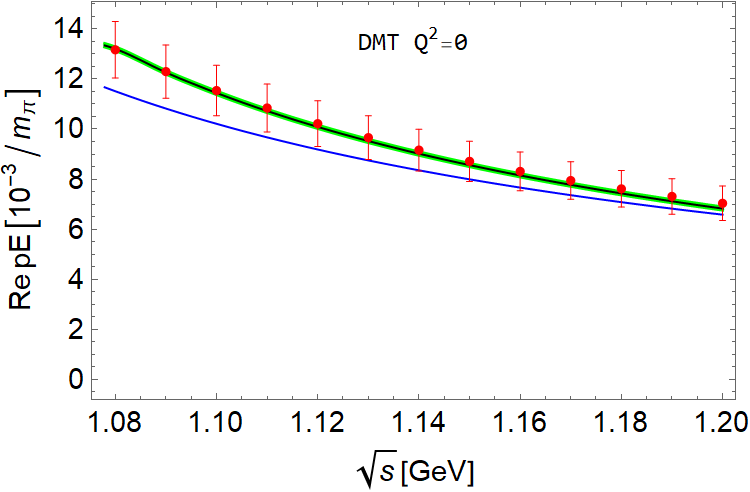}
    \vspace{0.02cm}
    \hspace{0.02cm}
    \end{minipage}
    \begin{minipage}[b]{0.245\linewidth}
    \includegraphics[width=3.63cm]{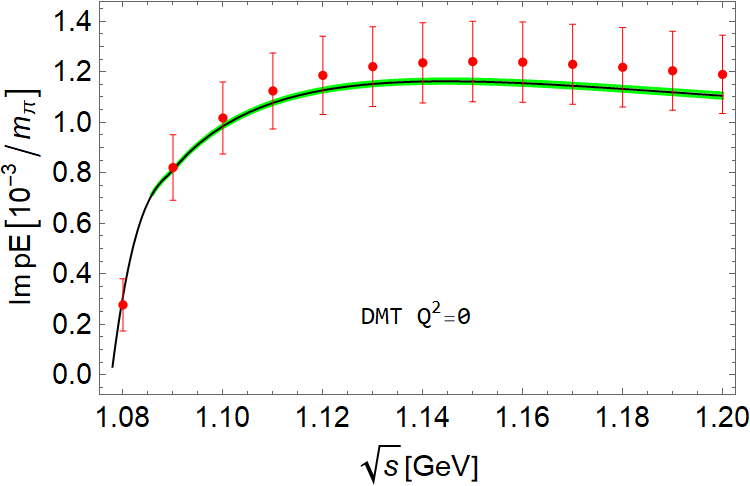}
    \vspace{0.02cm}
    \hspace{0.02cm}
    \end{minipage}
    
    \begin{minipage}[b]{0.24\linewidth}
    \includegraphics[width=3.63cm]{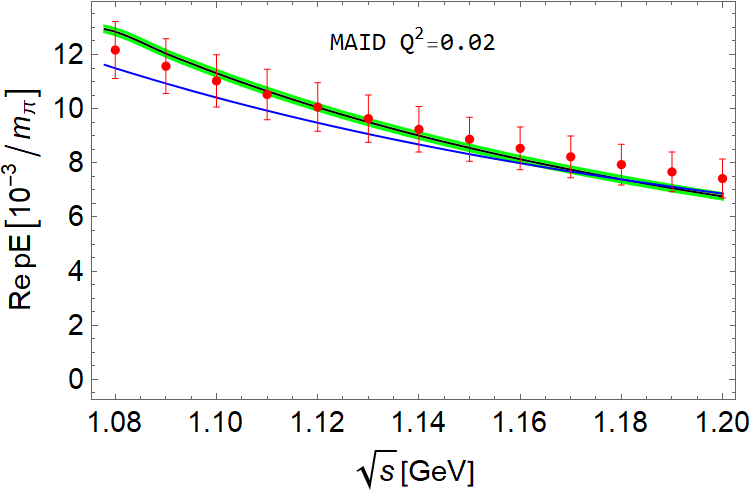}
    \vspace{0.02cm}
    \hspace{0.02cm}
    \end{minipage}
    \begin{minipage}[b]{0.24\linewidth}
    \includegraphics[width=3.63cm]{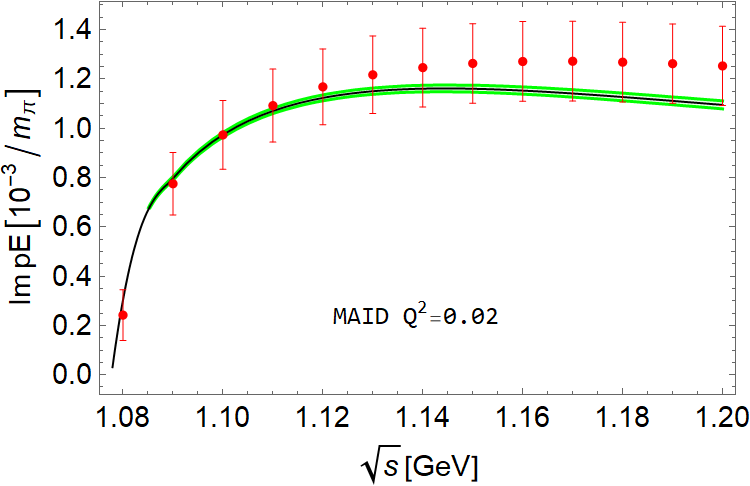}
    \vspace{0.02cm}
    \hspace{0.02cm}
    \end{minipage}
    \begin{minipage}[b]{0.24\linewidth}
    \includegraphics[width=3.63cm]{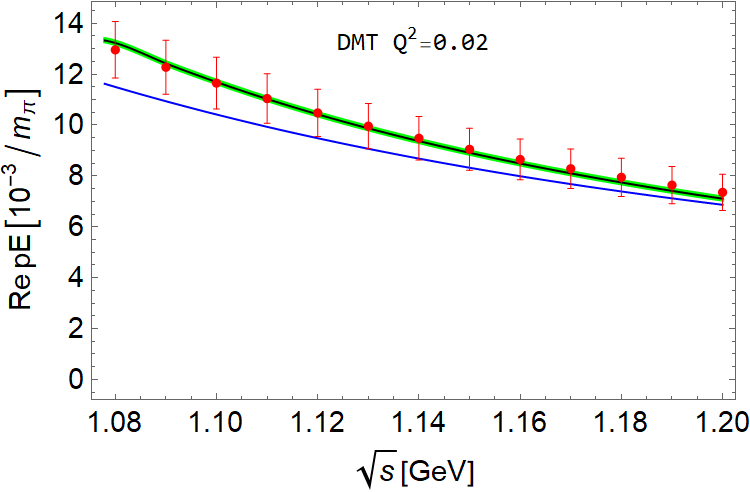}
    \vspace{0.02cm}
    \hspace{0.02cm}
    \end{minipage}
    \begin{minipage}[b]{0.24\linewidth}
    \includegraphics[width=3.63cm]{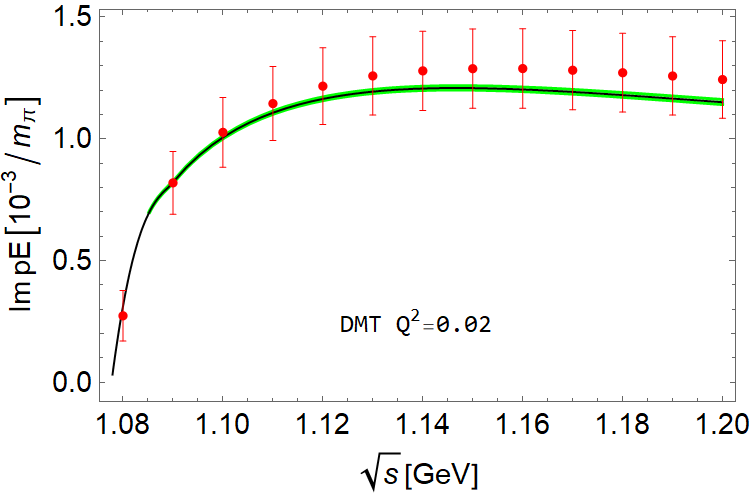}
    \vspace{0.02cm}
    \hspace{0.02cm}
    \end{minipage}
    
    \begin{minipage}[b]{0.245\linewidth}
    \includegraphics[width=3.63cm]{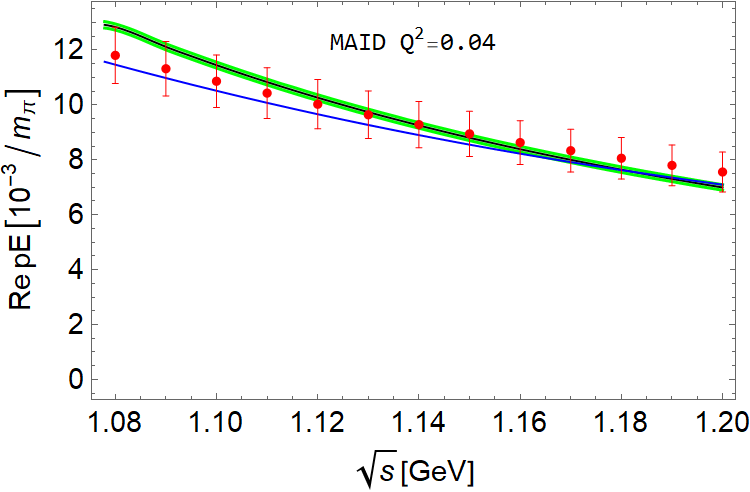}
    \vspace{0.02cm}
    \hspace{0.02cm}
    \end{minipage}
    \begin{minipage}[b]{0.245\linewidth}
    \includegraphics[width=3.63cm]{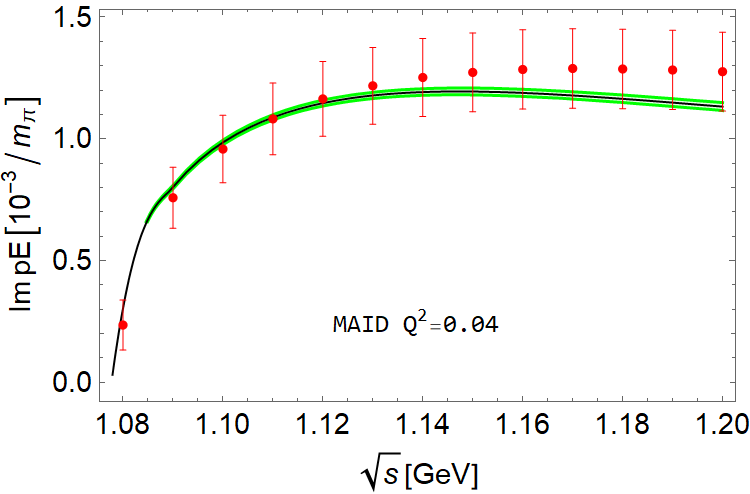}
    \vspace{0.02cm}
    \hspace{0.02cm}
    \end{minipage}
    \begin{minipage}[b]{0.245\linewidth}
    \includegraphics[width=3.63cm]{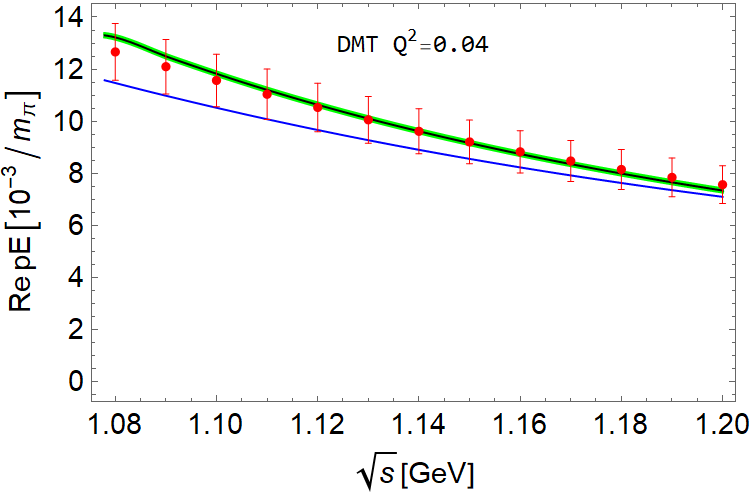}
    \vspace{0.02cm}
    \hspace{0.02cm}
    \end{minipage}
    \begin{minipage}[b]{0.245\linewidth}
    \includegraphics[width=3.63cm]{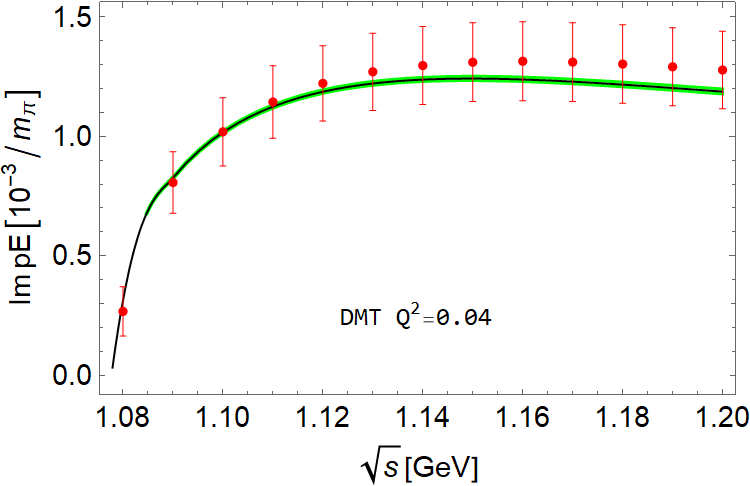}
    \vspace{0.02cm}
    \hspace{0.02cm}
    \end{minipage}
    
    \begin{minipage}[b]{0.245\linewidth}
    \includegraphics[width=3.63cm]{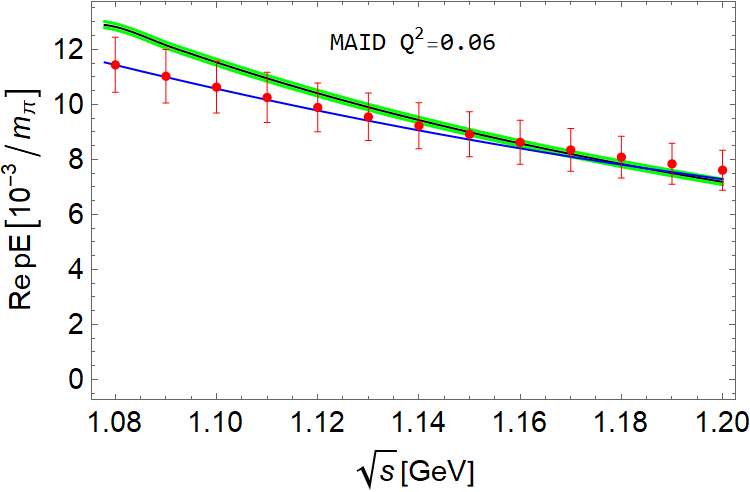}
    \vspace{0.02cm}
    \hspace{0.02cm}
    \end{minipage}
    \begin{minipage}[b]{0.245\linewidth}
    \includegraphics[width=3.63cm]{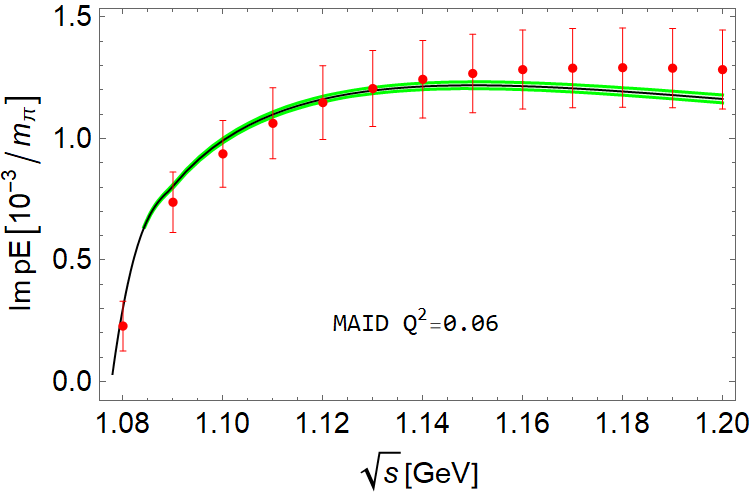}
    \vspace{0.02cm}
    \hspace{0.02cm}
    \end{minipage}
    \begin{minipage}[b]{0.245\linewidth}
    \includegraphics[width=3.63cm]{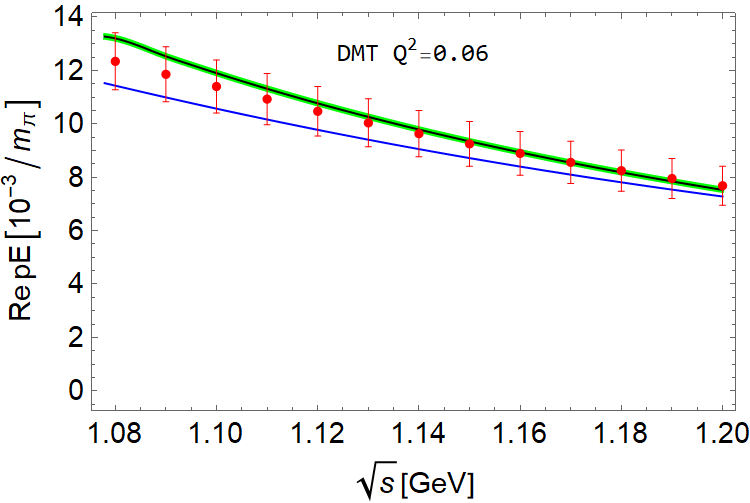}
    \vspace{0.02cm}
    \hspace{0.02cm}
    \end{minipage}
    \begin{minipage}[b]{0.245\linewidth}
    \includegraphics[width=3.63cm]{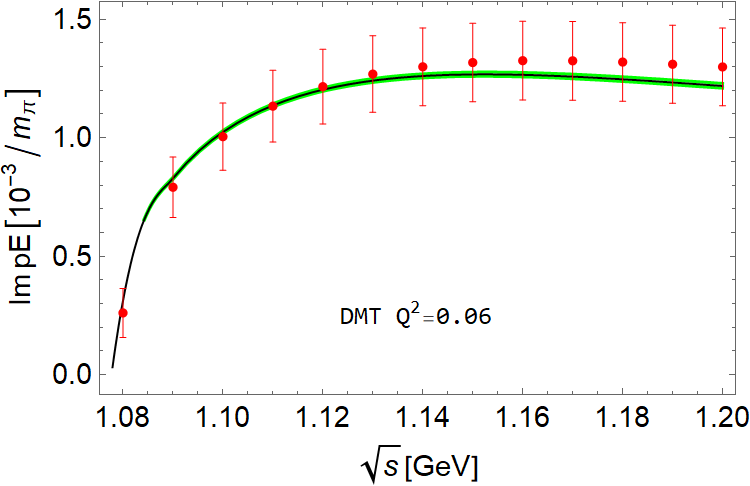}
    \vspace{0.02cm}
    \hspace{0.02cm}
    \end{minipage}
    
    \begin{minipage}[b]{0.245\linewidth}
    \includegraphics[width=3.63cm]{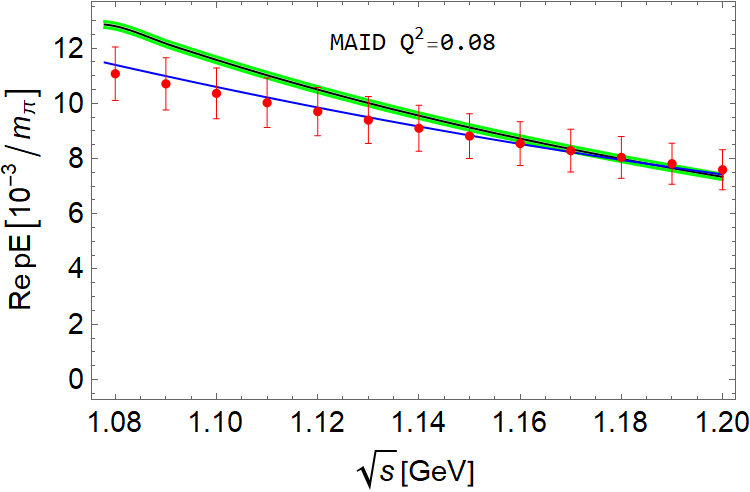}
    \vspace{0.02cm}
    \hspace{0.02cm}
    \end{minipage}
    \begin{minipage}[b]{0.245\linewidth}
    \includegraphics[width=3.63cm]{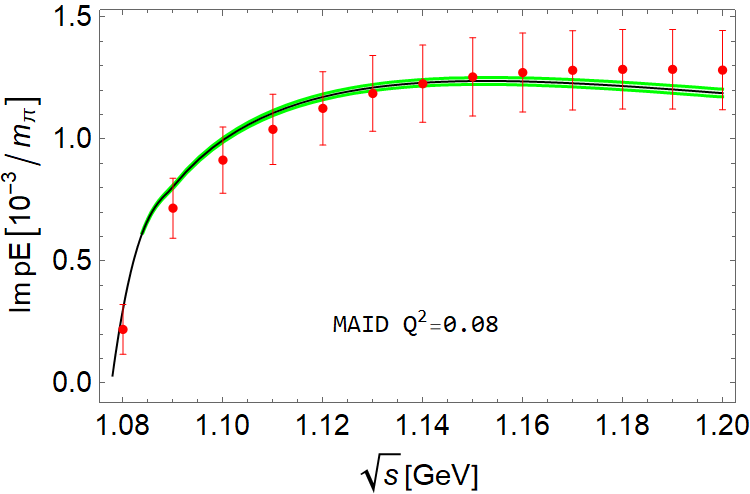}
    \vspace{0.02cm}
    \hspace{0.02cm}
    \end{minipage}
    \begin{minipage}[b]{0.245\linewidth}
    \includegraphics[width=3.63cm]{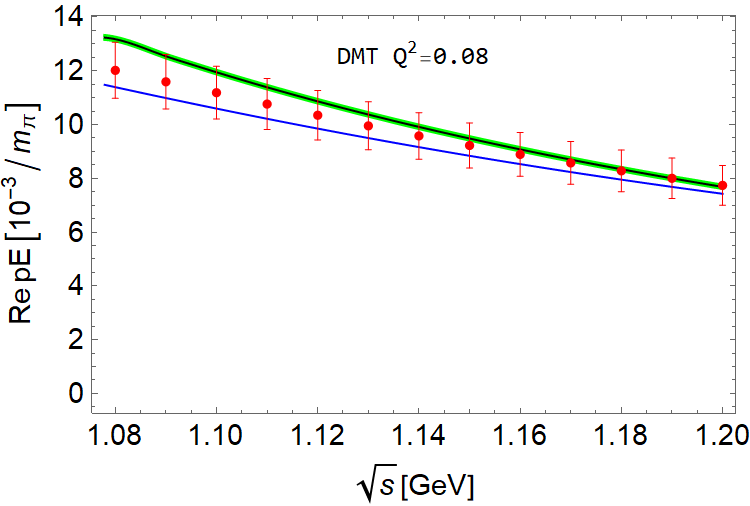}
    \vspace{0.02cm}
    \hspace{0.02cm}
    \end{minipage}
    \begin{minipage}[b]{0.245\linewidth}
    \includegraphics[width=3.63cm]{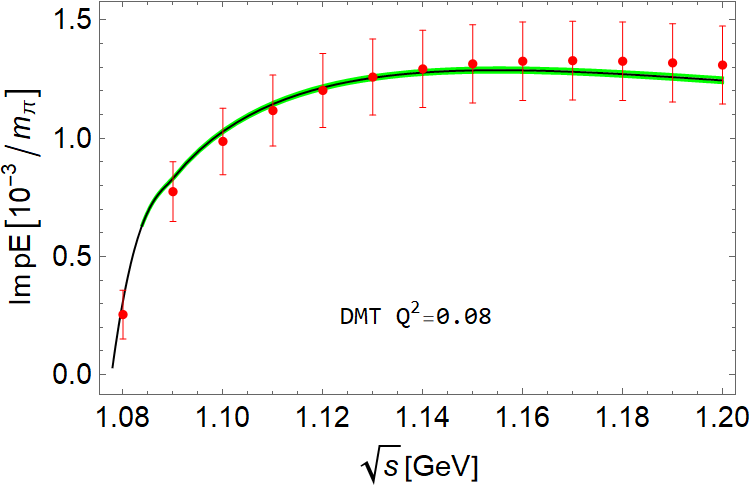}
    \vspace{0.02cm}
    \hspace{0.02cm}
    \end{minipage}
    
    \begin{minipage}[b]{0.245\linewidth}
    \includegraphics[width=3.63cm]{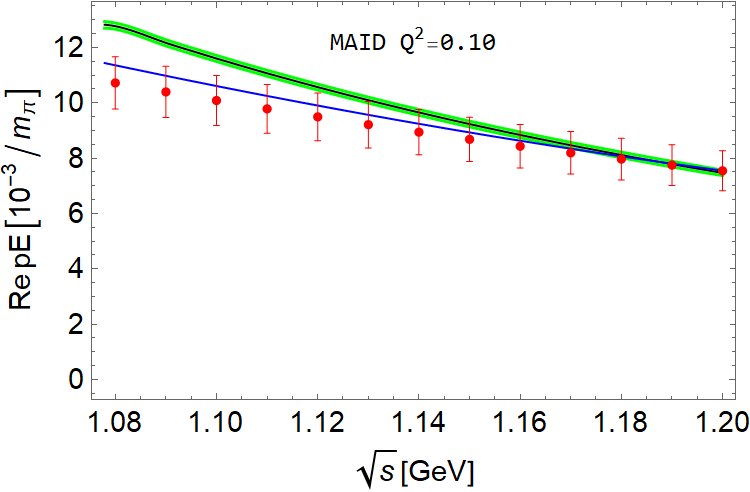}
    \vspace{0.02cm}
    \hspace{0.02cm}
    \end{minipage}
    \begin{minipage}[b]{0.245\linewidth}
    \includegraphics[width=3.63cm]{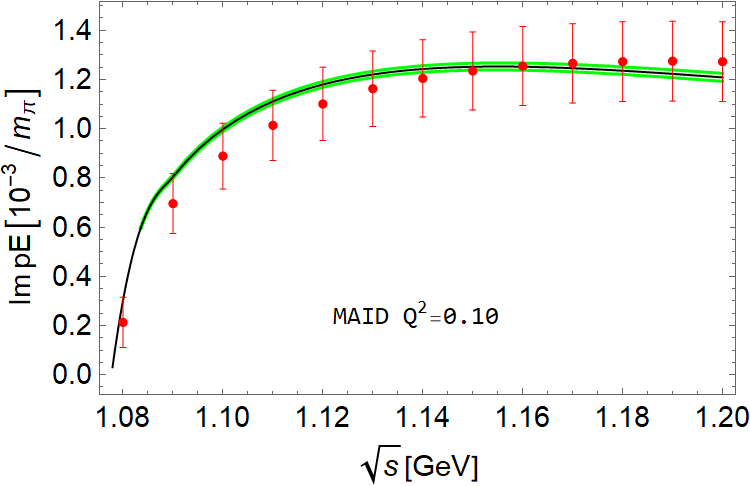}
    \vspace{0.02cm}
    \hspace{0.02cm}
    \end{minipage}
    \begin{minipage}[b]{0.245\linewidth}
    \includegraphics[width=3.63cm]{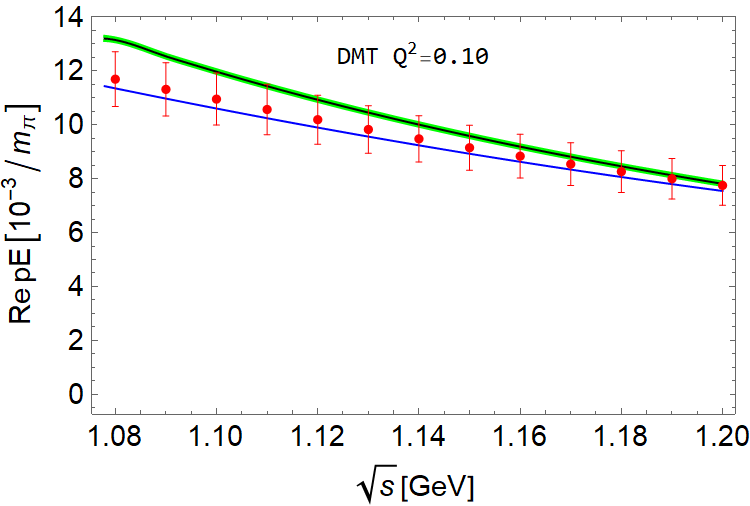}
    \vspace{0.02cm}
    \hspace{0.02cm}
    \end{minipage}
    \begin{minipage}[b]{0.245\linewidth}
    \includegraphics[width=3.63cm]{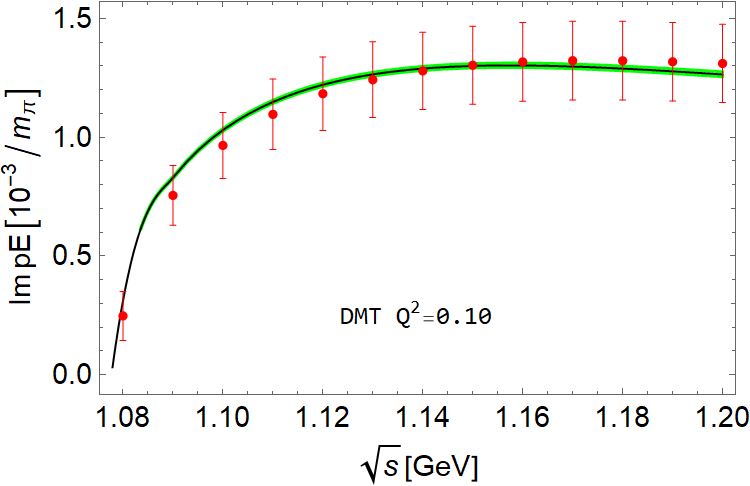}
    \vspace{0.02cm}
    \hspace{0.02cm}
    \end{minipage}
    
    \begin{minipage}[b]{0.245\linewidth}
    \includegraphics[width=3.63cm]{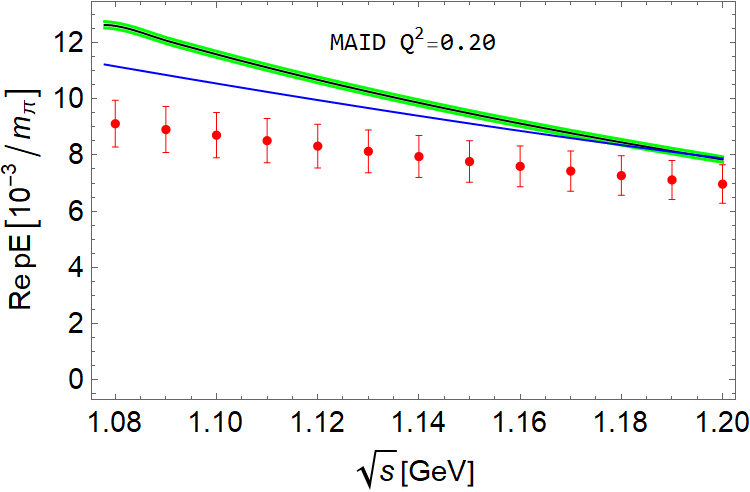}
    \vspace{0.02cm}
    \hspace{0.02cm}
    \end{minipage}
    \begin{minipage}[b]{0.245\linewidth}
    \includegraphics[width=3.63cm]{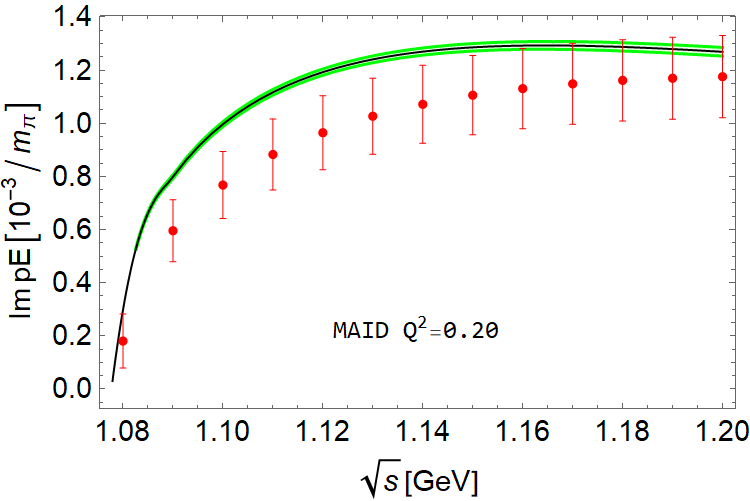}
    \vspace{0.02cm}
    \hspace{0.02cm}
    \end{minipage}
    \begin{minipage}[b]{0.245\linewidth}
    \includegraphics[width=3.63cm]{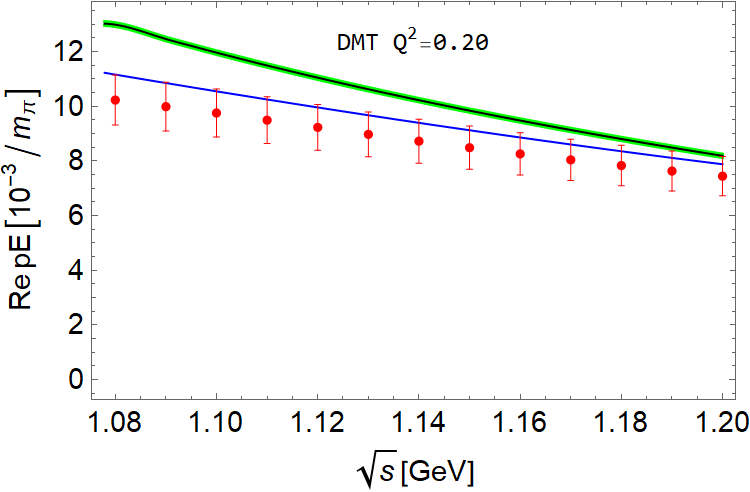}
    \vspace{0.02cm}
    \hspace{0.02cm}
    \end{minipage}
    \begin{minipage}[b]{0.245\linewidth}
    \includegraphics[width=3.63cm]{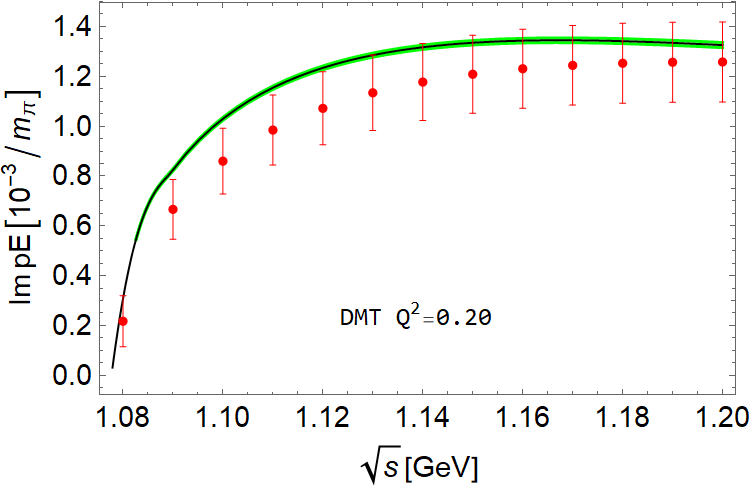}
    \vspace{0.02cm}
    \hspace{0.02cm}
    \end{minipage}    
    
    \end{minipage}
    }
    
\caption{ $S_{11}E_{0+}$ for proton (the abbreviation is $pE$): 
The `data' are from MAID (two left columns) and DMT (two right columns), respectively.
Moreover, the black lines represent our fit result. Meanwhile, 
we also show the green error band depicting the statistical error 
from the DR subtraction constant (variation within $2\sigma$ as in Table~\ref{pfrq}). 
For comparison, the chiral result is also shown by the blue lines.
}\label{f:pE}
\end{figure}

\begin{figure}[H]
\centering
\subfigure{
    \begin{minipage}[b]{0.985\linewidth}
    
    \begin{minipage}[b]{0.245\linewidth}
    \includegraphics[width=3.63cm]{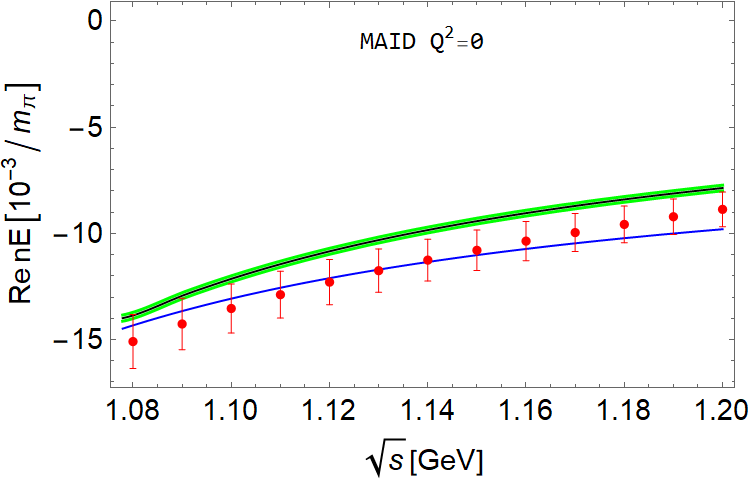}
    \vspace{0.02cm}
    \hspace{0.02cm}
    \end{minipage}
    \begin{minipage}[b]{0.245\linewidth}
    \includegraphics[width=3.63cm]{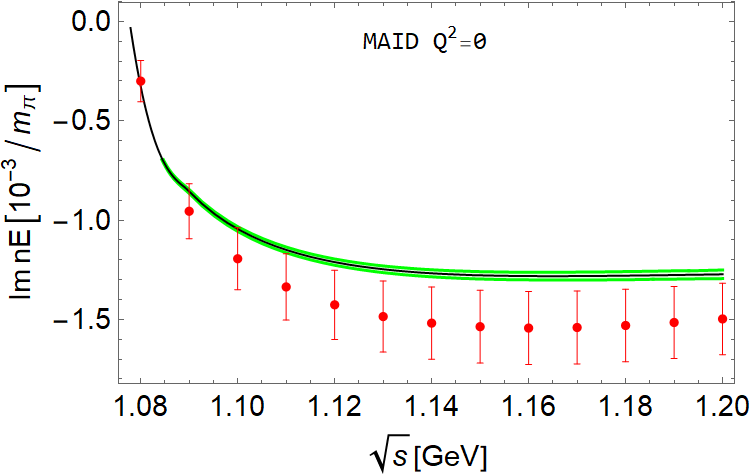}
    \vspace{0.02cm}
    \hspace{0.02cm}
    \end{minipage}
    \begin{minipage}[b]{0.245\linewidth}
    \includegraphics[width=3.63cm]{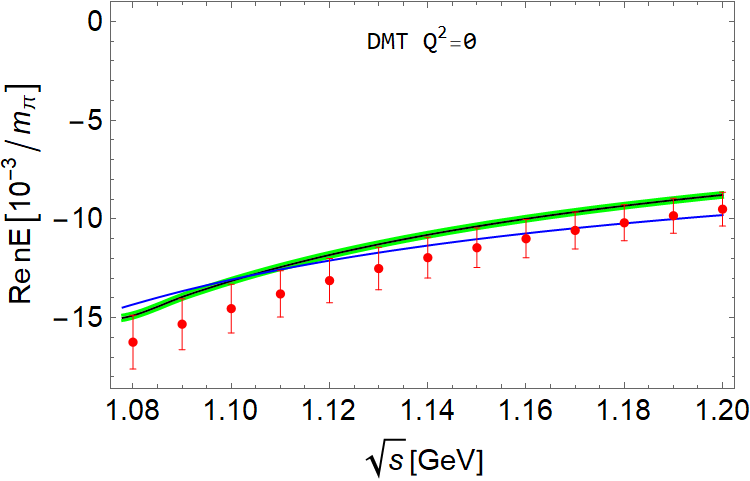}
    \vspace{0.02cm}
    \hspace{0.02cm}
    \end{minipage}
    \begin{minipage}[b]{0.245\linewidth}
    \includegraphics[width=3.63cm]{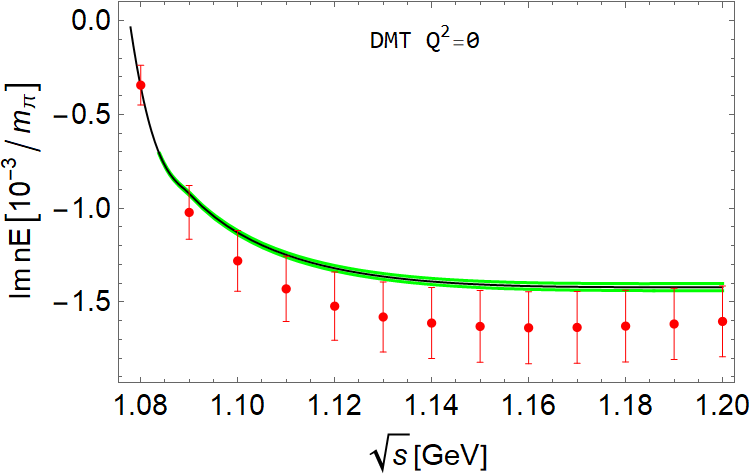}
    \vspace{0.02cm}
    \hspace{0.02cm}
    \end{minipage}
    
    \begin{minipage}[b]{0.24\linewidth}
    \includegraphics[width=3.63cm]{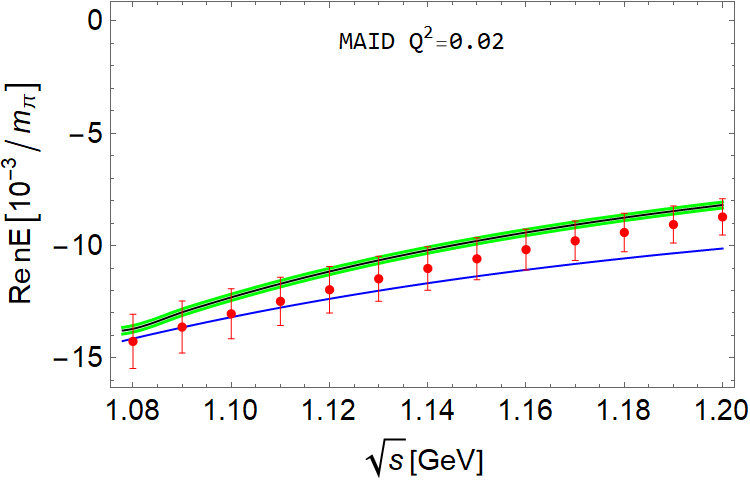}
    \vspace{0.02cm}
    \hspace{0.02cm}
    \end{minipage}
    \begin{minipage}[b]{0.24\linewidth}
    \includegraphics[width=3.63cm]{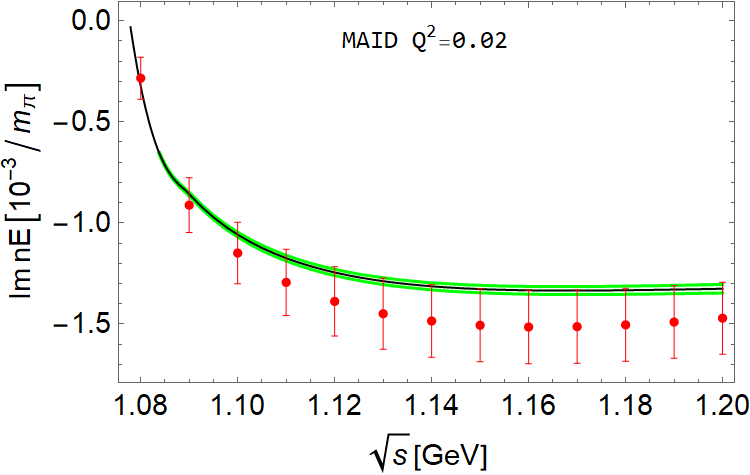}
    \vspace{0.02cm}
    \hspace{0.02cm}
    \end{minipage}
    \begin{minipage}[b]{0.24\linewidth}
    \includegraphics[width=3.63cm]{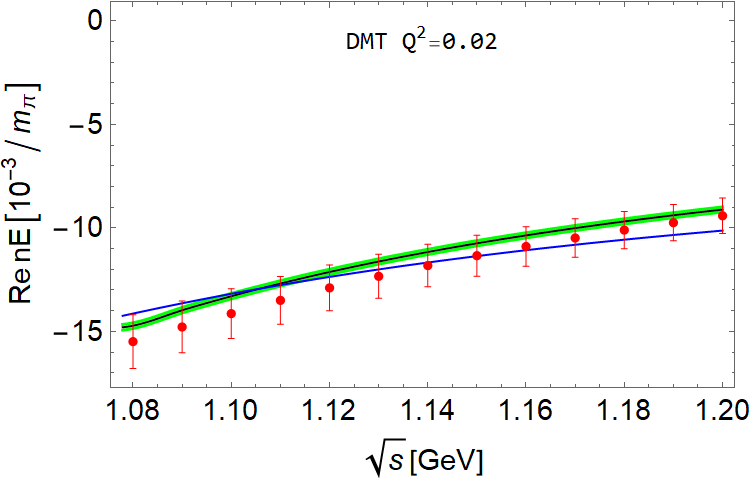}
    \vspace{0.02cm}
    \hspace{0.02cm}
    \end{minipage}
    \begin{minipage}[b]{0.24\linewidth}
    \includegraphics[width=3.63cm]{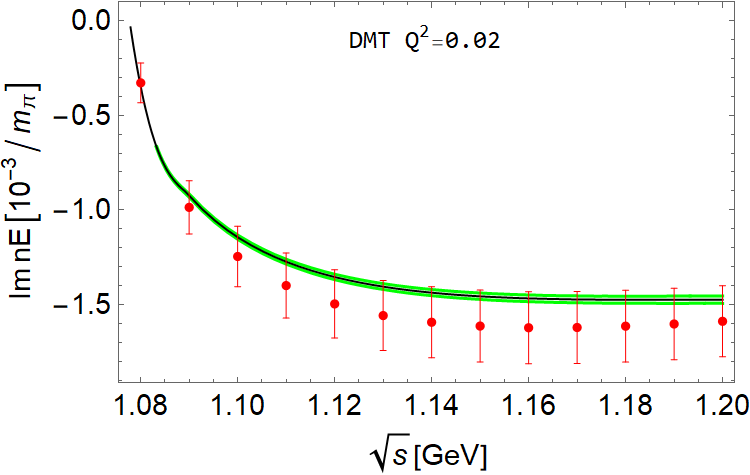}
    \vspace{0.02cm}
    \hspace{0.02cm}
    \end{minipage}
    
    \begin{minipage}[b]{0.245\linewidth}
    \includegraphics[width=3.63cm]{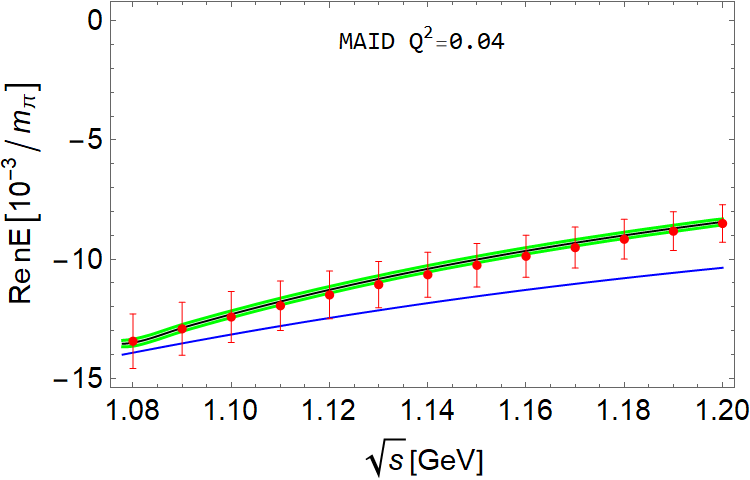}
    \vspace{0.02cm}
    \hspace{0.02cm}
    \end{minipage}
    \begin{minipage}[b]{0.245\linewidth}
    \includegraphics[width=3.63cm]{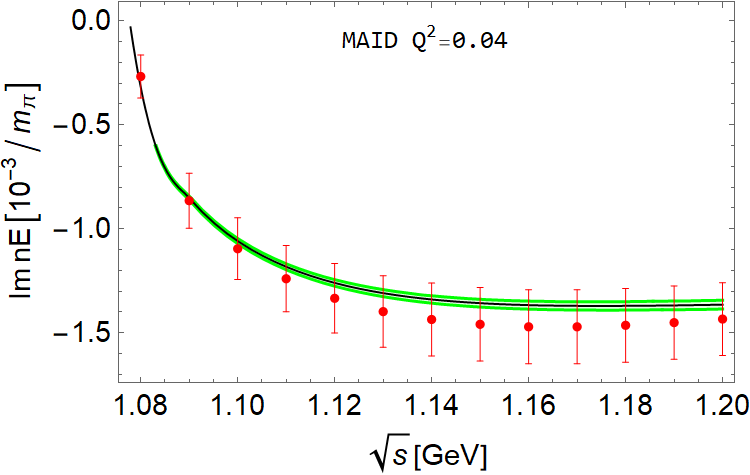}
    \vspace{0.02cm}
    \hspace{0.02cm}
    \end{minipage}
    \begin{minipage}[b]{0.245\linewidth}
    \includegraphics[width=3.63cm]{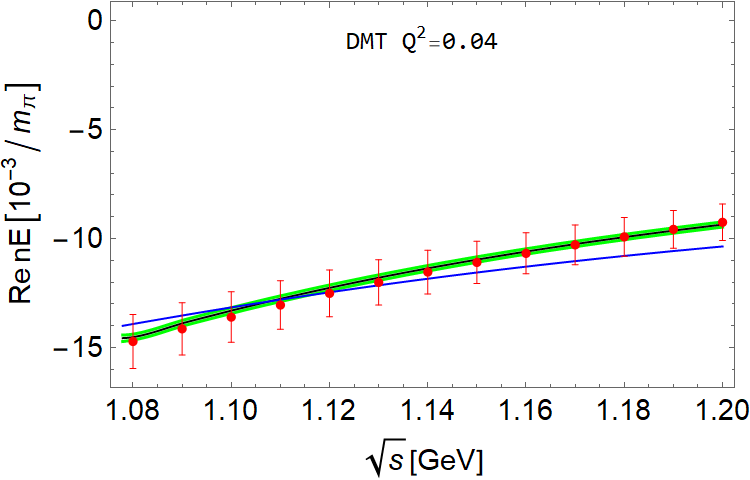}
    \vspace{0.02cm}
    \hspace{0.02cm}
    \end{minipage}
    \begin{minipage}[b]{0.245\linewidth}
    \includegraphics[width=3.63cm]{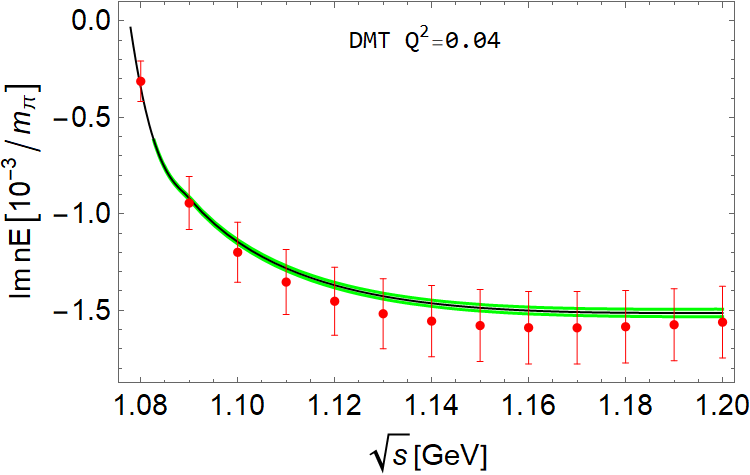}
    \vspace{0.02cm}
    \hspace{0.02cm}
    \end{minipage}
    
    \begin{minipage}[b]{0.245\linewidth}
    \includegraphics[width=3.63cm]{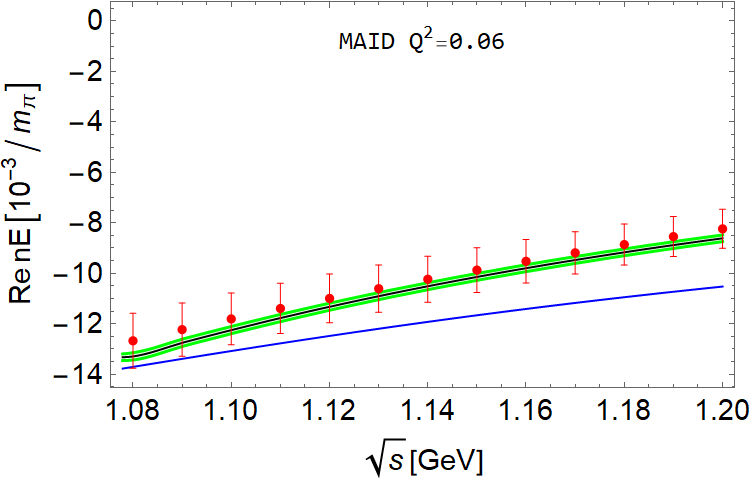}
    \vspace{0.02cm}
    \hspace{0.02cm}
    \end{minipage}
    \begin{minipage}[b]{0.245\linewidth}
    \includegraphics[width=3.63cm]{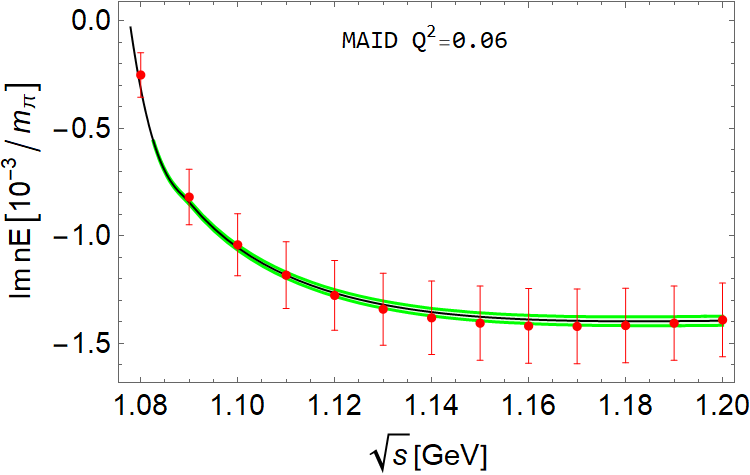}
    \vspace{0.02cm}
    \hspace{0.02cm}
    \end{minipage}
    \begin{minipage}[b]{0.245\linewidth}
    \includegraphics[width=3.63cm]{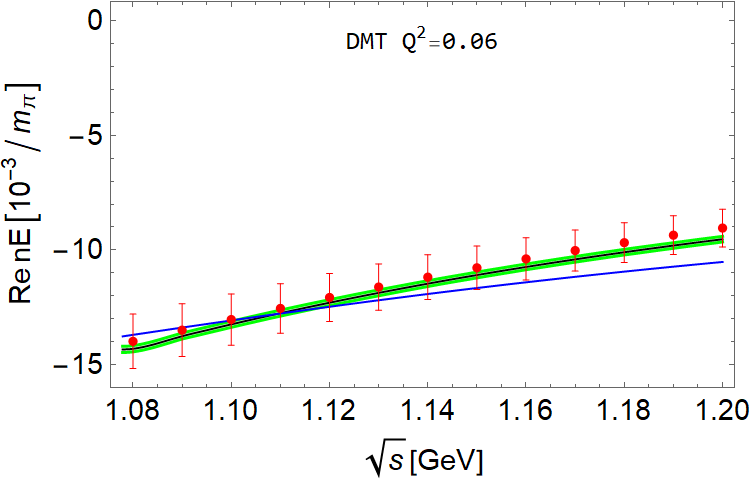}
    \vspace{0.02cm}
    \hspace{0.02cm}
    \end{minipage}
    \begin{minipage}[b]{0.245\linewidth}
    \includegraphics[width=3.63cm]{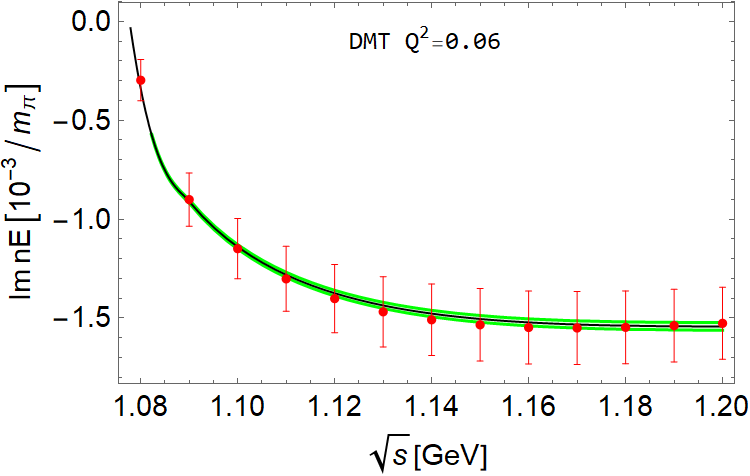}
    \vspace{0.02cm}
    \hspace{0.02cm}
    \end{minipage}
    
    \begin{minipage}[b]{0.245\linewidth}
    \includegraphics[width=3.63cm]{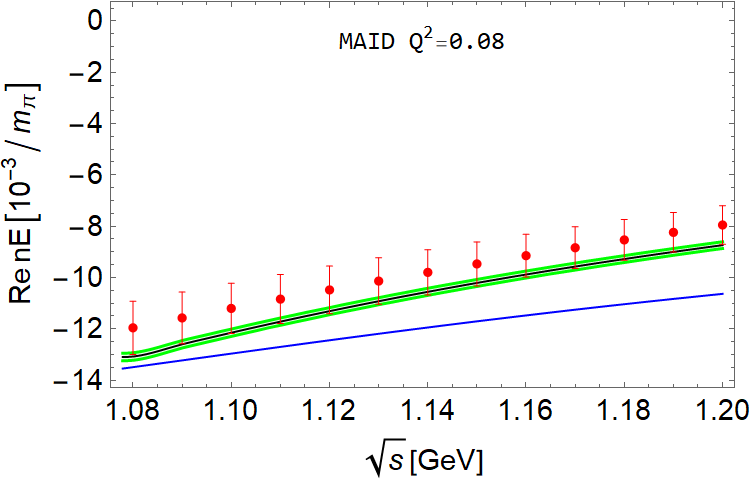}
    \vspace{0.02cm}
    \hspace{0.02cm}
    \end{minipage}
    \begin{minipage}[b]{0.245\linewidth}
    \includegraphics[width=3.63cm]{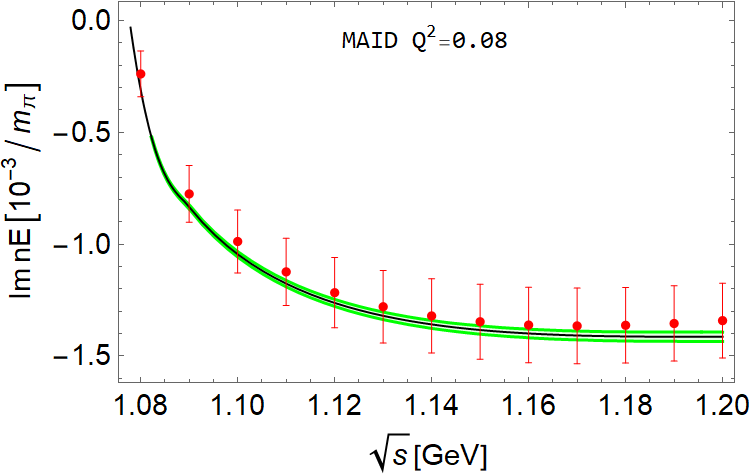}
    \vspace{0.02cm}
    \hspace{0.02cm}
    \end{minipage}
    \begin{minipage}[b]{0.245\linewidth}
    \includegraphics[width=3.63cm]{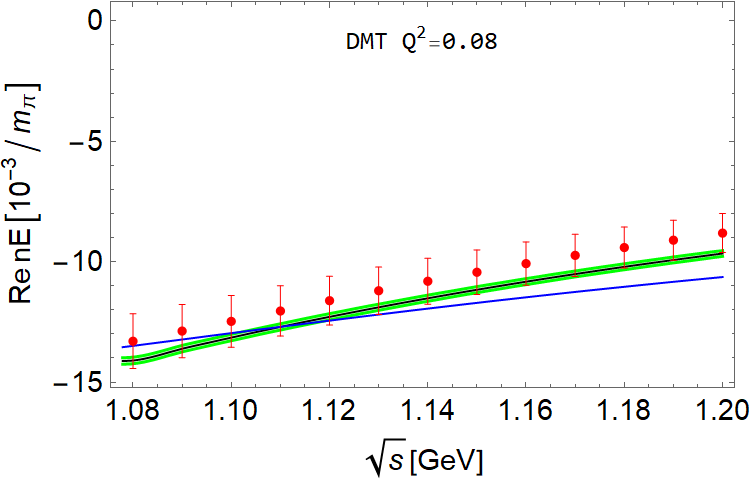}
    \vspace{0.02cm}
    \hspace{0.02cm}
    \end{minipage}
    \begin{minipage}[b]{0.245\linewidth}
    \includegraphics[width=3.63cm]{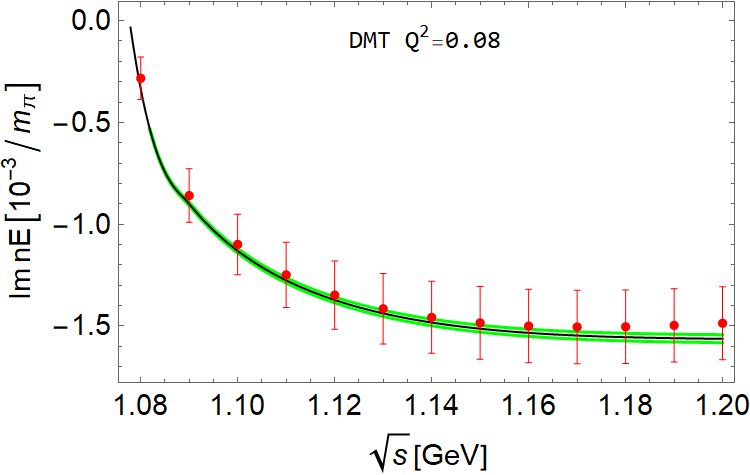}
    \vspace{0.02cm}
    \hspace{0.02cm}
    \end{minipage}
    
    \begin{minipage}[b]{0.245\linewidth}
    \includegraphics[width=3.63cm]{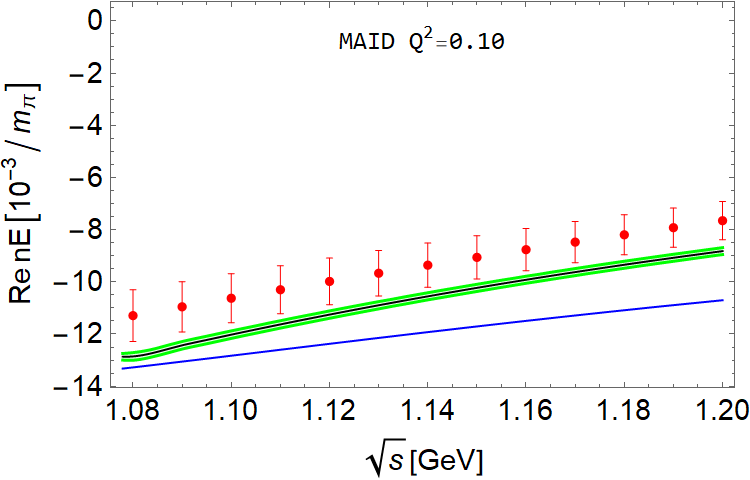}
    \vspace{0.02cm}
    \hspace{0.02cm}
    \end{minipage}
    \begin{minipage}[b]{0.245\linewidth}
    \includegraphics[width=3.63cm]{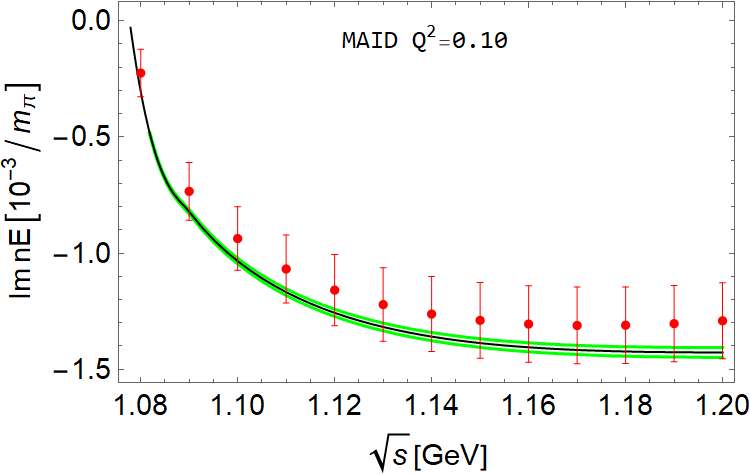}
    \vspace{0.02cm}
    \hspace{0.02cm}
    \end{minipage}
    \begin{minipage}[b]{0.245\linewidth}
    \includegraphics[width=3.63cm]{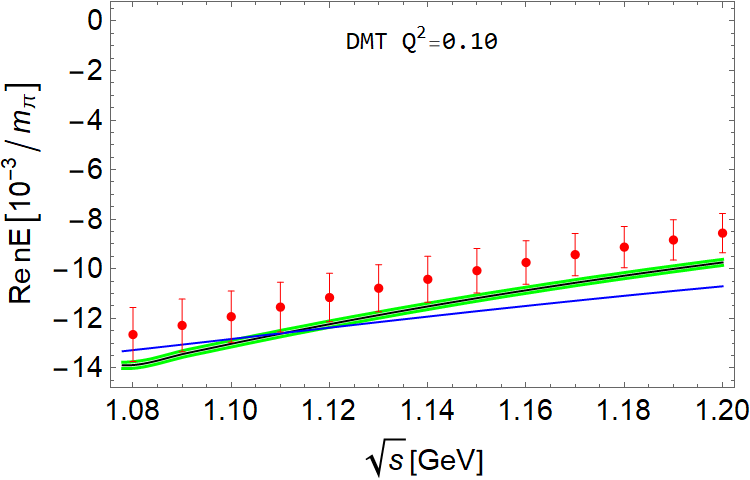}
    \vspace{0.02cm}
    \hspace{0.02cm}
    \end{minipage}
    \begin{minipage}[b]{0.245\linewidth}
    \includegraphics[width=3.63cm]{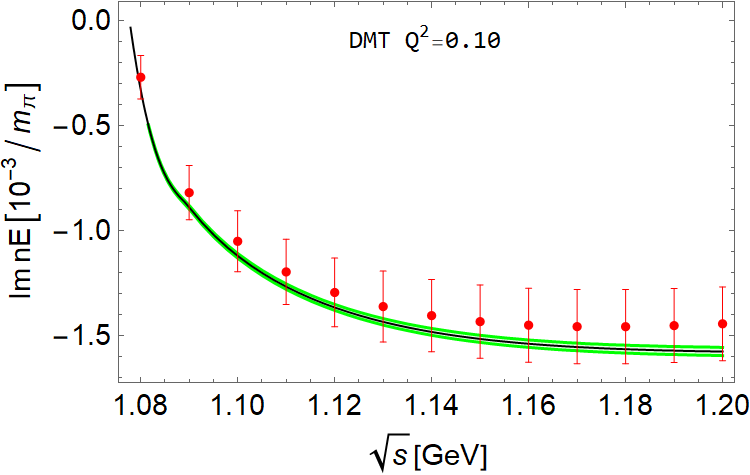}
    \vspace{0.02cm}
    \hspace{0.02cm}
    \end{minipage}
    
    \begin{minipage}[b]{0.245\linewidth}
    \includegraphics[width=3.63cm]{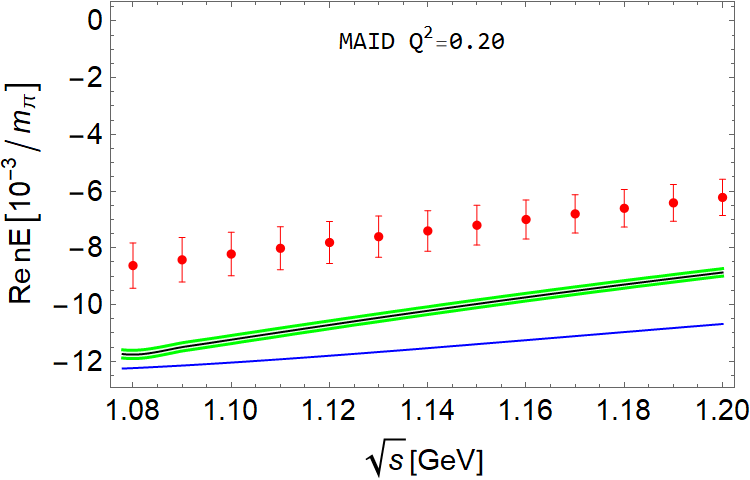}
    \vspace{0.02cm}
    \hspace{0.02cm}
    \end{minipage}
    \begin{minipage}[b]{0.245\linewidth}
    \includegraphics[width=3.63cm]{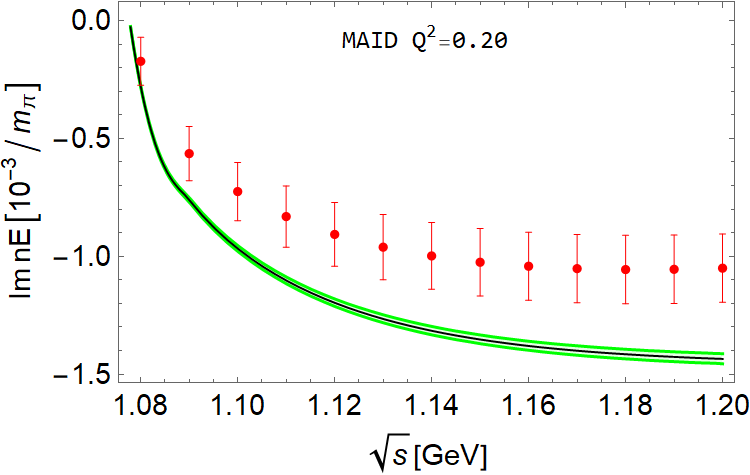}
    \vspace{0.02cm}
    \hspace{0.02cm}
    \end{minipage}
    \begin{minipage}[b]{0.245\linewidth}
    \includegraphics[width=3.63cm]{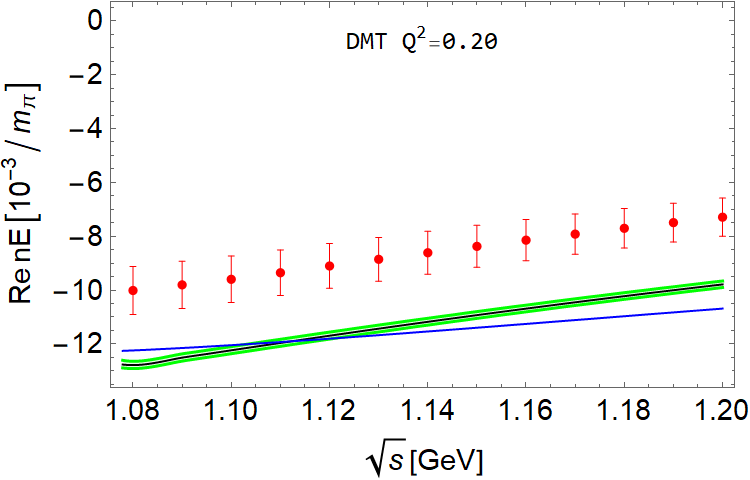}
    \vspace{0.02cm}
    \hspace{0.02cm}
    \end{minipage}
    \begin{minipage}[b]{0.245\linewidth}
    \includegraphics[width=3.63cm]{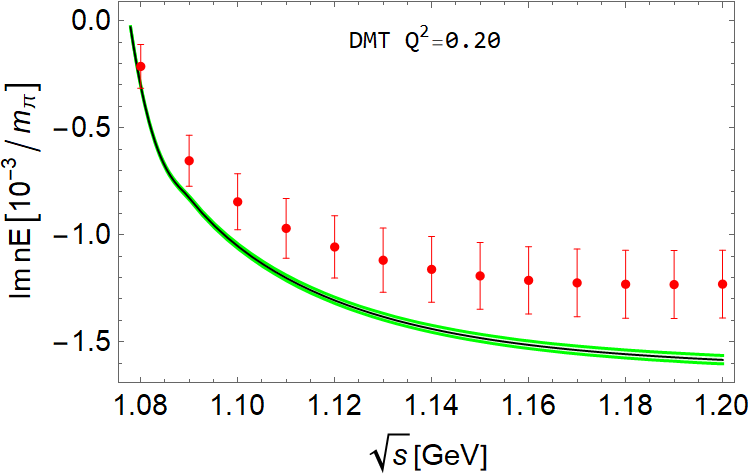}
    \vspace{0.02cm}
    \hspace{0.02cm}
    \end{minipage}    
    
    \end{minipage}
    }
    
\caption{$S_{11}E_{0+}$ for neutron~($nE$): descriptions the same as in Fig.~\ref{f:pE}}\label{f:nE}
\end{figure}

\begin{figure}[H]
\centering
\subfigure{
    \begin{minipage}[b]{0.985\linewidth}
    
    \begin{minipage}[b]{0.245\linewidth}
    \includegraphics[width=3.63cm]{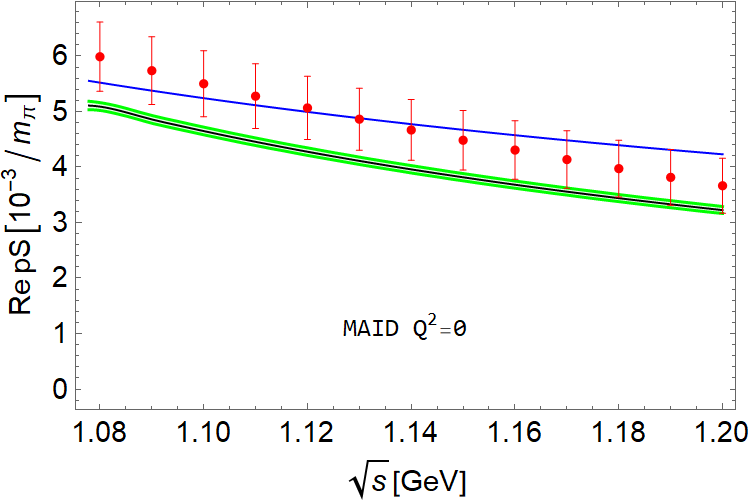}
    \vspace{0.02cm}
    \hspace{0.02cm}
    \end{minipage}
    \begin{minipage}[b]{0.245\linewidth}
    \includegraphics[width=3.63cm]{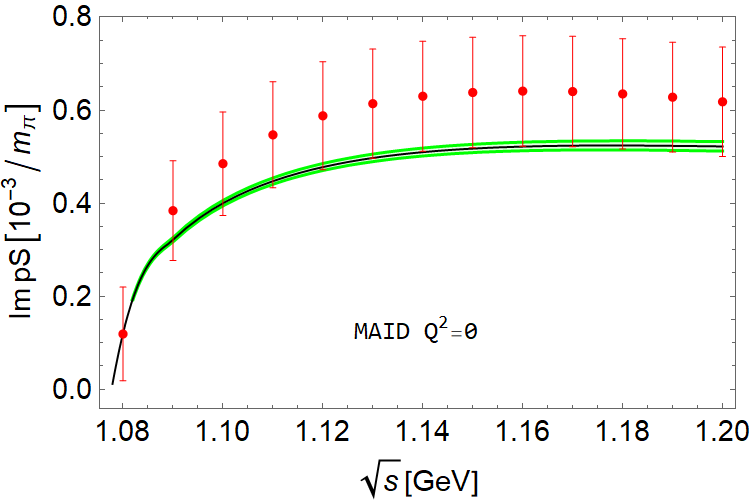}
    \vspace{0.02cm}
    \hspace{0.02cm}
    \end{minipage}
    \begin{minipage}[b]{0.245\linewidth}
    \includegraphics[width=3.63cm]{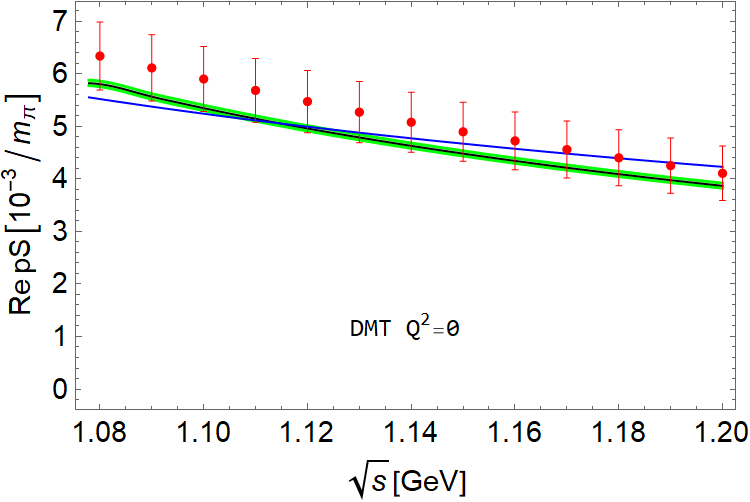}
    \vspace{0.02cm}
    \hspace{0.02cm}
    \end{minipage}
    \begin{minipage}[b]{0.245\linewidth}
    \includegraphics[width=3.63cm]{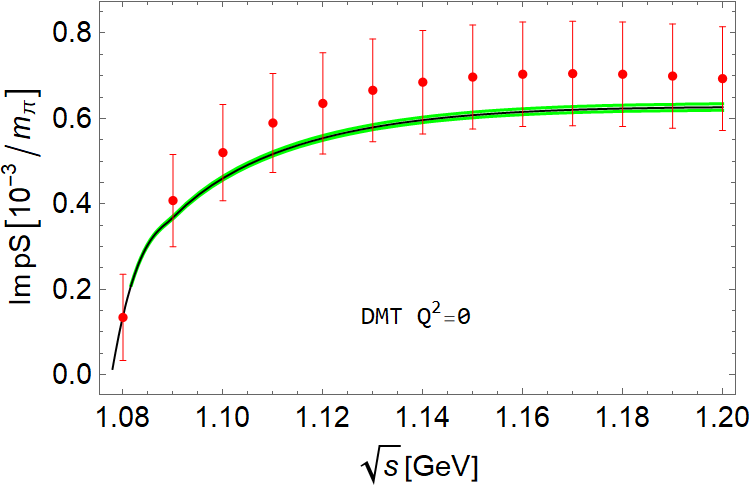}
    \vspace{0.02cm}
    \hspace{0.02cm}
    \end{minipage}
    
    \begin{minipage}[b]{0.24\linewidth}
    \includegraphics[width=3.63cm]{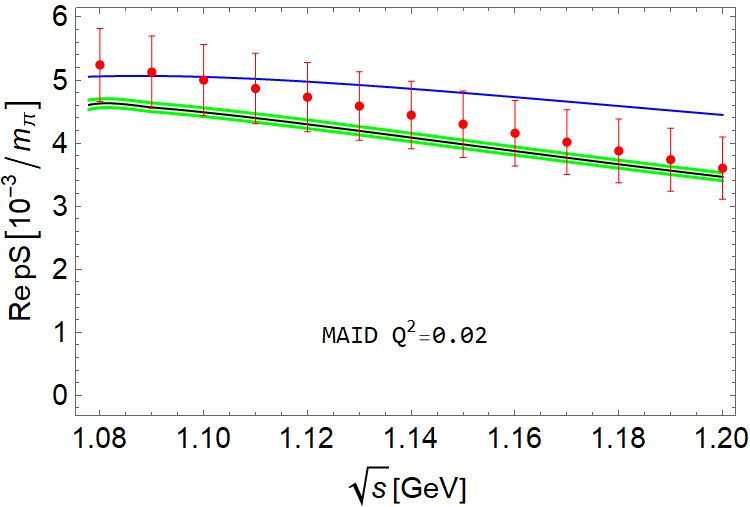}
    \vspace{0.02cm}
    \hspace{0.02cm}
    \end{minipage}
    \begin{minipage}[b]{0.24\linewidth}
    \includegraphics[width=3.63cm]{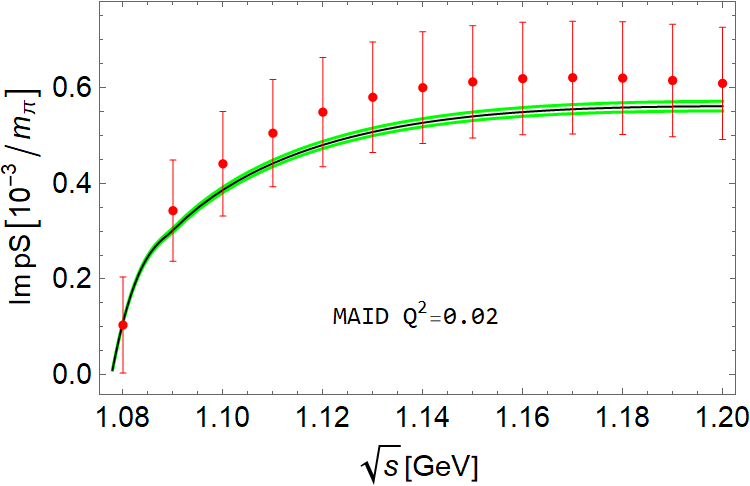}
    \vspace{0.02cm}
    \hspace{0.02cm}
    \end{minipage}
    \begin{minipage}[b]{0.24\linewidth}
    \includegraphics[width=3.63cm]{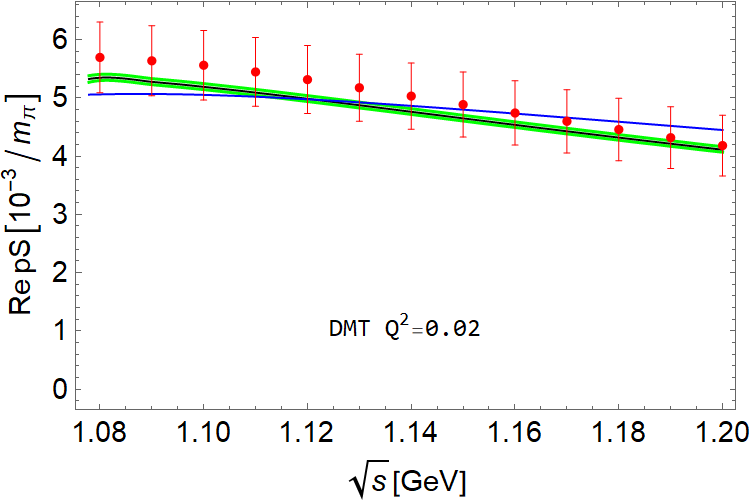}
    \vspace{0.02cm}
    \hspace{0.02cm}
    \end{minipage}
    \begin{minipage}[b]{0.24\linewidth}
    \includegraphics[width=3.63cm]{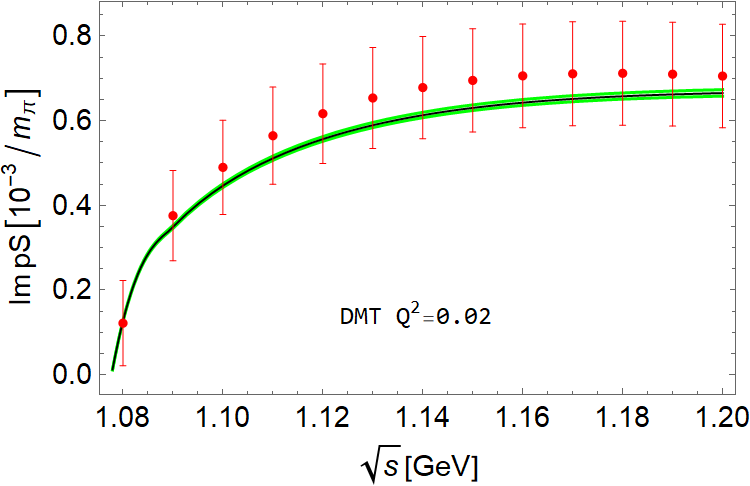}
    \vspace{0.02cm}
    \hspace{0.02cm}
    \end{minipage}
    
    \begin{minipage}[b]{0.245\linewidth}
    \includegraphics[width=3.63cm]{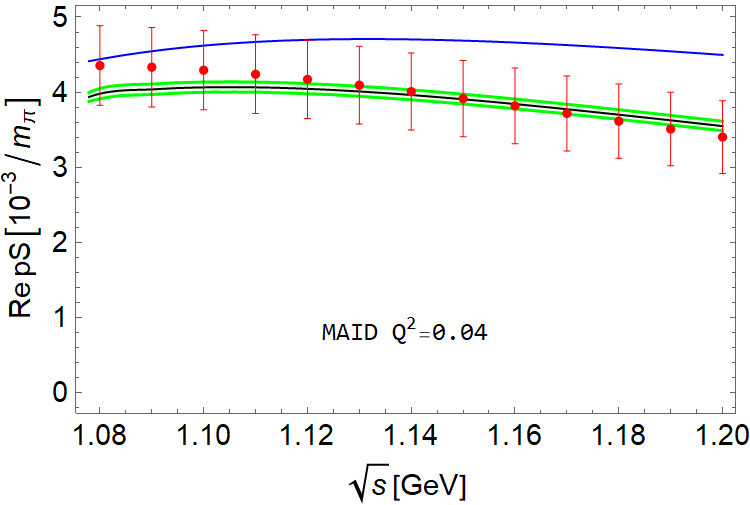}
    \vspace{0.02cm}
    \hspace{0.02cm}
    \end{minipage}
    \begin{minipage}[b]{0.245\linewidth}
    \includegraphics[width=3.63cm]{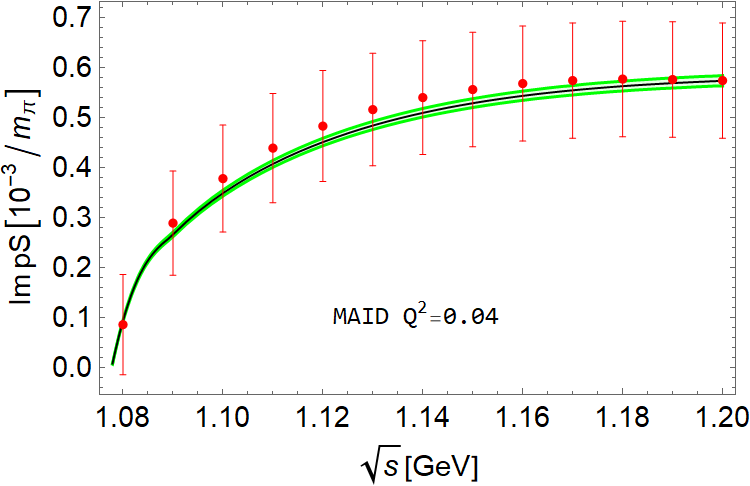}
    \vspace{0.02cm}
    \hspace{0.02cm}
    \end{minipage}
    \begin{minipage}[b]{0.245\linewidth}
    \includegraphics[width=3.63cm]{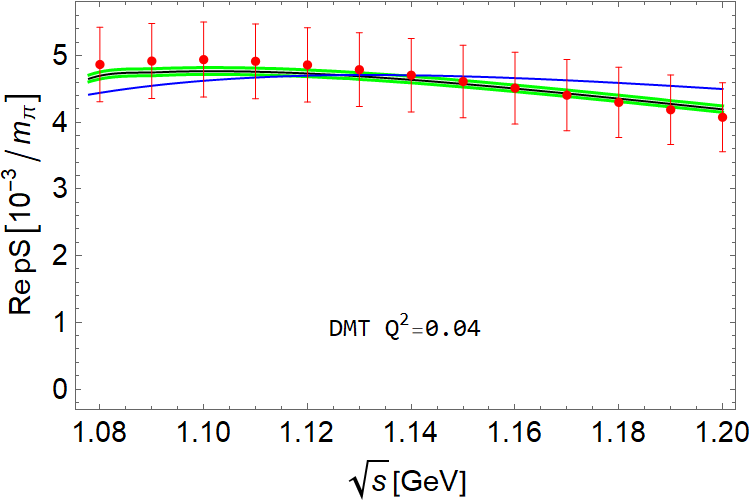}
    \vspace{0.02cm}
    \hspace{0.02cm}
    \end{minipage}
    \begin{minipage}[b]{0.245\linewidth}
    \includegraphics[width=3.63cm]{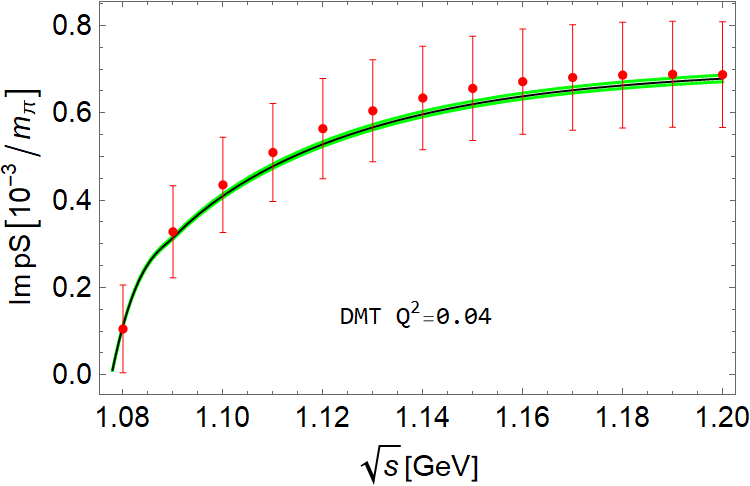}
    \vspace{0.02cm}
    \hspace{0.02cm}
    \end{minipage}
    
    \begin{minipage}[b]{0.245\linewidth}
    \includegraphics[width=3.63cm]{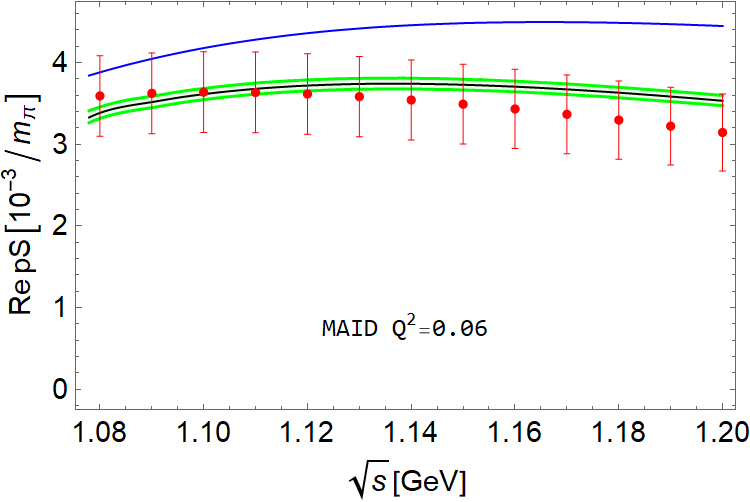}
    \vspace{0.02cm}
    \hspace{0.02cm}
    \end{minipage}
    \begin{minipage}[b]{0.245\linewidth}
    \includegraphics[width=3.63cm]{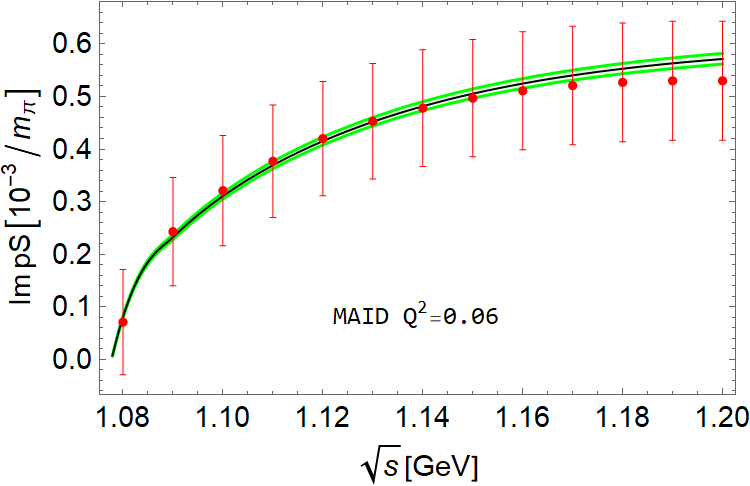}
    \vspace{0.02cm}
    \hspace{0.02cm}
    \end{minipage}
    \begin{minipage}[b]{0.245\linewidth}
    \includegraphics[width=3.63cm]{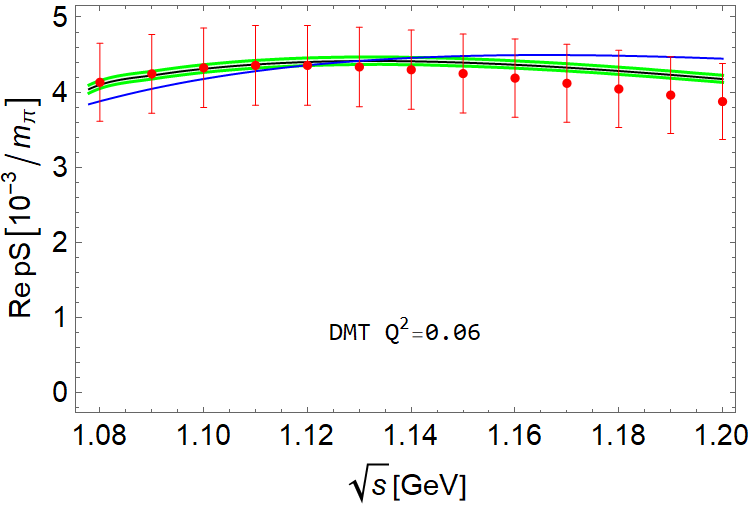}
    \vspace{0.02cm}
    \hspace{0.02cm}
    \end{minipage}
    \begin{minipage}[b]{0.245\linewidth}
    \includegraphics[width=3.63cm]{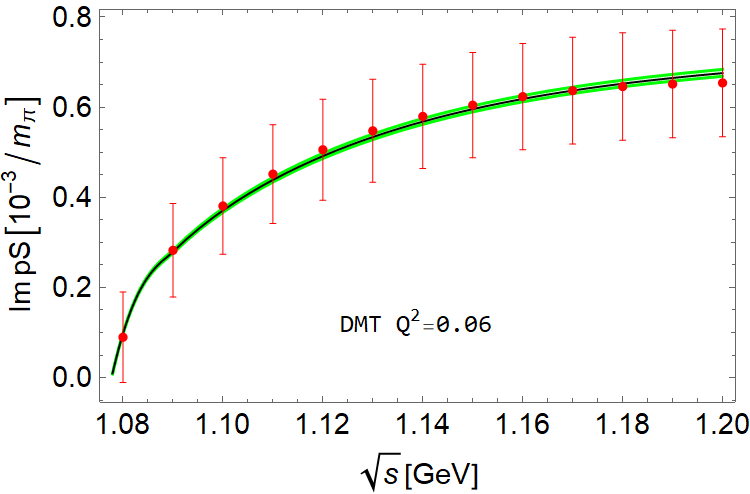}
    \vspace{0.02cm}
    \hspace{0.02cm}
    \end{minipage}
    
    \begin{minipage}[b]{0.245\linewidth}
    \includegraphics[width=3.63cm]{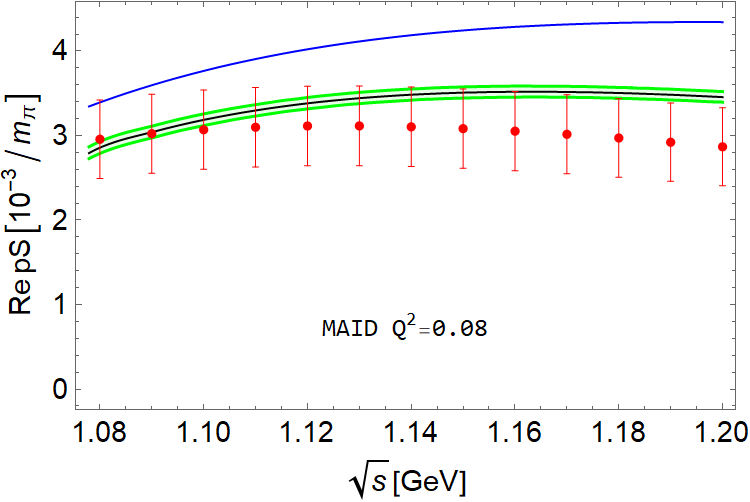}
    \vspace{0.02cm}
    \hspace{0.02cm}
    \end{minipage}
    \begin{minipage}[b]{0.245\linewidth}
    \includegraphics[width=3.63cm]{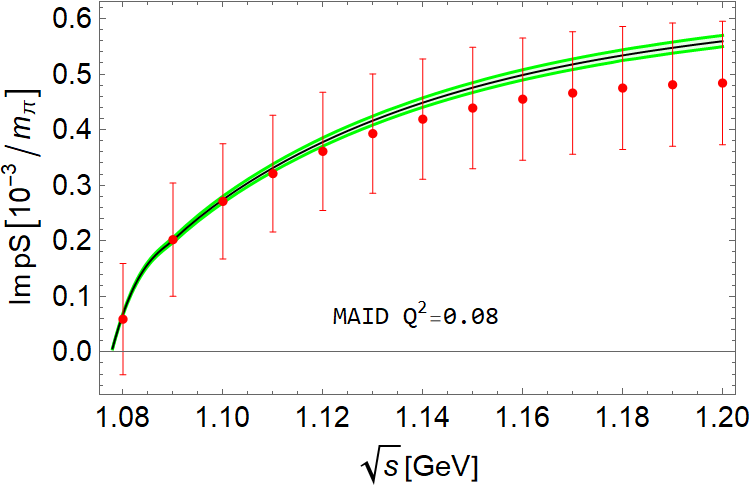}
    \vspace{0.02cm}
    \hspace{0.02cm}
    \end{minipage}
    \begin{minipage}[b]{0.245\linewidth}
    \includegraphics[width=3.63cm]{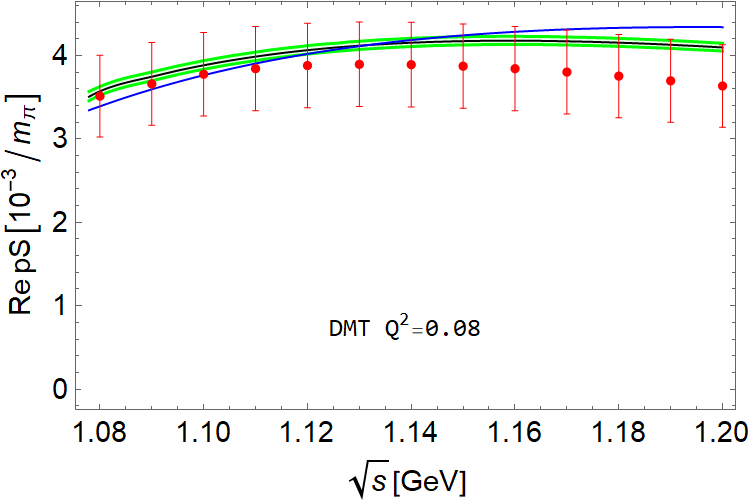}
    \vspace{0.02cm}
    \hspace{0.02cm}
    \end{minipage}
    \begin{minipage}[b]{0.245\linewidth}
    \includegraphics[width=3.63cm]{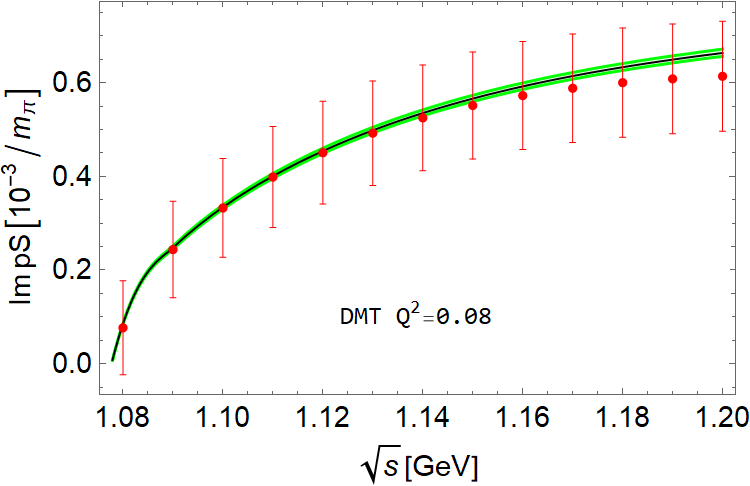}
    \vspace{0.02cm}
    \hspace{0.02cm}
    \end{minipage}
    
    \begin{minipage}[b]{0.245\linewidth}
    \includegraphics[width=3.63cm]{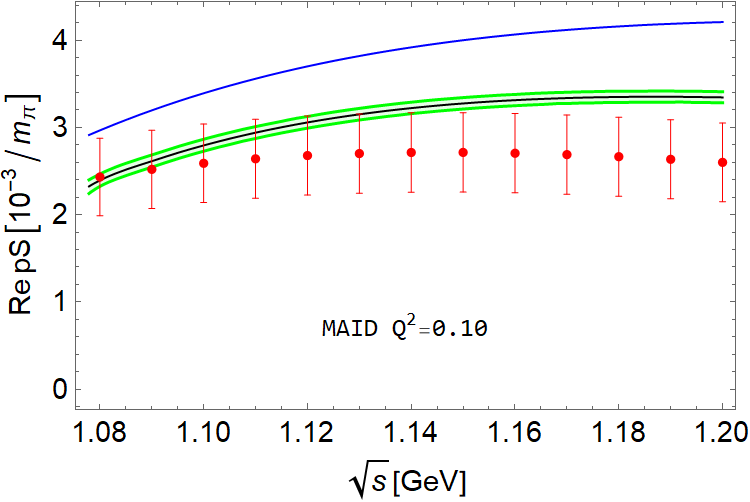}
    \vspace{0.02cm}
    \hspace{0.02cm}
    \end{minipage}
    \begin{minipage}[b]{0.245\linewidth}
    \includegraphics[width=3.63cm]{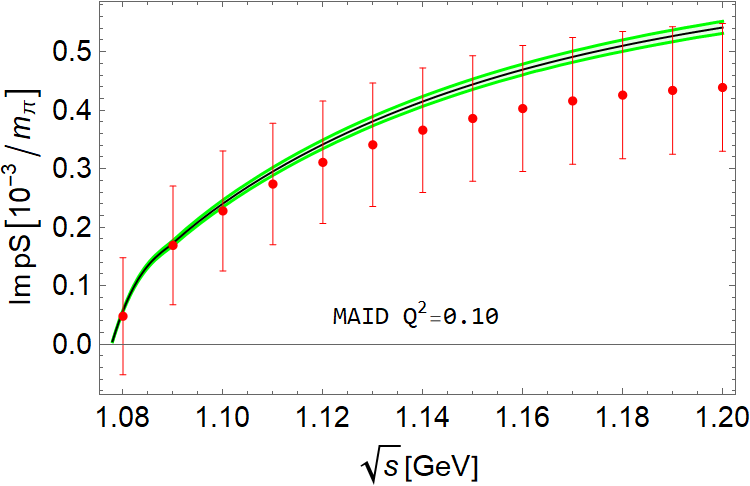}
    \vspace{0.02cm}
    \hspace{0.02cm}
    \end{minipage}
    \begin{minipage}[b]{0.245\linewidth}
    \includegraphics[width=3.63cm]{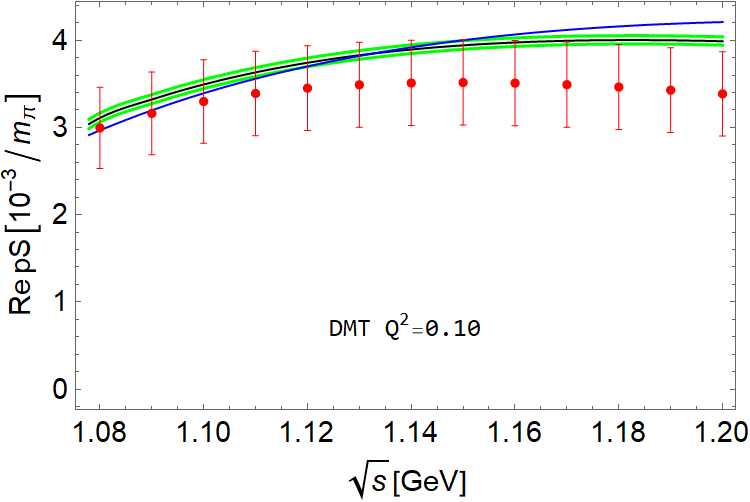}
    \vspace{0.02cm}
    \hspace{0.02cm}
    \end{minipage}
    \begin{minipage}[b]{0.245\linewidth}
    \includegraphics[width=3.63cm]{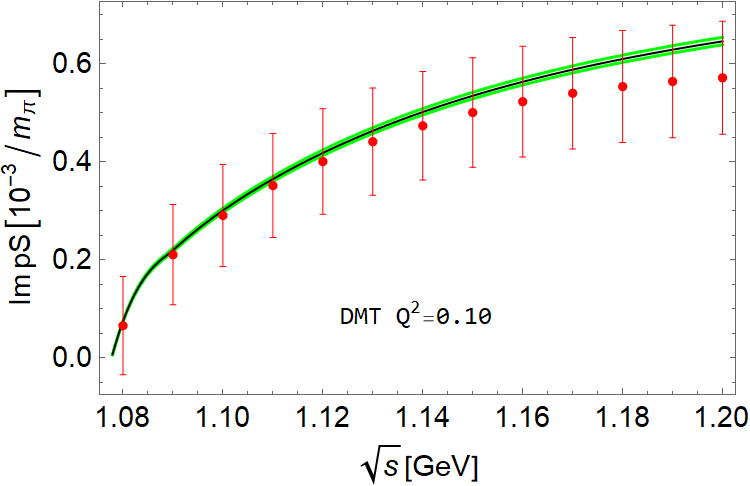}
    \vspace{0.02cm}
    \hspace{0.02cm}
    \end{minipage}
    
    \begin{minipage}[b]{0.245\linewidth}
    \includegraphics[width=3.63cm]{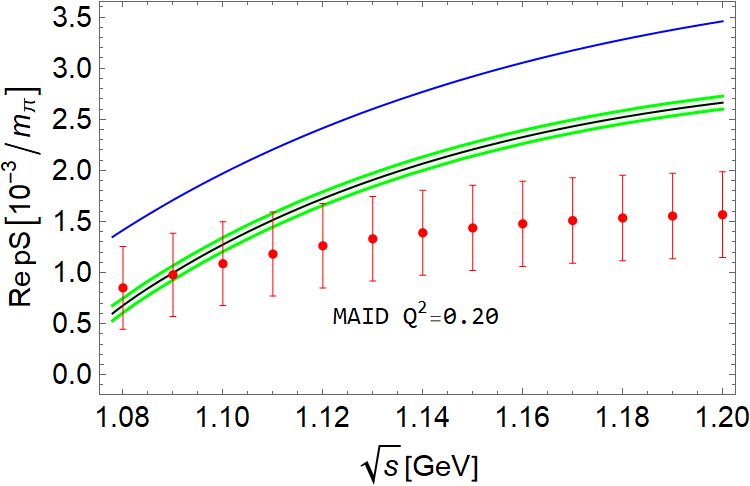}
    \vspace{0.02cm}
    \hspace{0.02cm}
    \end{minipage}
    \begin{minipage}[b]{0.245\linewidth}
    \includegraphics[width=3.63cm]{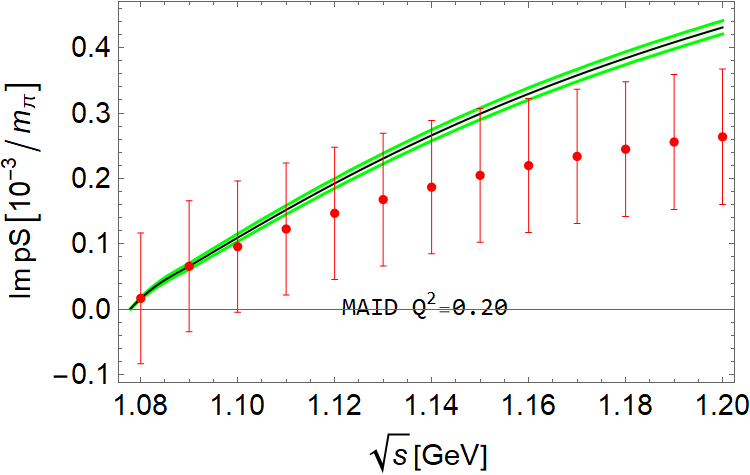}
    \vspace{0.02cm}
    \hspace{0.02cm}
    \end{minipage}
    \begin{minipage}[b]{0.245\linewidth}
    \includegraphics[width=3.63cm]{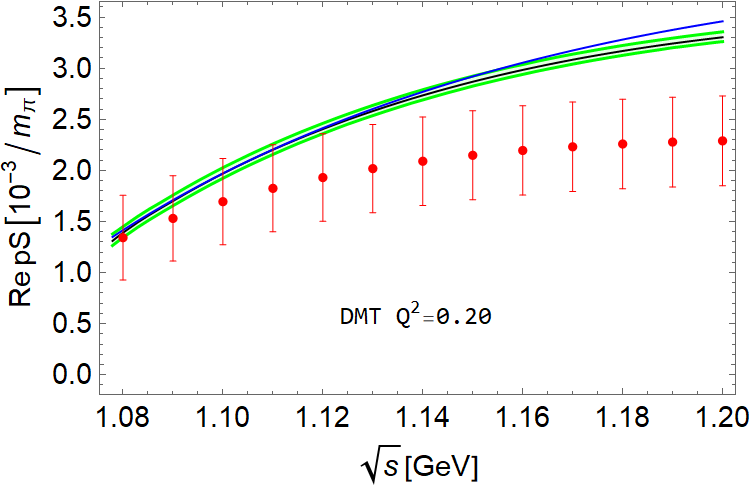}
    \vspace{0.02cm}
    \hspace{0.02cm}
    \end{minipage}
    \begin{minipage}[b]{0.245\linewidth}
    \includegraphics[width=3.63cm]{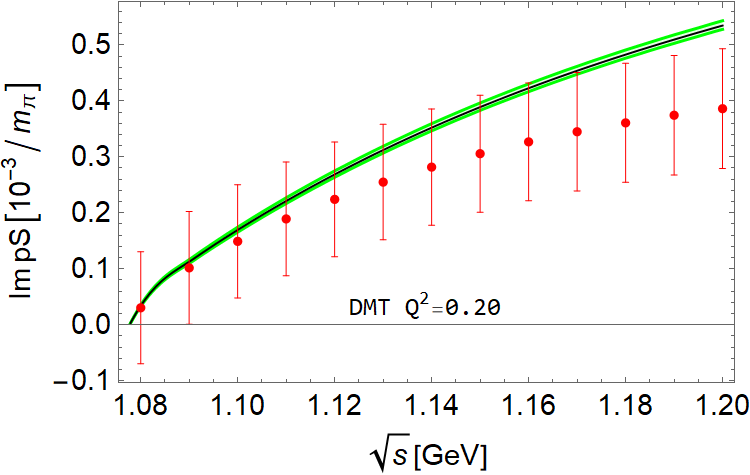}
    \vspace{0.02cm}
    \hspace{0.02cm}
    \end{minipage}    
    
    \end{minipage}
    }
    
\caption{$S_{11}S_{0+}$ for proton~($pS$): descriptions the same as in Fig.~\ref{f:pE}}\label{f:pS}
\end{figure}

\begin{figure}[H]
\centering
\subfigure{
    \begin{minipage}[b]{0.985\linewidth}
    
    \begin{minipage}[b]{0.245\linewidth}
    \includegraphics[width=3.63cm]{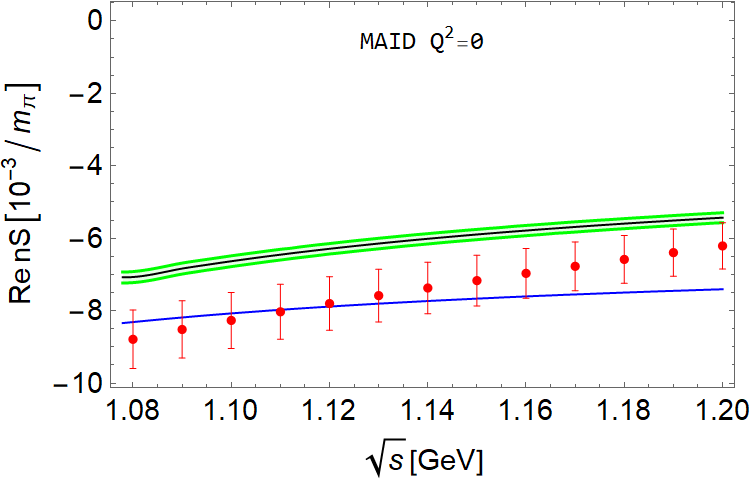}
    \vspace{0.02cm}
    \hspace{0.02cm}
    \end{minipage}
    \begin{minipage}[b]{0.245\linewidth}
    \includegraphics[width=3.63cm]{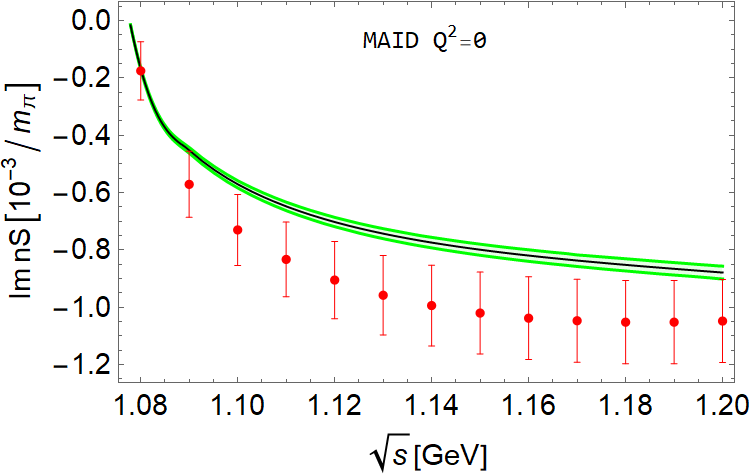}
    \vspace{0.02cm}
    \hspace{0.02cm}
    \end{minipage}
    \begin{minipage}[b]{0.245\linewidth}
    \includegraphics[width=3.63cm]{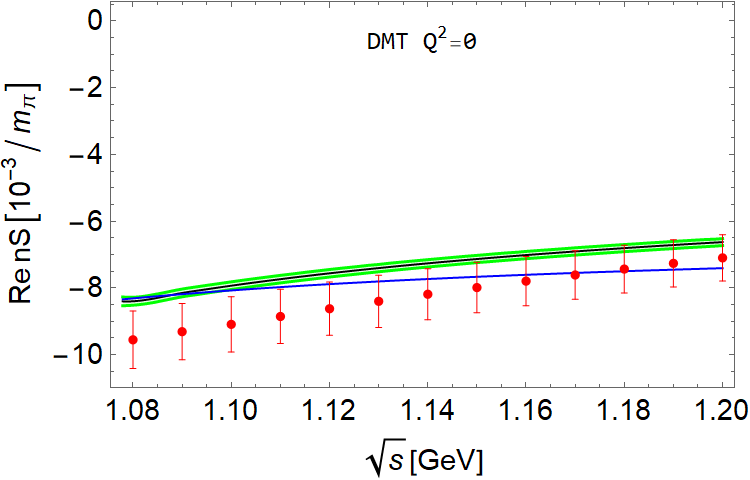}
    \vspace{0.02cm}
    \hspace{0.02cm}
    \end{minipage}
    \begin{minipage}[b]{0.245\linewidth}
    \includegraphics[width=3.63cm]{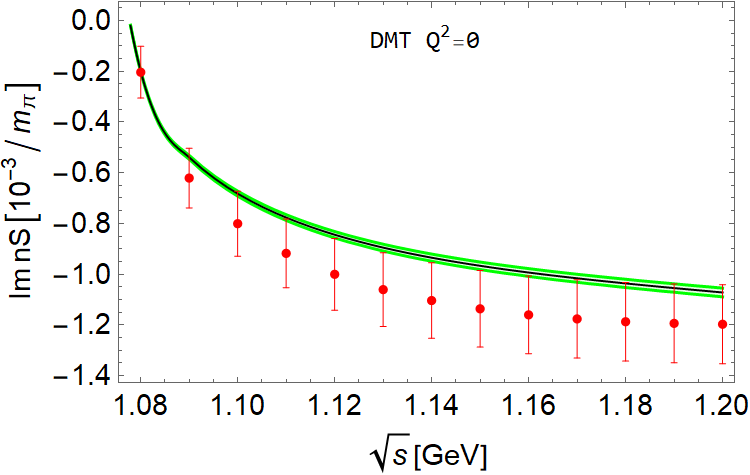}
    \vspace{0.02cm}
    \hspace{0.02cm}
    \end{minipage}
    
    \begin{minipage}[b]{0.24\linewidth}
    \includegraphics[width=3.63cm]{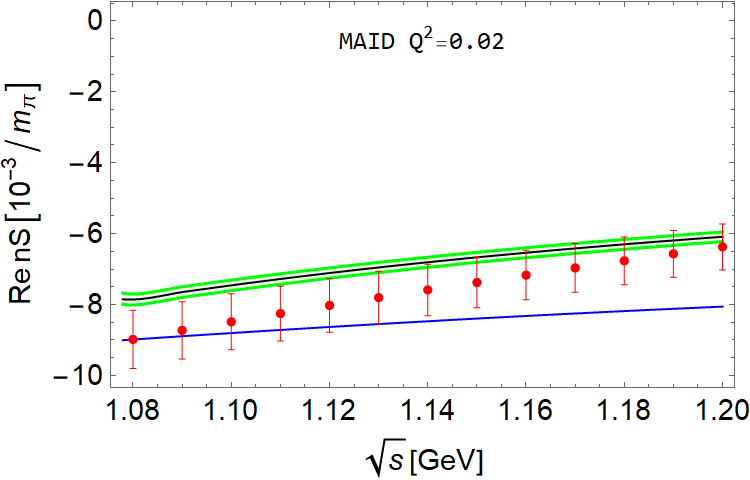}
    \vspace{0.02cm}
    \hspace{0.02cm}
    \end{minipage}
    \begin{minipage}[b]{0.24\linewidth}
    \includegraphics[width=3.63cm]{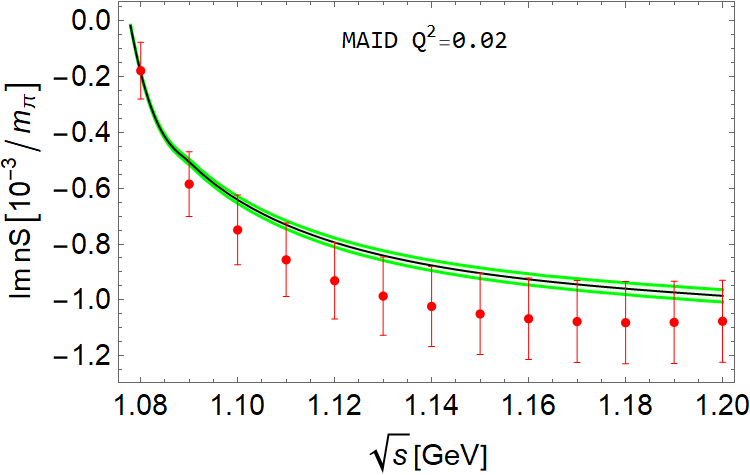}
    \vspace{0.02cm}
    \hspace{0.02cm}
    \end{minipage}
    \begin{minipage}[b]{0.24\linewidth}
    \includegraphics[width=3.63cm]{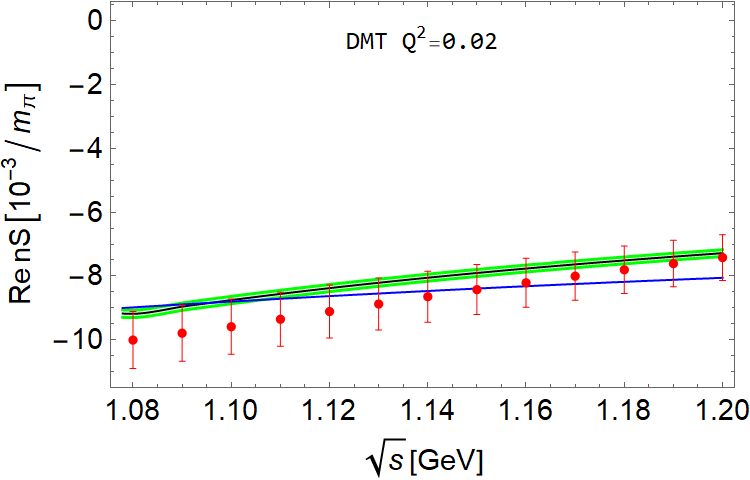}
    \vspace{0.02cm}
    \hspace{0.02cm}
    \end{minipage}
    \begin{minipage}[b]{0.24\linewidth}
    \includegraphics[width=3.63cm]{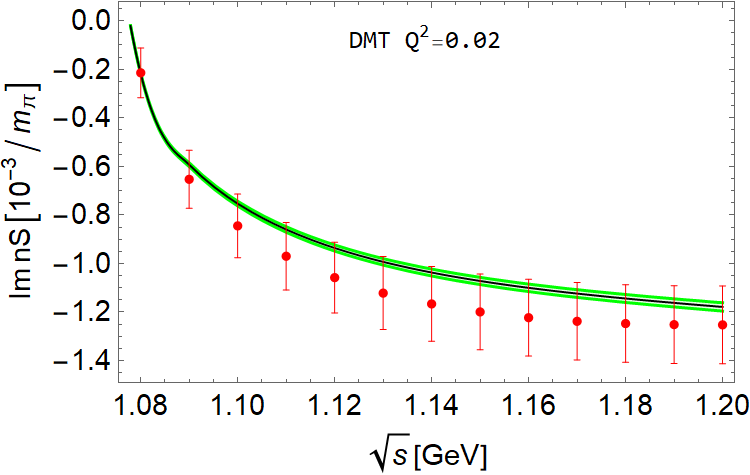}
    \vspace{0.02cm}
    \hspace{0.02cm}
    \end{minipage}
    
    \begin{minipage}[b]{0.245\linewidth}
    \includegraphics[width=3.63cm]{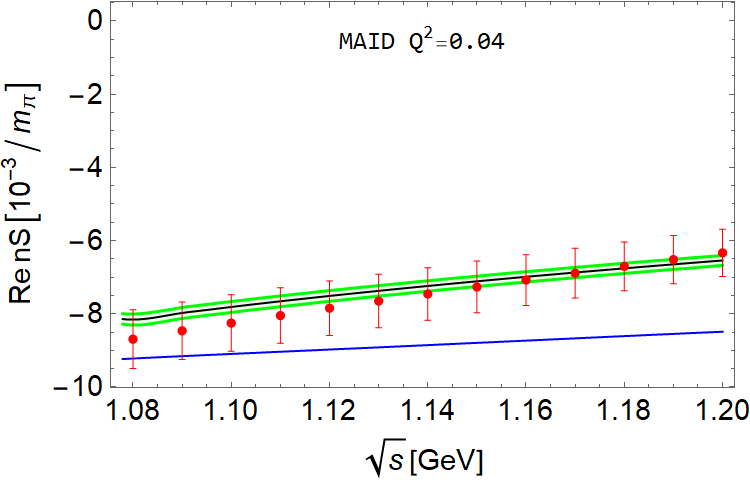}
    \vspace{0.02cm}
    \hspace{0.02cm}
    \end{minipage}
    \begin{minipage}[b]{0.245\linewidth}
    \includegraphics[width=3.63cm]{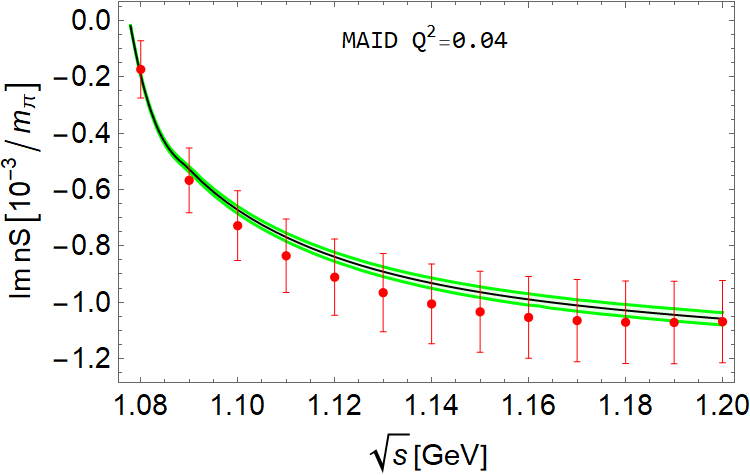}
    \vspace{0.02cm}
    \hspace{0.02cm}
    \end{minipage}
    \begin{minipage}[b]{0.245\linewidth}
    \includegraphics[width=3.63cm]{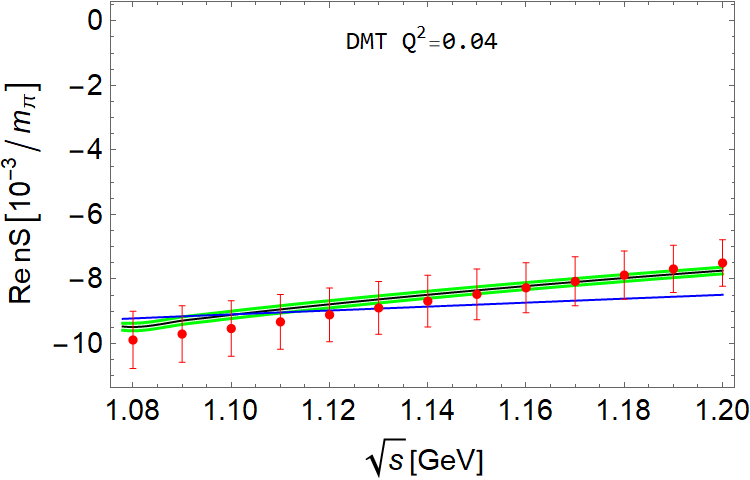}
    \vspace{0.02cm}
    \hspace{0.02cm}
    \end{minipage}
    \begin{minipage}[b]{0.245\linewidth}
    \includegraphics[width=3.63cm]{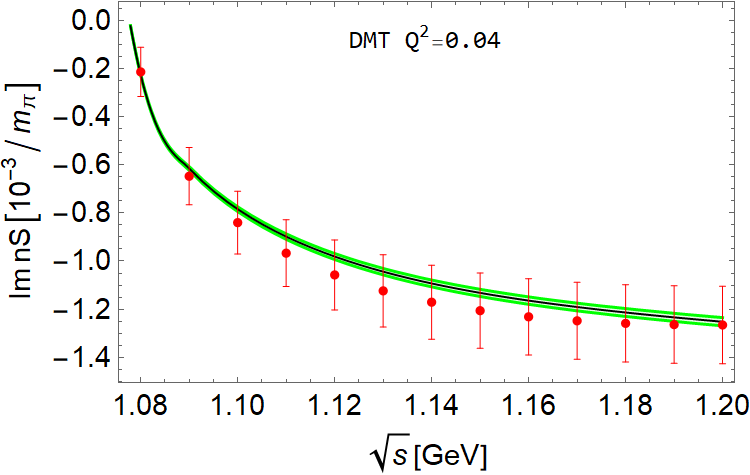}
    \vspace{0.02cm}
    \hspace{0.02cm}
    \end{minipage}
    
    \begin{minipage}[b]{0.245\linewidth}
    \includegraphics[width=3.63cm]{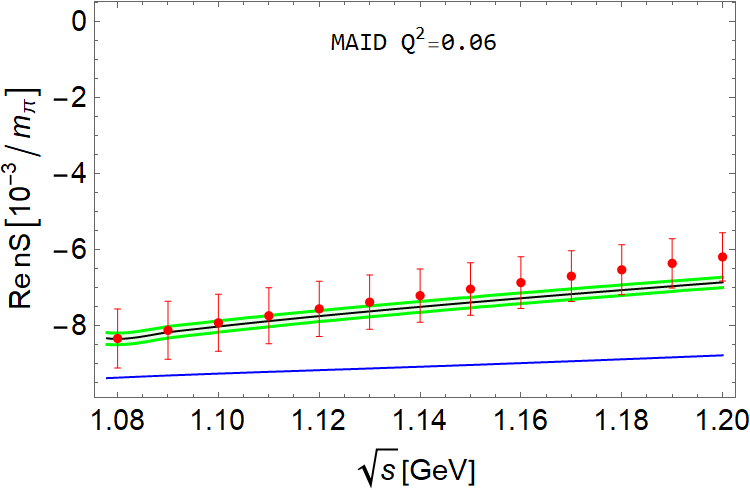}
    \vspace{0.02cm}
    \hspace{0.02cm}
    \end{minipage}
    \begin{minipage}[b]{0.245\linewidth}
    \includegraphics[width=3.63cm]{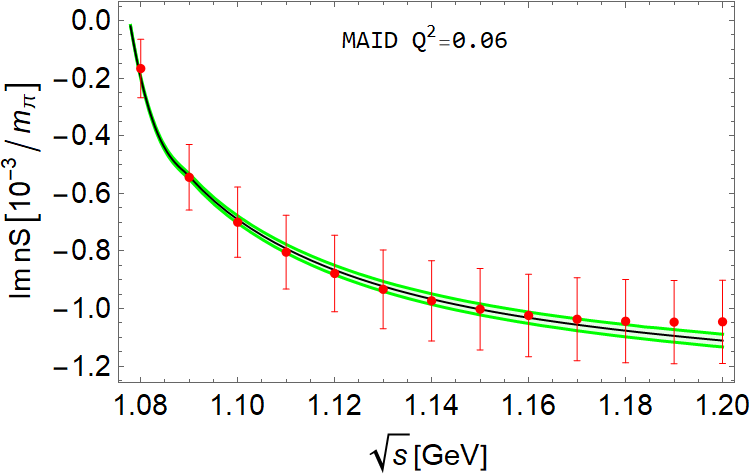}
    \vspace{0.02cm}
    \hspace{0.02cm}
    \end{minipage}
    \begin{minipage}[b]{0.245\linewidth}
    \includegraphics[width=3.63cm]{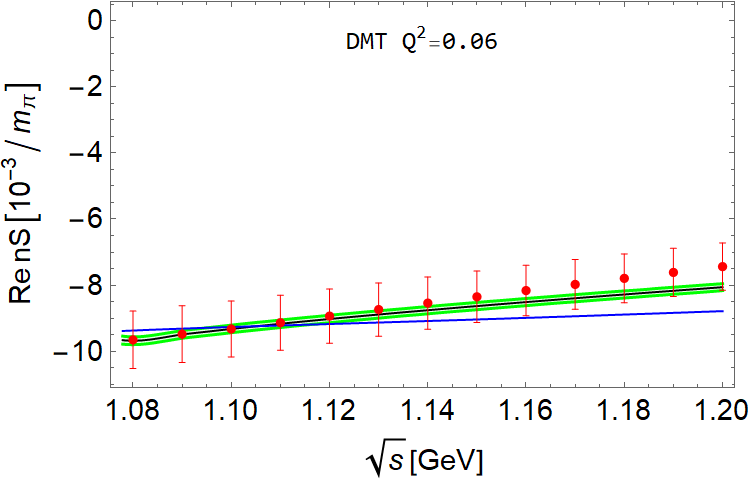}
    \vspace{0.02cm}
    \hspace{0.02cm}
    \end{minipage}
    \begin{minipage}[b]{0.245\linewidth}
    \includegraphics[width=3.63cm]{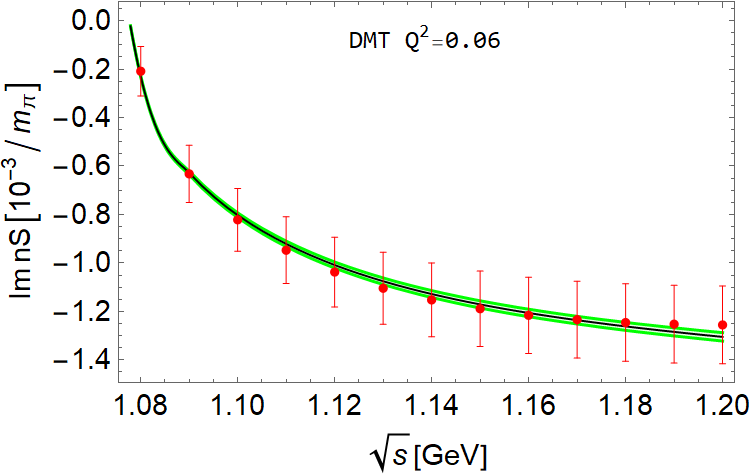}
    \vspace{0.02cm}
    \hspace{0.02cm}
    \end{minipage}
    
    \begin{minipage}[b]{0.245\linewidth}
    \includegraphics[width=3.63cm]{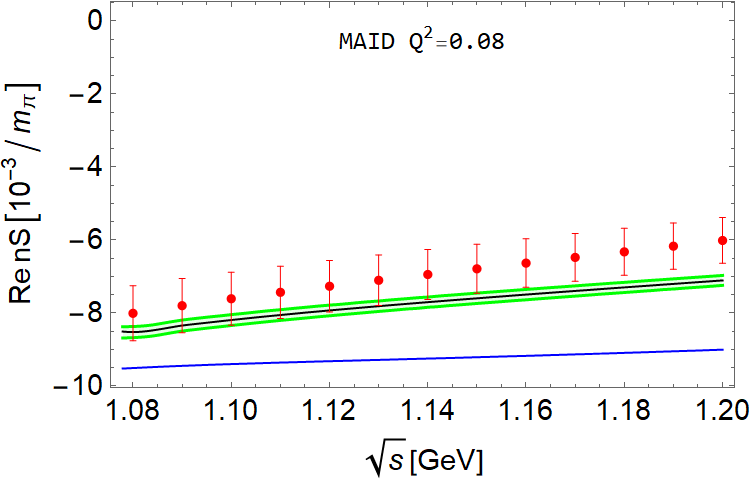}
    \vspace{0.02cm}
    \hspace{0.02cm}
    \end{minipage}
    \begin{minipage}[b]{0.245\linewidth}
    \includegraphics[width=3.63cm]{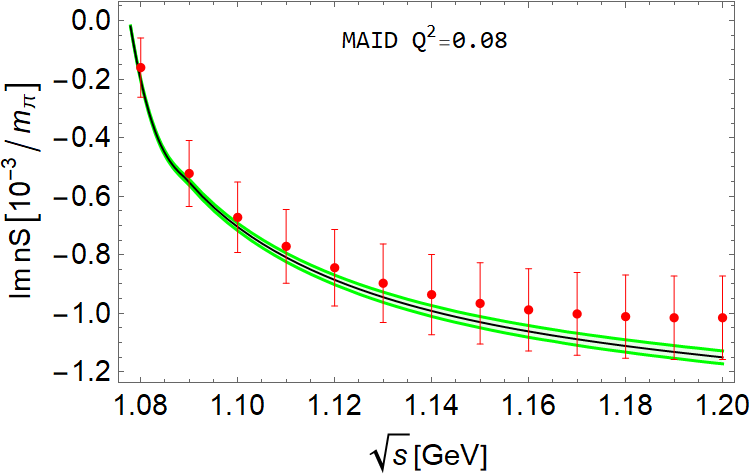}
    \vspace{0.02cm}
    \hspace{0.02cm}
    \end{minipage}
    \begin{minipage}[b]{0.245\linewidth}
    \includegraphics[width=3.63cm]{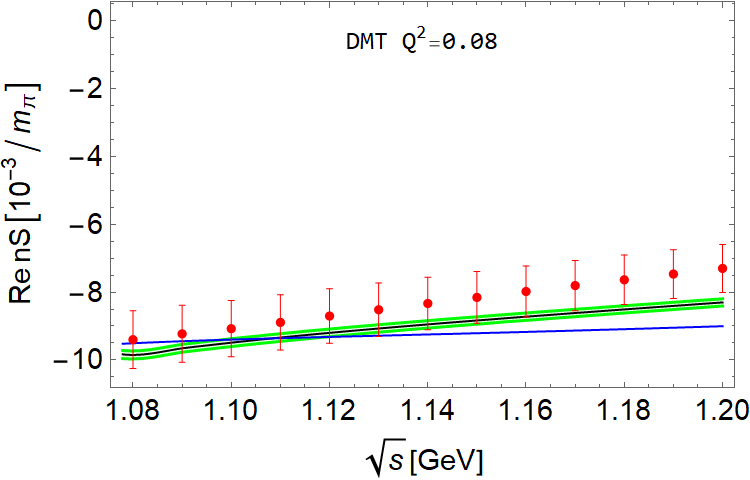}
    \vspace{0.02cm}
    \hspace{0.02cm}
    \end{minipage}
    \begin{minipage}[b]{0.245\linewidth}
    \includegraphics[width=3.63cm]{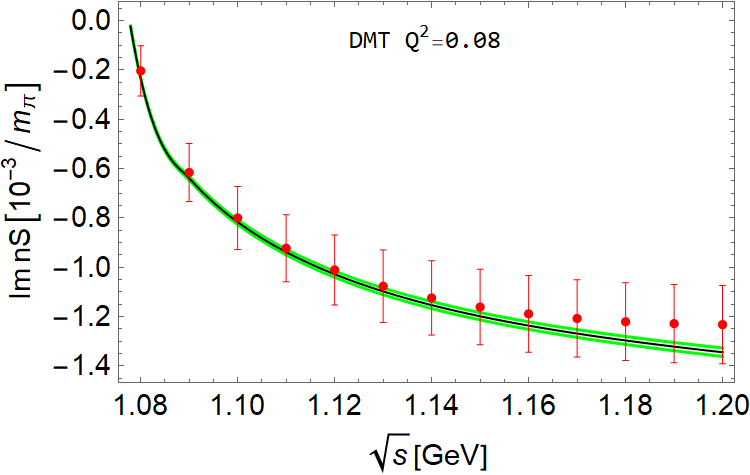}
    \vspace{0.02cm}
    \hspace{0.02cm}
    \end{minipage}
    
    \begin{minipage}[b]{0.245\linewidth}
    \includegraphics[width=3.63cm]{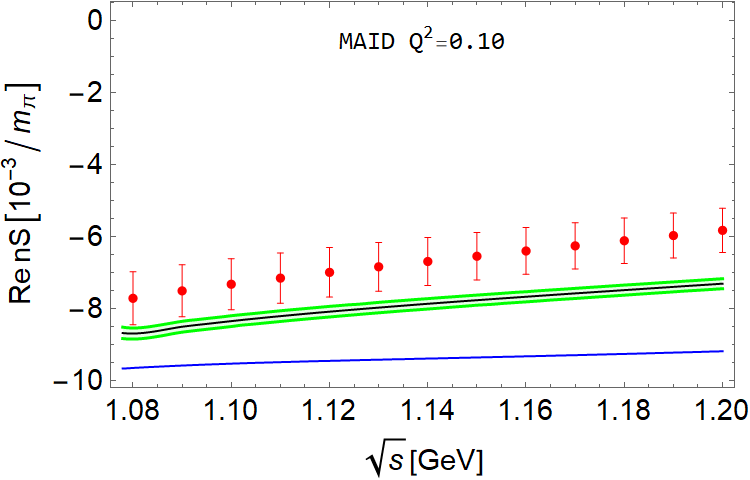}
    \vspace{0.02cm}
    \hspace{0.02cm}
    \end{minipage}
    \begin{minipage}[b]{0.245\linewidth}
    \includegraphics[width=3.63cm]{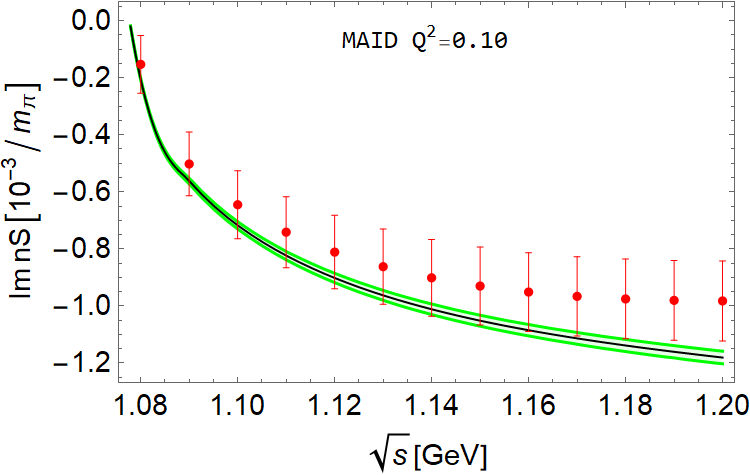}
    \vspace{0.02cm}
    \hspace{0.02cm}
    \end{minipage}
    \begin{minipage}[b]{0.245\linewidth}
    \includegraphics[width=3.63cm]{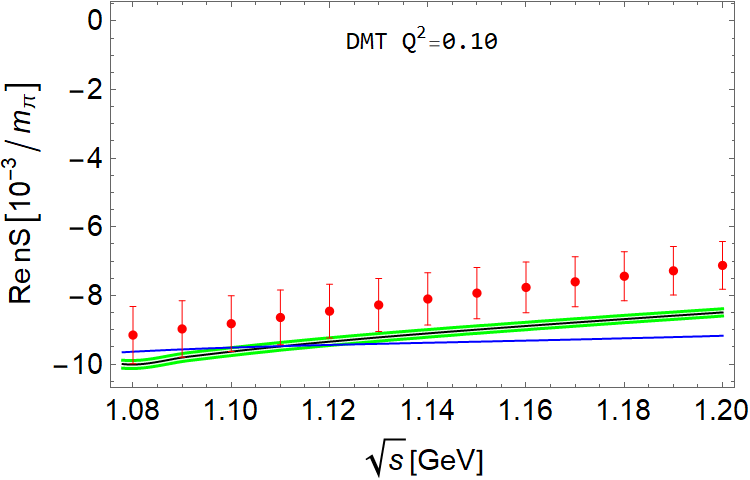}
    \vspace{0.02cm}
    \hspace{0.02cm}
    \end{minipage}
    \begin{minipage}[b]{0.245\linewidth}
    \includegraphics[width=3.63cm]{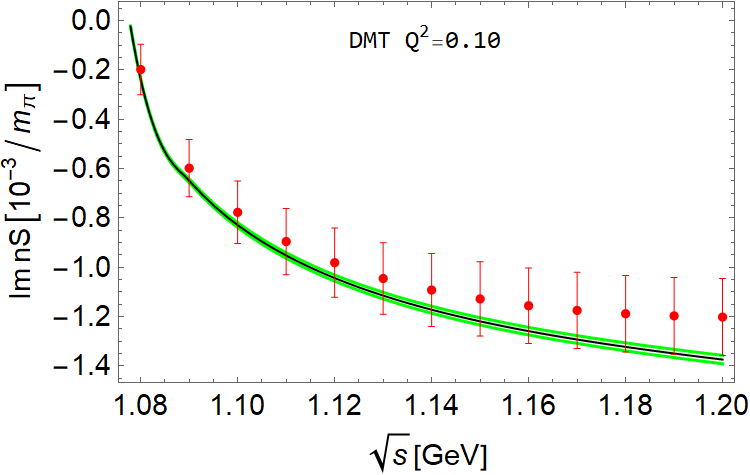}
    \vspace{0.02cm}
    \hspace{0.02cm}
    \end{minipage}
    
    \begin{minipage}[b]{0.245\linewidth}
    \includegraphics[width=3.63cm]{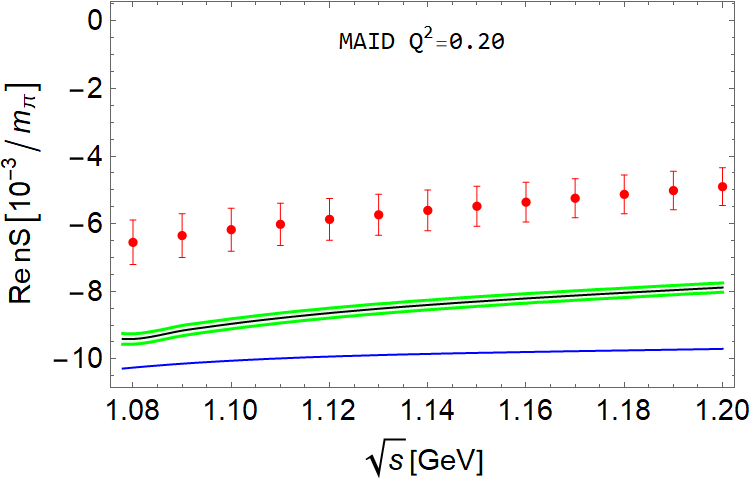}
    \vspace{0.02cm}
    \hspace{0.02cm}
    \end{minipage}
    \begin{minipage}[b]{0.245\linewidth}
    \includegraphics[width=3.63cm]{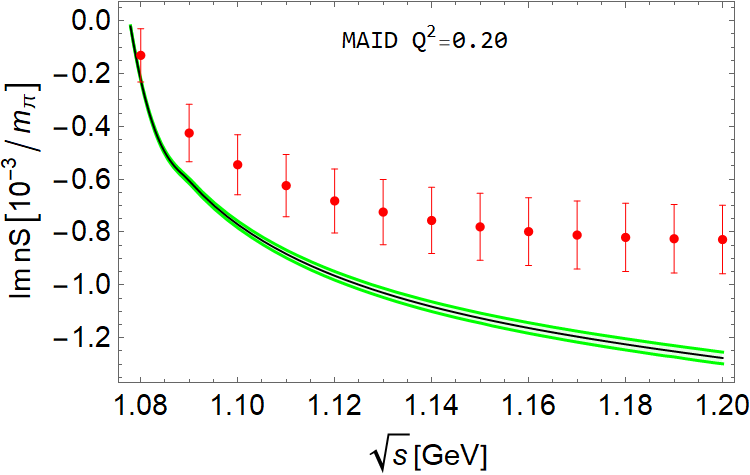}
    \vspace{0.02cm}
    \hspace{0.02cm}
    \end{minipage}
    \begin{minipage}[b]{0.245\linewidth}
    \includegraphics[width=3.63cm]{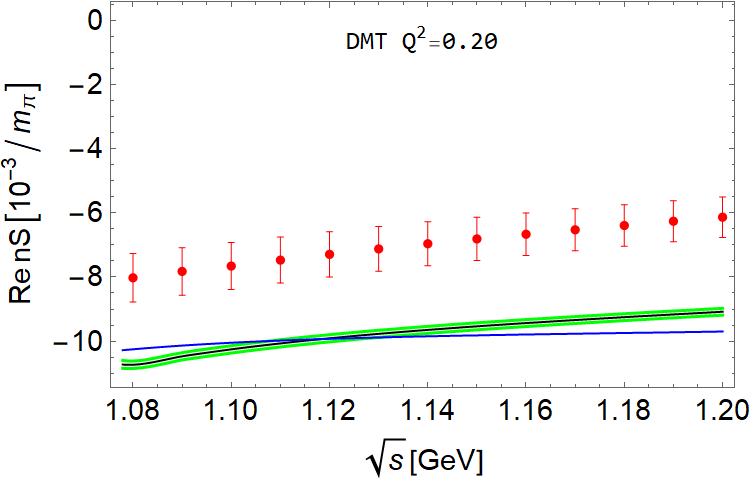}
    \vspace{0.02cm}
    \hspace{0.02cm}
    \end{minipage}
    \begin{minipage}[b]{0.245\linewidth}
    \includegraphics[width=3.63cm]{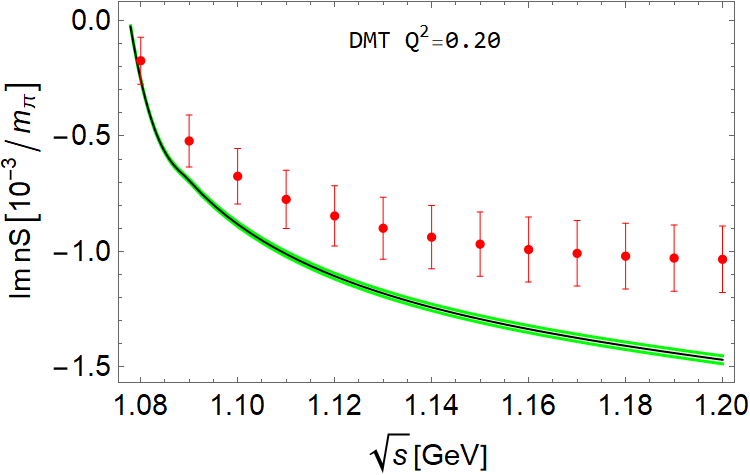}
    \vspace{0.02cm}
    \hspace{0.02cm}
    \end{minipage}    
    
    \end{minipage}
    }
\caption{$S_{11}S_{0+}$ for neutron~($nS$): descriptions the same as in Fig.~\ref{f:pE}}\label{f:nS}
\end{figure}

In the numerical analyses, we fit the multipole amplitudes with the $S_{11}$ channel from \(\pi N\) threshold to \(1.440~\rm GeV^2\) just below the resonance \(\Delta(1232)\).
Since no error bars are given, we assign them according to Refs.~\cite{Chen:2012nx, Wang2019},\begin{align}
\operatorname{err}(\mathcal{M}_l^I)=\sqrt{\left(e_{s}^{R,I}\right)^2+\left(e_{r}^{R,I}\right)^2 \left(\mathcal{M}_l^I\right)^{2}}\ .
\end{align}
Here the superscripts $R,I$ represent the real and imaginary parts of the amplitude.
We choose $e_{s}^{R,I}=0.4,0.1 [10^{-3}/m_\pi],\ e_{r}^{R,I}=10\%$.
We take into account the errors caused by the model dependence of the partial wave data as much as possible~\cite{Drechsel:2007if, Pascalutsa:2006up}.
The fit results to MAID2007 and DMT2001 data are displayed in Figs.~\ref{f:pE},~\ref{f:nE} and \ref{f:pS},~\ref{f:nS}, respectively.
For comparison, we also show the \(\mathcal{O}(p^2)\)  chiral results of multipole amplitudes. 
As expected, the chiral results only describe the data at low energies close to threshold and in low $Q^2$.
The values of the fit parameters are collected in Table~\ref{pfrq}.

\begin{table}[H]\footnotesize
    \centering
    \caption{Fit results of once subtraction ($\mathcal{P}=a$). The parameter \(a\) is given in unit of $[10^{-3}/m_\pi]$.}\label{pfrq}
    \begin{threeparttable}
        \begin{tabular}{@{}ccccc}
            \toprule
            Multipole&Target&Case&Value&\(\chi^2/d.o.f\) \\
            \midrule
            \multirow{4}*{$E_{0+}$}&\multirow{2}*{p}&MAID&\(-0.12 \pm 0.05\)&\(0.46\) \\
            \cline{3-5}
            &&DMT&\(0.21 \pm 0.03\)&\(0.17\) \\
            \cline{2-5}
            &\multirow{2}*{n}&MAID&\(2.11 \pm 0.07\)&\(0.71\) \\
            \cline{3-5}
            &&DMT&\(1.25 \pm 0.06\)&\(0.49\) \\
            \midrule
            \multirow{4}*{$S_{0+}$}&\multirow{2}*{p}&MAID&\(-1.07 \pm 0.03\)&\(0.49\) \\
            \cline{3-5}
            &&DMT&\(-0.46 \pm 0.02\)&\(0.23\) \\
            \cline{2-5}
            &\multirow{2}*{n}&MAID&\(2.25 \pm 0.06\)&\(1.14\) \\
            \cline{3-5}
            &&DMT&\(1.13 \pm 0.05\)&\(0.60\) \\
            \bottomrule
        \end{tabular}
    \end{threeparttable}
\end{table}

In Figs.~\ref{f:pE},~\ref{f:nE} and \ref{f:pS},~\ref{f:nS}, we fit amplitudes from $Q^2=0$ to $Q^2=0.1 \mathrm{GeV^2}$ in the increments of $0.02\mathrm{GeV^2}$.
We also draw the result where $Q^2=0.2\mathrm{GeV^2}$.
It can be seen that, except that the fit to $pS_{0+}$ is rather good, the other fit results do not improve much when $Q^2=0.2\mathrm{GeV^2}$. This is within the expectation that we did not consider corrections of vector meson exchanges~\cite{Kubis:2000aa, Kubis:2000zd}.
In general, it can be seen in the Table~\ref{pfrq} that our results are in good agreement with the experimental data.
It is observed that the imaginary part is an order of magnitude smaller than the real part since it is of higher orders in $\chi$PT expansions.
Moreover, the agreement is a direct consequence of unitarity and follows automatically from Watson's theorem~\cite{Watson:1954uc}.
Meanwhile, the central value of $a$ is very small in any case. That can be understood by the fact that multipoles calculated from $\chi$PT and the unitarity method can already well describe the experimental data.
We also do the fit which uses a twice subtraction polynomial; i.e., $\mathcal{P}=a+bs$.
However, the fit parameters \(a\) and \(b\) are found to be highly negative correlated.
Thus, once subtraction is more advisable.

In a $\chi$PT calculation, the source of error is the systematical one of the theory due to the truncation of the chiral expansion at a given $\mathcal{O}(p^n)$.
Using the method of \cite{Navarro:2019iqj, Navarro:2020zqn}, for an order $n$ calculation $\mathcal{O}(p^n)$,
we estimate this systematical error as
\begin{align}
    \delta O_{\mathrm{Th}}^{(n)}=\max \left(\left|O^{\left(n_{L O}\right)}\right| B^{n-n_{L O}+1},\left\{\left|O^{(k)}-O^{(l)}\right| B^{n-l}\right\}\right), \quad n_{L O} \leq l \leq k \leq n\ ,
  \end{align}
where $B=m_{\pi} / \Lambda_{b}$, and $\Lambda_{b}=4 \pi F_{\pi} \sim 1 \mathrm{GeV}$ is the breakdown scale of the chiral expansion.
Here, we set $n_{L O}=1$, and $n=2$.
The error is roughly estimated to be less than $5\%$.

Furthermore, the possible errors caused by the truncation of dispersion integration can be estimated, 
\begin{align}\label{truncation}
  I_{\alpha}=-\frac{s}{\pi} \int_{s_R}^{\Lambda} \frac{\left(\operatorname{Im} \Omega\left(s^{\prime}\right)^{-1}\right) \mathcal{M}_{\alpha}\left(s^{\prime}\right)}{s^{\prime}\left(s^{\prime}-s\right)} \mathrm{d} s^{\prime}\ ,
\end{align}
where $\alpha=pE,nE,pS,nS$, represent muitipoles in $p,n$ targets, respectively. 
We set the upper limit of the dispersion integrals, $\Lambda$, to $2.1\mathrm{GeV}^2$. The upper
limit in DRs has less physical significance, and it is rather an indicator of how well one estimates
the remainder of the integral. As a comparison, we list the integration results 
with the cutoff set to $2.2\mathrm{GeV}^2$ and $2.5\mathrm{GeV}^2$. 
The imaginary part of the dispersion integral function $I_\alpha$ is 
almost insensitive to truncation.
However, the real part of $I_\alpha$ is somewhat sensitive 
to truncation, and has a weak linear dependence on truncation as a whole, but this dependence 
can be compensated by the subtraction constant of the dispersion integral. 
Finally, the overall physical results are not sensitive to the dispersion integral truncation.
\begin{figure}[H]
    \centering
    \subfigure{
        \begin{minipage}[b]{0.985\linewidth}
          
        \begin{minipage}[b]{0.45\linewidth}
        \includegraphics[width=6.5cm]{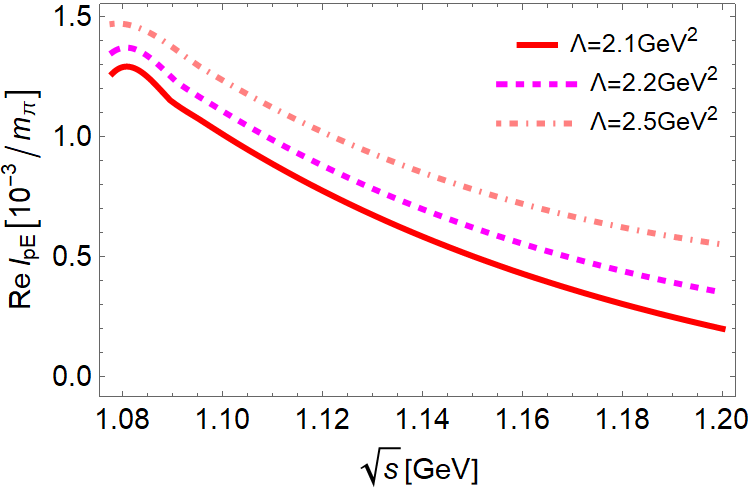}
        \vspace{0.02cm}
        \hspace{0.02cm}
        \end{minipage}
        \begin{minipage}[b]{0.45\linewidth}
        \includegraphics[width=6.5cm]{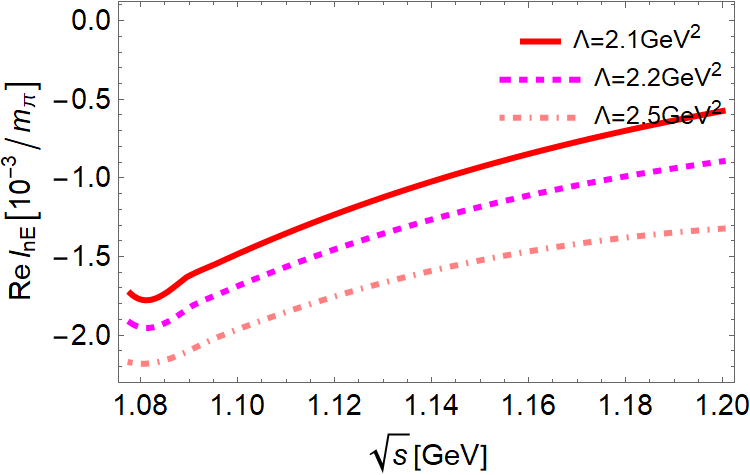}
        \vspace{0.02cm}
        \hspace{0.02cm}
        \end{minipage}
    
        \begin{minipage}[b]{0.45\linewidth}
        \includegraphics[width=6.5cm]{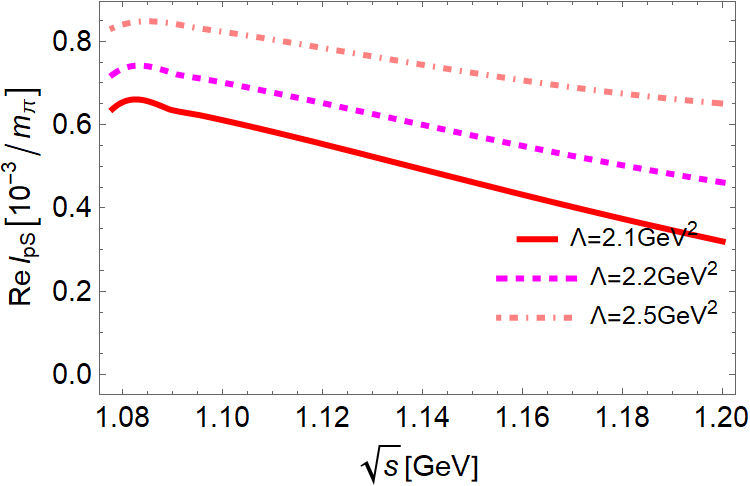}
        \vspace{0.02cm}
        \hspace{0.02cm}
        \end{minipage}  
        \begin{minipage}[b]{0.45\linewidth}
        \includegraphics[width=6.5cm]{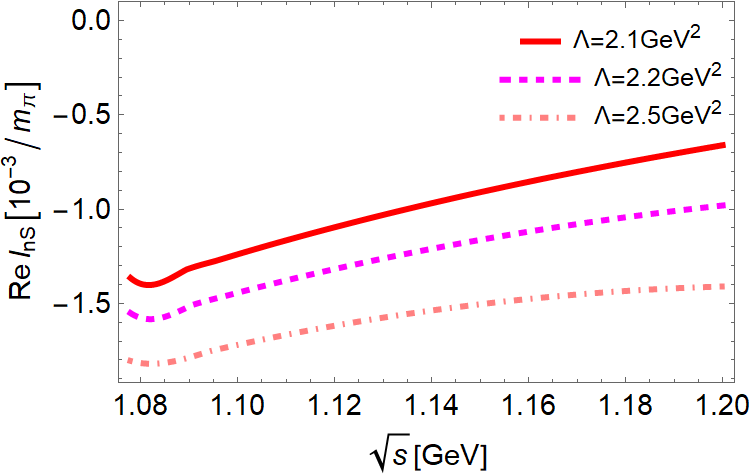}
        \vspace{0.02cm}
        \hspace{0.02cm}
        \end{minipage}
   
        \end{minipage}
        }
    \caption{Tests of cutoff dependence. $\Lambda$ corresponds 
    to truncation of integrand in Eq.~(\ref{truncation}). }
    \end{figure}

It is convincing that no matter what data are used, the fit results of multipole amplitudes are very similar.
The results can illustrate that our unitarity method is very powerful and effective in low energy regions and low $Q^2$ regions.
But our results do not fit well in the case of high $Q^2$.
In our future work, we will consider using resonance $\chi$PT to improve the description of this method at high $Q^2$.

At last, we make a brief comparison of our work and the work of Hilt \emph{et al}.
Here we find that the $\mathcal{O}(p^4)$ results~\cite{Hilt:2013fda} and our amplitudes of multipole $S_{11}pE_{0+}$ are different in higher $Q^2$ regions.
Furthermore, $S_{11}nE_{0+},~S_{11}pS_{0+}$, and $S_{11}nS_{0+}$ are even different from our calculations at lower $Q^2$.
This needs to be clarified in the future.
For comparison, we list MAID , DMT model, our results, and that of~\cite{Hilt:2013fda}.
\begin{figure}[H]
\centering
\subfigure{
    \begin{minipage}[b]{0.985\linewidth}
    
    \begin{minipage}[b]{0.45\linewidth}
    \includegraphics[width=6.5cm]{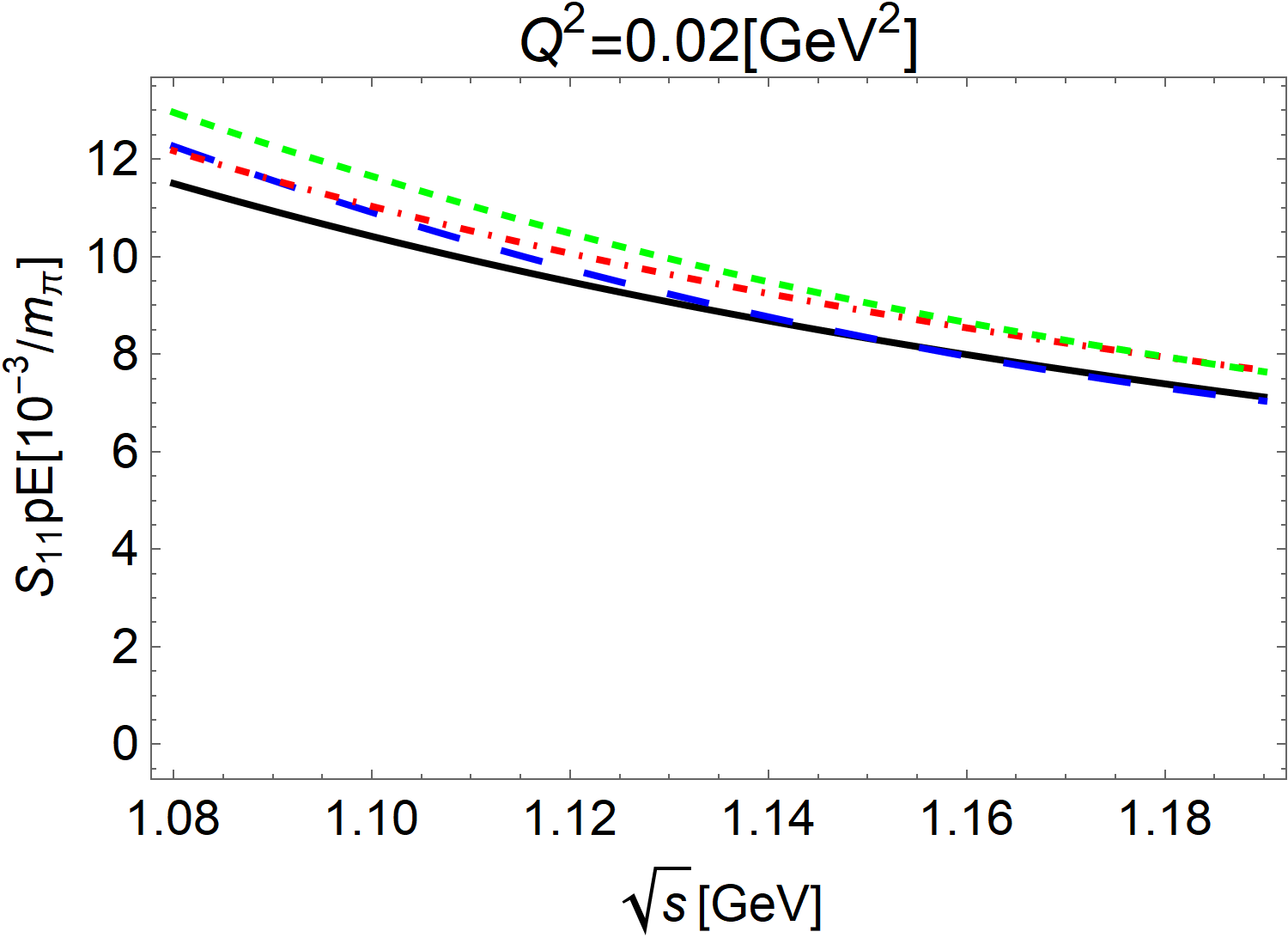}
    \vspace{0.02cm}
    \hspace{0.02cm}
    \end{minipage}
    \begin{minipage}[b]{0.45\linewidth}
    \includegraphics[width=6.7cm]{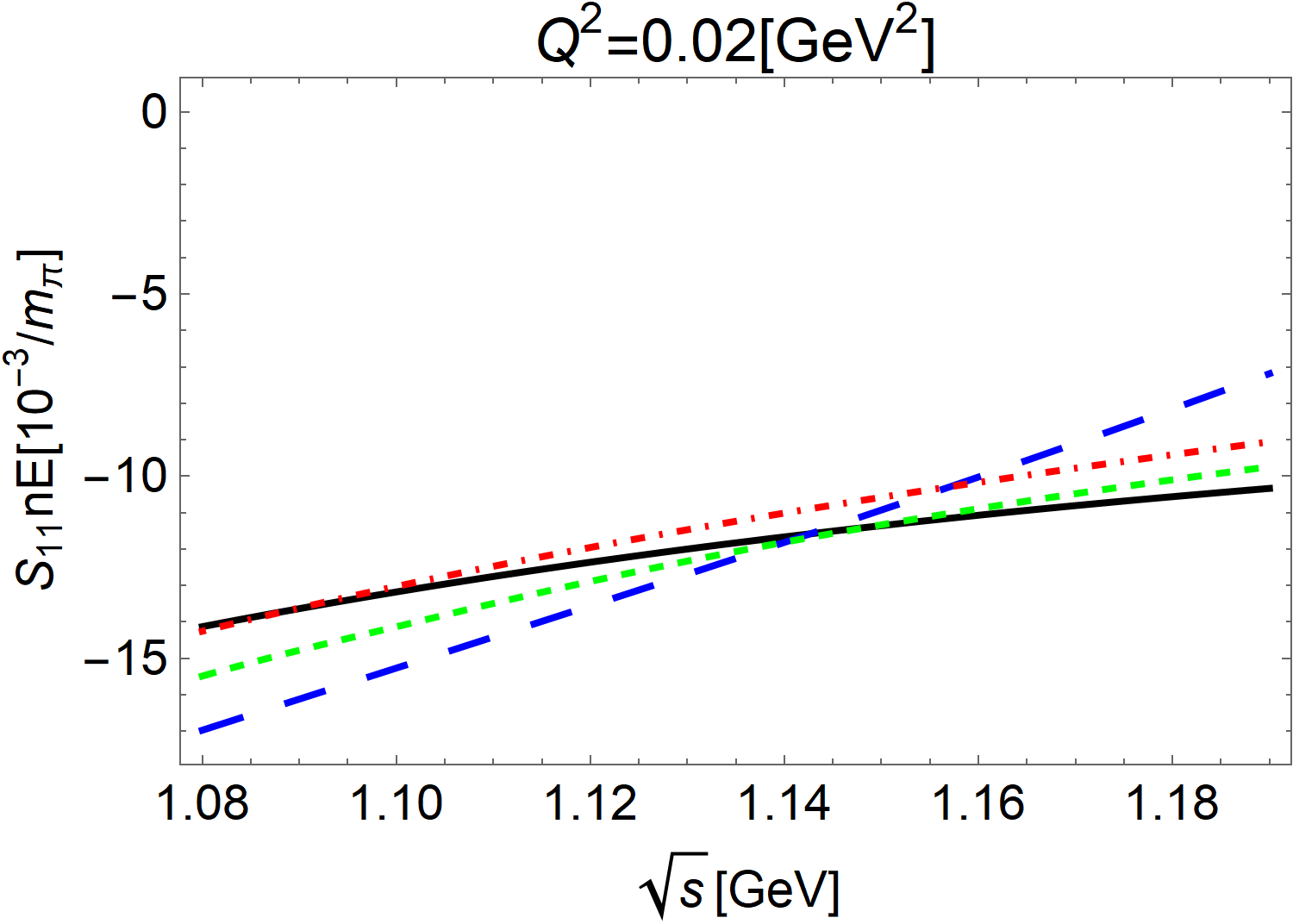}
    \vspace{0.02cm}
    \hspace{0.02cm}
    \end{minipage}

    \begin{minipage}[b]{0.45\linewidth}
    \includegraphics[width=6.5cm]{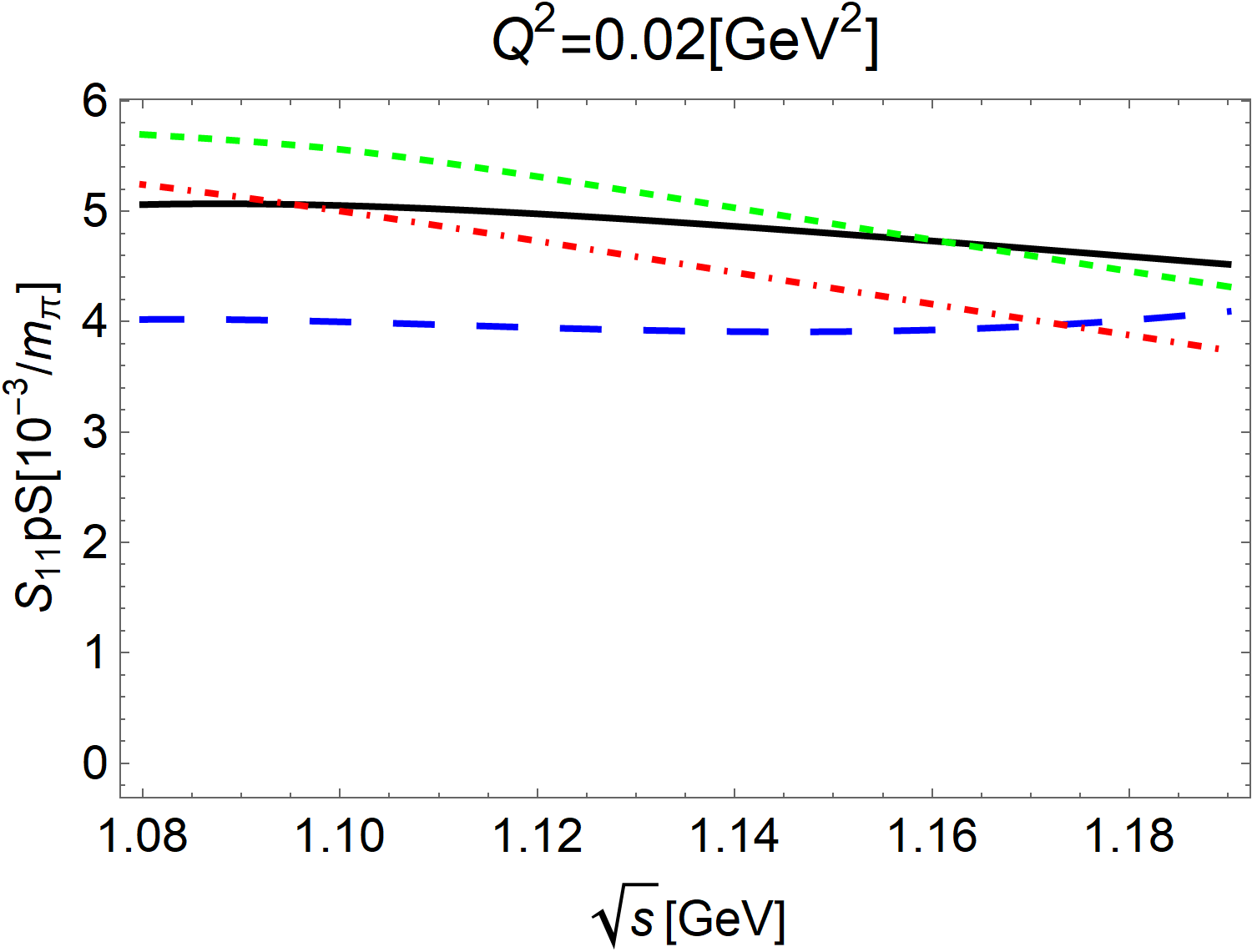}
    \vspace{0.02cm}
    \hspace{0.02cm}
    \end{minipage}  
    \begin{minipage}[b]{0.45\linewidth}
    \includegraphics[width=6.7cm]{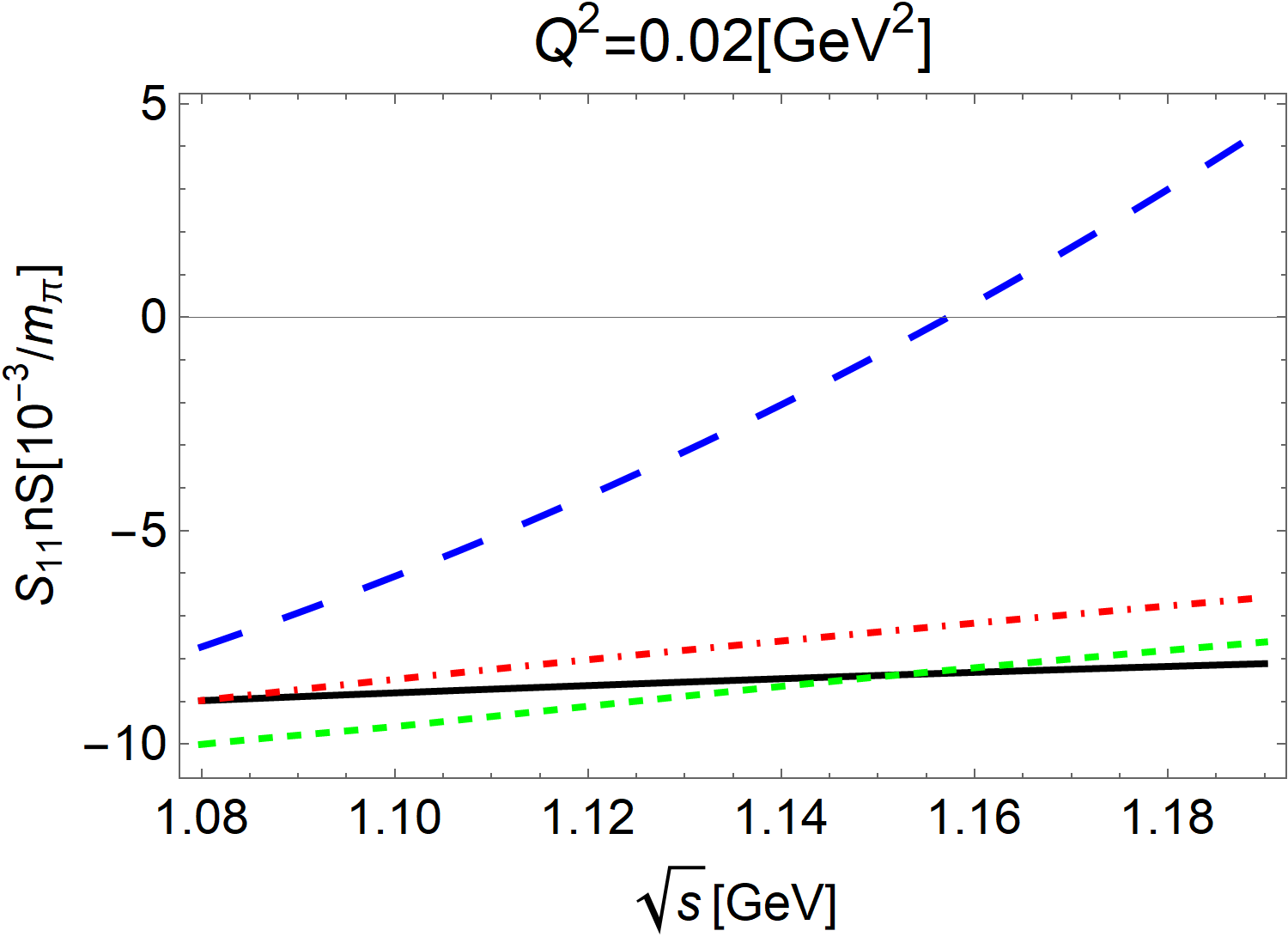}
    \vspace{0.02cm}
    \hspace{0.02cm}
    \end{minipage}
    \end{minipage}
    }
\caption{The black solid curves show our calculations at $\mathcal{O}(p^2)$ and the blue long-dashed curves are the outputs of the $\mathcal{O}(p^4)$ results~\cite{Hilt:2013fda}. The red dot-dashed and green short-dashed curves are the predictions of MAID and DMT model, respectively.}
\end{figure}

\section{Summary}\label{sec:Clu}

In this paper, we have performed a careful dispersive analysis about the process of single pion electroproduction off the nucleon in the \(S_{11}\) channel of the final $\pi N$ system.
In the dispersive representation, the right-hand cut contribution can be related to an Omn\`es solution, which takes the elastic \(\pi N\) phase shifts as inputs.
At the same time, we estimate the left-hand cut contribution by making use of the \(\mathcal{O}(p^2)\) amplitudes taken from $\chi$PT.
A detailed discussion on a virtual photoproduction amplitude at the level of multipoles is presented. 
Different from Refs.~\cite{Hilt:2013fda, Navarro:2020zqn}, here we go beyond pure $\chi$PT calculations by applying the final state interaction theorem to partial wave amplitudes.
To pin down the free parameters in the dispersive amplitude, we perform fits to the experimental data of multipole amplitudes \(E_{0+}\) and \(S_{0+}\) for the energies ranging from \(\pi N\) threshold to \(1.440~\rm GeV^2\). 
It is found that the experimental data can be well described by the dispersive amplitude with only one free subtraction parameter, when $Q^2\leq0.1\mathrm{GeV}^2$. 
As $Q^2$ further increases to $0.2\mathrm{GeV}^2$, the fit fails, similar to what happened in the literature~\cite{Kubis:2000aa, Kubis:2000zd, Hilt:2013fda}.

Our dispersive approach does not always do better as compared with the pure $\chi$PT results apparently, for fitting the real parts of multipole amplitudes.
However, the power of dispersion relations is nicely visible in the imaginary parts of multipoles.
In this situation, even at low energies, the $\mathcal{O}(p^2)$ perturbative calculation is not sufficient. Therefore, our method is superior to $\mathcal{O}(p^2)$ perturbation theory in the sense that DRs can generate the corresponding imaginary parts. 
Further, it is hard to compare the $\mathcal{O}(p^4)$ $\chi$PT results and our calculations.
In our calculation, we only use the left-hand part contribution extracted from the $\mathcal{O}(p^2)$ amplitude.
In principle, an $\mathcal{O}(p^4)$ calculation is advantageous compared with an $\mathcal{O}(p^2)$ calculation.
But in a perturbation calculation, unitarization effects are not taken into account, which are automatically fulfilled in our scheme.

\section*{Acknowledgments}
The authors would like to thank De-Liang Yao, Yu-Fei Wang, and Wen-Qi Niu for helpful discussions.
This work is supported in part by National Nature Science Foundations of China (NSFC) under Contracts No.11975028 and No.10925522.

\begin{appendices}
 
\section{Invariant amplitudes}\label{ap:amp}

\begin{align}
    \begin{aligned}
    A_{1}^{(+)} &=-\frac{e g_{A} m_{N}}{2 F}\left(\frac{1}{s-m_{N}^{2}}+\frac{1}{u-m_{N}^{2}}\right)-\frac{e g_{A} c_{6}}{F}, \\
    A_{2}^{(+)} &=\frac{- e g_{A} m_{N}}{F} \frac{1}{t-m_{\pi}^{2}}\left(\frac{1}{s-m_{N}^{2}}+\frac{1}{u-m_{N}^{2}}\right), \\
    A_{3}^{(+)} &=\frac{e g_{A} m_{N} c_{6}}{F}\left(\frac{1}{s-m_{N}^{2}}-\frac{1}{u-m_{N}^{2}}\right),  \\
    A_{4}^{(+)} &=\frac{e g_{A} m_{N} c_{6}}{F}\left(\frac{1}{s-m_{N}^{2}}+\frac{1}{u-m_{N}^{2}}\right), \\
    A_{5}^{(+)} &=-\frac{e g_{A} m_{N}}{2 F} \frac{1}{t-m_{\pi}^{2}}\left(\frac{1}{s-m_{N}^{2}}-\frac{1}{u-m_{N}^{2}}\right), \\
    A_{6}^{(+)} &=0,  \\
    A_{1}^{(-)} &=\frac{- e g_{A} m_{N}}{2 F}\left(\frac{1}{s-m_{N}^{2}}-\frac{1}{u-m_{N}^{2}}\right), \\
    A_{2}^{(-)} &=\frac{- e g_{A} m_{N}}{F} \frac{1}{t-m_{\pi}^{2}}\left(\frac{1}{s-m_{N}^{2}}-\frac{1}{u-m_{N}^{2}}\right), \\
    A_{3}^{(-)} &=\frac{e g_{A} m_{N} c_{6}}{F}\left(\frac{1}{s-m_{N}^{2}}+\frac{1}{u-m_{N}^{2}}\right),  \\
    A_{4}^{(-)} &=\frac{e g_{A} m_{N} c_{6}}{F}\left(\frac{1}{s-m_{N}^{2}}-\frac{1}{u-m_{N}^{2}}\right), \\
    A_{5}^{(-)} &=-\frac{e g_{A} m_{N}}{2 F} \frac{1}{t-m_{\pi}^{2}}\left(\frac{1}{s-m_{N}^{2}}+\frac{1}{u-m_{N}^{2}}\right), \\
    A_{6}^{(-)} &=0,  \\
    A_{1}^{(0)} &=\frac{- e g_{A} m_{N}}{2 F}\left(\frac{1}{s-m_{N}^{2}}+\frac{1}{u-m_{N}^{2}}\right)-\frac{e g_{A} c_{7}}{2F}, \\
    A_{2}^{(0)} &=\frac{- e g_{A} m_{N}}{F} \frac{1}{t-m_{\pi}^{2}}\left(\frac{1}{s-m_{N}^{2}}+\frac{1}{u-m_{N}^{2}}\right), \\
    A_{3}^{(0)} &=\frac{e g_{A} m_{N} c_{7}}{2F}\left(\frac{1}{s-m_{N}^{2}}-\frac{1}{u-m_{N}^{2}}\right),  \\
    A_{4}^{(0)} &=\frac{e g_{A} m_{N} c_{7}}{2F}\left(\frac{1}{s-m_{N}^{2}}+\frac{1}{u-m_{N}^{2}}\right), \\
    A_{5}^{(0)} &=-\frac{e g_{A} m_{N}}{2 F} \frac{1}{t-m_{\pi}^{2}}\left(\frac{1}{s-m_{N}^{2}}-\frac{1}{u-m_{N}^{2}}\right), \\
    A_{6}^{(0)} &=0\ ,
    \end{aligned}
\end{align}
where the two LECs $F$ and $g_A$ denote the chiral limit of pion decay constant and the axial-vector coupling constant, respectively. 
Here \(c_{6}\) and \(c_7\) are LECs of the \(\mathcal{O}(p^2)\) chiral Lagrangian.

\section{The relations between CGLN amplitudes and invariant amplitudes}\label{CGLN}
The functions $A_i$ and $\mathcal{F}_i$ are connected with each other as the following relation~\cite{Pasquini:2007fw, Borasoy:2007ku}:

\begin{align}
   \mathcal{F}_{1} &=\left(\sqrt{s}-m_{N}\right) \frac{N_{1} N_{2}}{8 \pi \sqrt{s}} \nonumber\\
   &\times \left[A_{1}+\frac{k \cdot q}{\sqrt{s}-m_{N}} A_{3}+\left(\sqrt{s}-m_{N}-\frac{k \cdot q}{\sqrt{s}-m_{N}}\right) A_{4}-\frac{k^{2}}{\sqrt{s}-m_{N}} A_{6}\right]\ , \\
    \mathcal{F}_{2} &=\left(\sqrt{s}+m_{N}\right) \frac{N_{1} N_{2}}{8 \pi \sqrt{s}} \frac{|\mathbf{q}||\mathbf{k}|}{\left(E_{1}+m_{N}\right)\left(E_{2}+m_{N}\right)} \nonumber\\    
    &\times \left[-A_{1}+\frac{k \cdot q}{\sqrt{s}+m_{N}}A_3+\left(\sqrt{s}+m_{N}-\frac{k \cdot q}{\sqrt{s}+m_{N}}\right) A_{4}-\frac{k^{2}}{\sqrt{s}+m_{N}} A_{6}\right]\ ,  \\
    \mathcal{F}_{3} &=\left(\sqrt{s}+m_{N}\right) \frac{N_{1} N_{2}}{8 \pi \sqrt{s}} \frac{|\mathbf{q}||\mathbf{k}|}{E_{1}+m_{N}}\left[\frac{m_{N}^{2}-s+\frac{1}{2} k^{2}}{\sqrt{s}+m_{N}} A_{2}+A_{3}-A_{4}-\frac{k^{2}}{\sqrt{s}+m_{N}} A_{5}\right]\ ,  \\
    \mathcal{F}_{4} &=\left(\sqrt{s}-m_{N}\right) \frac{N_{1} N_{2}}{8 \pi \sqrt{s}} \frac{|\mathbf{q}|^{2}}{E_{2}+m_{N}}\left[\frac{s-m_{N}^{2}-\frac{1}{2} k^{2}}{\sqrt{s}-m_{N}} A_{2}+A_{3}-A_{4}+\frac{k^{2}}{\sqrt{s}-m_{N}} A_{5}\right]\ , \\
    \mathcal{F}_{7} &=\frac{N_{1} N_{2}}{8 \pi \sqrt{s}} \frac{|\mathbf{q}|}{E_{2}+m_{N}}\Bigg[ \left(m_{N}-E_{1}\right) A_{1}-\left(\frac{|\mathbf{k}|^{2}}{2 k_{0}}\left(2k_0\sqrt{s}-3k\cdot q\right)-\frac{\mathbf{q} \cdot \mathbf{k}}{2 k_{0}}\left(2 s-2 m_{N}^{2}-k^{2}\right)\right) A_{2} \nonumber\\
    &+\left(q_{0}\left(\sqrt{s}-m_{N}\right)-k \cdot q\right) A_{3}+\left(k \cdot q-q_{0}\left(\sqrt{s}-m_{N}\right)+\left(E_{1}-m_{N}\right)\left(\sqrt{s}+m_{N}\right)\right) A_{4} \nonumber\\
    &+\left(q_{0} k^{2}-k_{0} k \cdot q\right) A_{5}-\left(E_{1}-m_{N}\right)\left(\sqrt{s}+m_{N}\right) A_{6} \Bigg]\ ,  \\
     \mathcal{F}_{8} &=\frac{N_{1} N_{2}}{8 \pi \sqrt{s}} \frac{|\mathbf{k}|}{E_{2}+m_{N}}\Bigg[ \left(m_{N}+E_{1}\right) A_{1}+\left(\frac{|\mathbf{k}|^{2}}{2 k_{0}}\left(2k_0\sqrt{s}-3k\cdot q\right)-\frac{\mathbf{q} \cdot \mathbf{k}}{2 k_{0}}\left(2 s-2 m_{N}^{2}-k^{2}\right)\right) A_{2} \nonumber\\
     &+\left(q_{0}\left(\sqrt{s}+m_{N}\right)-k \cdot q\right) A_{3}+\left(k \cdot q-q_{0}\left(\sqrt{s}+m_{N}\right)+\left(E_{1}+m_{N}\right)\left(\sqrt{s}-m_{N}\right)\right) A_{4} \nonumber\\
    &+\left(q_{0} k^{2}-k_{0} k \cdot q\right) A_{5}-\left(E_{1}+m_{N}\right)\left(\sqrt{s}-m_{N}\right) A_{6} \Bigg]\ ,
\end{align}
with
\begin{align}
N_i=\sqrt{E_i+m_N},\quad E_i=\sqrt{\bp_i^2+m_N^2},\quad i=1,2\ .
\end{align}

\section{Partial wave helicity amplitudes}\label{ap:Helicity}

In the following part, we introduce the partial wave helicity amplitude method of pion photo- and electroproduction~\cite{Walker:1968xu}.

It is convenient to perform partial wave projection using the helicity formalism proposed in Refs.~\cite{Jacob:1959at, Walker:1968xu}.
Here, we define \(\lambda_i\) (\(i=1,2,3,4\)), which stand for the helicity of photon, initial nucleon, pion, and final nucleon. For each set of helicity quantum numbers, denoted by \(H_s\equiv \{\lambda_1\lambda_2\lambda_3\lambda_4\} \), there is a helicity amplitude \(\mathcal{A}_{H_s}\), which can be expanded as
\begin{align}\label{AHs}
    A_{H_s}\left(s,t(\theta)\right)=16\pi\sum_{J=M}^\infty(2J+1)A_{H_s}^{J}(s)\,{d}_{\lambda \mu}^J(\theta)\ ,
\end{align}
where \(M=\mathrm{max}\{|\lambda|,|\mu|\} \), \(\lambda\equiv\lambda_1-\lambda_2\) and \(\mu\equiv\lambda_3-\lambda_4=-\lambda_4\)\ ,
and \(d^J(\theta)\) is the standard Wigner function. 
By imposing the orthogonal properties of the \(d\) functions, the partial wave helicity amplitudes \(A_{H_s}^{J}(s)\) in the above equation may be projected; i.e,
\begin{align}\label{pwamp}
    A^{J}_{H_s}(s)=\frac{1}{32\pi}\int_{-1}^{1} \mathrm{d} \cos\theta A_{H_s}(s,t)d_{\lambda,\lambda'}^{J}(\theta)\ .
\end{align}
  \begin{table}[tp]  
  \centering  
  \fontsize{10}{6}\selectfont  
    \begin{threeparttable}
    \caption{Helicity amplitudes $\{A_{\mu\lambda}(\theta)\}=\{A_{H_s}$\}.}
      \label{H}
      \begin{tabular}{ccccccc}
      \toprule
      \multirow{2}{*}{\diagbox{$\mu$}{$\lambda$}}&
      \multicolumn{2}{c}{ $\lambda_1=+1$}&\multicolumn{2}{c}{ $\lambda_1=-1$}
      &\multicolumn{2}{c}{ $\lambda_1=0$}\\  
      \cmidrule(lr){2-3} \cmidrule(lr){4-5}  \cmidrule(lr){6-7}  
      &$\frac{3}{2}$&$\frac{1}{2}$&$-\frac{1}{2}$&$-\frac{3}{2}$
      &$\frac{1}{2}$&$-\frac{1}{2}$\\ 
      \midrule  
      $\frac{1}{2}$&$H_1$&$H_2$&$H_4$&$-H_3$&$H_5$&$H_6$\\ 
      $-\frac{1}{2}$&$H_3$&$H_4$&$-H_2$&$H_1$&$H_6$&$-H_5$\\ 
      \bottomrule  
      \end{tabular}
    \end{threeparttable}  
  \end{table}
In particular, we use $H_i (i=1 \sim 6)$ as symbols to define the helicity amplitude. 
The relations between $A_{\mu \lambda}$ and $H_i$ are listed in Table~\ref{H} \cite{Walker:1968xu}.

The differential scattering cross section can be written as
\begin{align}\label{Sig}
\frac{\mathrm{d}\sigma}{\mathrm{d}\Omega}=\frac{1}{2} \frac{|\bq|}{k^\mathrm{cm}} \sum_{\lambda_i }\left|A_{\mu \lambda}\right|^{2}\ .
\end{align}
From Eq.~(\ref{Sig}) and Table~\ref{H}, we can integrate the angle dependence,
\begin{align}
\sigma=2 \pi\frac{|\bq|}{k^\mathrm{cm}} \sum_{J} \sum_{i=1}^{6}(2 j+1)\left|H_{i}^{J}\right|^{2}\ .
\end{align}

$A_{\mu\lambda}^J (H_i^J)$ has definite angular momentum but cannot be determined in parity.
Therefore, we can add the final state with the opposite helicity $\mu,-\mu$ to obtain the so-called partial wave helicity parity eigenstates,
\begin{align}
    \begin{aligned}
    A_{l+} &=-\frac{1}{\sqrt{2}}\left(A_{\frac{1}{2},\frac{1}{2}(\lambda_1=1)}^{J}+A_{-\frac{1}{2},\frac{1}{2}(\lambda_1=1)}^{J}\right)\ , \\
    A_{(l+1)-} &=\frac{1}{\sqrt{2}}\left(A_{\frac{1}{2},\frac{1}{2}(\lambda_1=1)}^{J}-A_{-\frac{1}{2},\frac{1}{2}(\lambda_1=1)}^{J}\right)\ , \\
    B_{l+} &=\sqrt{\frac{2}{l(l+2)}} \left(A_{\frac{1}{2},\frac{3}{2}}^{J}+A_{-\frac{1}{2},\frac{3}{2}}^{J}\right)\quad \ell \geq 1\ , \\
    B_{(l+1)-} &=-\sqrt{\frac{2}{l(l+2)}} \left(A_{\frac{1}{2},\frac{3}{2}}^{J}-A_{-\frac{1}{2},\frac{3}{2}}^{J}\right)\quad \ell \geq 1\ ,   \\
    S_{l+} &=-\frac{Q}{2|\bk|}(l+1) \left(A_{\frac{1}{2},\frac{1}{2}(\lambda_1=0)}^{J}+A_{-\frac{1}{2},\frac{1}{2}(\lambda_1=0)}^{J}\right)\ , \\
    S_{(l+1)-} &=-\frac{Q}{2|\bk|}(l+1) \left(A_{\frac{1}{2},\frac{1}{2}(\lambda_1=0)}^{J}-A_{-\frac{1}{2},\frac{1}{2}(\lambda_1=0)}^{J}\right)\ .
    \end{aligned}
\end{align}
Notice that the normalization coefficients we use here are different from those in Refs.~\cite{Berends:1978ta,Arndt:1990ej}, and $J=l+1/2$ for `$+$' amplitudes and $J=l-1/2$ for `$-$' amplitudes.
$A,\ B$, and $S$ represent amplitudes with initial helicity of $1/2,\ 3/2,\ 1/2$,  respectively, so it can also be written as $\mathcal{A}^{1/2}, \mathcal{A}^{3/2}$, and $\mathcal{S}^{1/2}$ up to some normalization factors; see Eq.~(\ref{AAS}).

Furthermore, with the definitions of Eqs.~(\ref{AHs}) and (\ref{pwamp}), then we can obtain
\begin{align}
    \begin{aligned}\label{HAB}
    H_{1}&=\frac{1}{\sqrt{2}} \sin \theta \cos \frac{\theta}{2} \sum\left(B_{l+}-B_{(l+1)-}\right)\left(P_{l}^{\prime \prime}-P_{l+1}^{\prime \prime}\right)\ , \\
    H_{2}&=\sqrt{2} \cos \frac{\theta}{2} \sum\left(A_{l+}-A_{(l+1)-}\right)\left(P_{l}^{\prime}-P_{l+1}^{\prime}\right)\ , \\
    H_{3}&=\frac{1}{\sqrt{2}} \sin \theta \sin \frac{\theta}{2} \sum\left(B_{l+}+B_{(l+1)-}\right)\left(P_{l}^{\prime \prime}+P_{l+1}^{\prime \prime}\right)\ , \\
    H_{4}&=\sqrt{2} \sin \frac{\theta}{2} \sum\left(A_{l+}+A_{(l+1)-}\right)\left(P_{l}^{\prime}+P_{l+1}^{\prime}\right)\ ,  \\
    H_{5} &=\frac{Q}{|\mathbf{k}|} \cos \frac{\theta}{2} \sum(l+1)\left(S_{l+}+S_{(l+1)-}\right)\left(P_{l}^{\prime}-P_{l+1}^{\prime}\right)\ ,  \\
    H_{6} &=\frac{Q}{|\mathbf{k}|} \sin \frac{\theta}{2} \sum(l+1)\left(S_{l+}-S_{(l+1)-}\right)\left(P_{l}^{\prime}+P_{l+1}^{\prime}\right)\ .
    \end{aligned}
\end{align}

According to the expansion method of CGLN~\cite{Chew:1957tf}, the relationship between helicity amplitudes and CGLN multipole amplitudes can also be obtained~\cite{Chew:1957tf,Tiator:2017cde},
\begin{align}
    \begin{aligned}\label{HF}
    H_{1}&=-\frac{1}{\sqrt{2}}\sin \theta \cos\frac{\theta}{2} \left(\mathcal{F}_{3}+\mathcal{F}_{4}\right)\ ,  \\
    H_{2}&=\sqrt{2} \cos\frac{\theta}{2} \left[\left(\mathcal{F}_{2}-\mathcal{F}_{1}\right)+\frac{1}{2}(1-\cos \theta)(\mathcal{F}_{3}-\mathcal{F}_{4}) \right]\ ,  \\
    H_{3}&=\frac{1}{\sqrt{2}}\sin \theta \sin \frac{\theta}{2} \left(\mathcal{F}_{3}-\mathcal{F}_{4}\right)\ ,  \\
    H_{4}&=\sqrt{2} \sin \frac{\theta}{2} \left[\left(\mathcal{F}_{1}+\mathcal{F}_{2}\right)+\frac{1}{2}(1+\cos \theta)(\mathcal{F}_{3}+\mathcal{F}_{4}) \right]\ ,  \\
    H_{5} &=\cos \frac{\theta}{2}\left(\mathcal{F}_{5}+\mathcal{F}_{6}\right)\ , \\
    H_{6} &=-\sin \frac{\theta}{2}\left(\mathcal{F}_{5}-\mathcal{F}_{6}\right)\ .
    \end{aligned}
\end{align}
Compare Eqs.~(\ref{HAB}) and (\ref{HF}) with the CGLN expansion, we have
\begin{align}
\begin{aligned}
A_{l+} &=\frac{1}{2}\left[(l+2) E_{l+}+l M_{l+}\right]\ , \\
B_{l+} &=E_{l+}-M_{l+}\ , \\
A_{(l+1)-} &=-\frac{1}{2}\left[l E_{(l+1)-}-(l+2) M_{(l+1)-}\right]\ , \\
B_{(l+1)-} &=E_{(l+1)-}+M_{(l+1)-} \ .
\end{aligned}
\end{align}
$\mathcal{A}^h,\mathcal{S}^{1/2}$ can be related to the resonant part of the corresponding multipole amplitudes at the pole position in the following way:
\begin{align}
    \begin{aligned}\label{AAS}
    \mathcal{A}^{1/2}_{l+} &=-\frac{1}{2}\left[(l+2) E_{l+}+l M_{l+}\right]\ , \\
    \mathcal{A}^{3/2}_{l+} &=\frac{1}{2}\sqrt{l(l+2)} \left(E_{l+}-M_{l+}\right)\ , \\
    \mathcal{S}^{1/2}_{l+} &=-\frac{l+1}{\sqrt{2}} S_{l+}\ ,  \\
    \mathcal{A}^{1/2}_{(l+1)-} &=-\frac{1}{2}\left[l E_{(l+1)-}-(l+2) M_{(l+1)-}\right]\ ,  \\
    \mathcal{A}^{3/2}_{(l+1)-} &=-\frac{1}{2}\sqrt{l(l+2)} \left(E_{(l+1)-}+M_{(l+1)-} \right)\ ,  \\
    \mathcal{S}^{1/2}_{(l+1)-} &=-\frac{l+1}{\sqrt{2}} S_{(l+1)-}\ .
    \end{aligned}
\end{align}

The scattering cross section is written in terms of $\mathcal{A}_\alpha^h$ as
\begin{align}
    \begin{aligned}
    \sigma_{T} &=\left(\sigma_{T}^{1 / 2}+\sigma_{T}^{3 / 2}\right)+\epsilon\sigma_L\ ,  \\
    \sigma_{T}^{h} &=2 \pi \frac{|\bq|}{k^\mathrm{cm}} \sum_{\alpha(\ell, J)}(2 J+1)\left|\mathcal{A}_{\alpha}^{h}\right|^{2}\ ,  \\
    \sigma_{L} &=2 \pi \frac{|\bq|}{k^\mathrm{cm}} \frac{Q^{2}}{k^{2}} \sum_{\alpha(\ell, J)}(2 J+1)\left|\mathcal{S}_{\alpha}^{1 / 2}\right|^{2}\ ,
    \end{aligned}
\end{align}
where superscript $h$ stands for helicity.
Expand the above formula; it can be obtained,
\begin{align}
    \begin{aligned}
    \sigma_T^{1 / 2} &=2 \pi \frac{|\bq|}{k^\mathrm{cm}} \sum 2(l+1)\left[\left|A_{l+}\right|^{2}+\left|A_{(1+1)-}\right|^{2}\right]\ , \\
    \sigma_T^{3 / 2} &=2 \pi \frac{|\bq|}{k^\mathrm{cm}} \sum \frac{l}{2}(l+1)(l+2)\left[\left|B_{l+}\right|^{2}+\left|B_{(l+1)-}\right|^{2}\right]\ ,  \\
     \sigma_{L} &=4 \pi \frac{|\bq|}{k^\mathrm{cm}} \sum \frac{Q^{2}}{\bk^{2}}(l+1)^{3}\left[\left|C_{l+}\right|^{2}+\left|C_{(l+1)-}\right|^{2}\right]\ .
    \end{aligned}
\end{align}

\section{Electromagnetic couplings of the subthreshold resonance}\label{sec:EMc}

    In Refs.~\cite{Wang2018,Wang2019,Wang2018a}, evidences are found on the possible existence of a sub-thresthod resonance named \(N^\ast(890)\) in the \(S_{11}\) channel using the method proposed in Refs.~\cite{Xiao:2000kx, He:2002ut, Zheng:2003rw, Zhou:2004ms, Zhou:2006wm}, assisted by chiral amplitudes obtained in Refs.~\cite{Chen:2012nx,Alarcon:2012kn,Yao:2016vbz,Siemens:2017opr}.
    In this appendix, further results are provided on $\gamma^* N$ coupling to this resonance for future reference.    
    In the main text, all the involved parameters in the partial wave virtual photoproduction amplitude \(\mathcal{M}_l(s)\) have been determined.
Since \(N^\ast(890)\), as a subthreshold resonance, is located on the second Riemann sheet, one needs to perform analytic continuation in order to extract its couplings to the \(\gamma^* N\) and \(\pi N\) systems.

It can be proved that the residue can be extracted from\cite{Ma:2020hpe}
\begin{align}\label{eq:gggp}
    g_{\gamma N}g_{\pi N} \simeq \frac{\mathcal{M}_l(s_{{\rm p}})}{\mathcal{S}_l^\prime(s_{{\rm p}})}\ ,
\end{align}
where \(\mathcal{S}_l(s)\) corresponds to partial wave \(S\) matrix of elastic \(\pi N\) scattering. 
Residues \(g_{\gamma N}\) and \(g_{\pi N}\) denote the \(\gamma N\) and \(\pi N\) couplings, respectively.  
The \(\pi N\) coupling can also be extracted from elastic \(\pi N\) scattering, i.e., $g_{\pi N}^2 \simeq \mathcal{T}_l(s_{{\rm p}})/\mathcal{S}_l^\prime(s_{{\rm p}})$, where \(\mathcal{T}_l\) is the corresponding partial wave \(\pi N\) scattering amplitude.
In order to compare the results with Refs.\cite{Svarc:2013laa,Svarc:2014sqa}, which are extracted directly from multipole amplitudes parameterized in the \(W(\sqrt{s})\) plane, so we can write
\begin{align}
    E_{0+}^{\mathrm{II}}(s\rightarrow s_p)\simeq\frac{g_{\gamma N}^E g_{\pi N}}{s-s_p}\simeq\frac{g_{\gamma N}^E g_{\pi N}}{2\sqrt{s_p}(\sqrt{s}-\sqrt{s_p})}=\frac{\left( g_{\gamma N}^E g_{\pi N}/2W_p \right)}{W-W_p}\ , \\
    S_{0+}^{\mathrm{II}}(s\rightarrow s_p)\simeq\frac{g_{\gamma N}^S g_{\pi N}}{s-s_p}\simeq\frac{g_{\gamma N}^S g_{\pi N}}{2\sqrt{s_p}(\sqrt{s}-\sqrt{s_p})}=\frac{\left( g_{\gamma N}^S g_{\pi N}/2W_p \right)}{W-W_p}\ ,
\end{align}
where subscript \(p\) stands for pole parameters.

Using the above formulas, we can calculate the virtual-photon decay amplitudes \(A^{\mathrm{pole}}_{h},S_{1/2}^{\mathrm{pole}}\) at the \(S_{11}N^\ast(890)\) pole position, which is Refs.~\cite{Suzuki:2010yn, Svarc:2013laa, Svarc:2014sqa}:
\begin{align}
A_{h}^{\mathrm{pole}} &=C \sqrt{\frac{|\bq_{p}|}{k^\mathrm{cm}_{p}} \frac{2 \pi(2 J+1) W_{p}}{m_{N} \operatorname{Res}{\mathcal{T}_{\pi N}}}} \operatorname{Res} \mathcal{A}_{\alpha}^{h}\ ,  \\
S_{1/2}^{\mathrm{pole}} &=C \sqrt{\frac{|\bq_{p|}}{k^\mathrm{cm}_{p}} \frac{2 \pi(2 J+1) W_{p}}{m_{N} \operatorname{Res}{\mathcal{T}_{\pi N}}}} \operatorname{Res} \mathcal{S}_{\alpha}^{1/2}\ .
\end{align}
Refer to appendix~\ref{ap:Helicity} for the definition of $\mathcal{A}^h_\alpha,\ \mathcal{S}^{1/2}_\alpha$, here we use $\mathcal{A}^{1/2}_{0+}=-E_{0+},\ \mathcal{S}^{1/2}_{0+}=-(1/\sqrt{2})S_{0+}$.
Intuitively, $\mathcal{A}^h_\alpha,\ \mathcal{S}^{1/2}_\alpha$ characterize the power of electromagnetic couplings and the amplitudes of the decay process $N^* \to \gamma^* N$.
$|\bq_p|$ is the pion momenta at the pole.
The factor $C$ is $\sqrt{2/3}$ for isospin $3/2$ and $-\sqrt{3}$ for isospin $1/2$.
So if we focus only on $S_{11}$ channel, the corresponding virtual-photon decay amplitudes are given by
\begin{align}
A^{\mathrm{pole}}_{1/2}(Q^2) =g_{\gamma N}^E \sqrt{\frac{3 \pi W_p}{m_N k^\mathrm{cm}_p}}\ ,\quad S^{\mathrm{pole}}_{1/2}(Q^2) =g_{\gamma N}^S \sqrt{\frac{3 \pi W_p}{2 m_N k^\mathrm{cm}_p}}\ .
\end{align}
It can be seen that the amplitudes, $\mathcal{A}_{\alpha}^{1/2}$ and $\mathcal{S}_{\alpha}^{1/2}$, as well as the residues, $A^{\mathrm{pole}}_{1/2}$ and $S^{\mathrm{pole}}_{1/2}$, are functions of the photon virtuality $Q^2$.

According to Eqs.~\eqref{eq:gggp}, \(N^*(890)\) residues or couplings, \(g_{\gamma N}g_{\pi N}\), can be extracted from multipole amplitudes $E_{0+},S_{0+}$.
In the meantime, \(g_{\pi N}^2\) can be computed by using $g_{\pi N}^2 \simeq \mathcal{T}_l(s_{{\rm p}})/\mathcal{S}_l^\prime(s_{{\rm p}})$, which was already obtained in Ref.~\cite{Wang2018}.
We employed the MAID solution of the fit (The result of DMT is similar), and chose central value \(\sqrt{s}=0.882-0.190i\) for pole position to extract pole residues.
\(\mathcal{T}(s_p)\) can be obtained through \(\mathcal{S}(s_p)=1+2i\rho_{\pi N}(s_p)\mathcal{T}(s_p)=0\) and \(\frac{1}{S'(s_p)}\) is just the residue of \(\mathcal{S}^{\rm II}\) as explained Ref.~\cite{Wang2018}.

\begin{figure}[H]
\centering
\subfigure{
    \begin{minipage}[b]{0.985\linewidth}
    
    \begin{minipage}[b]{0.48\linewidth}
    \includegraphics[width=7cm]{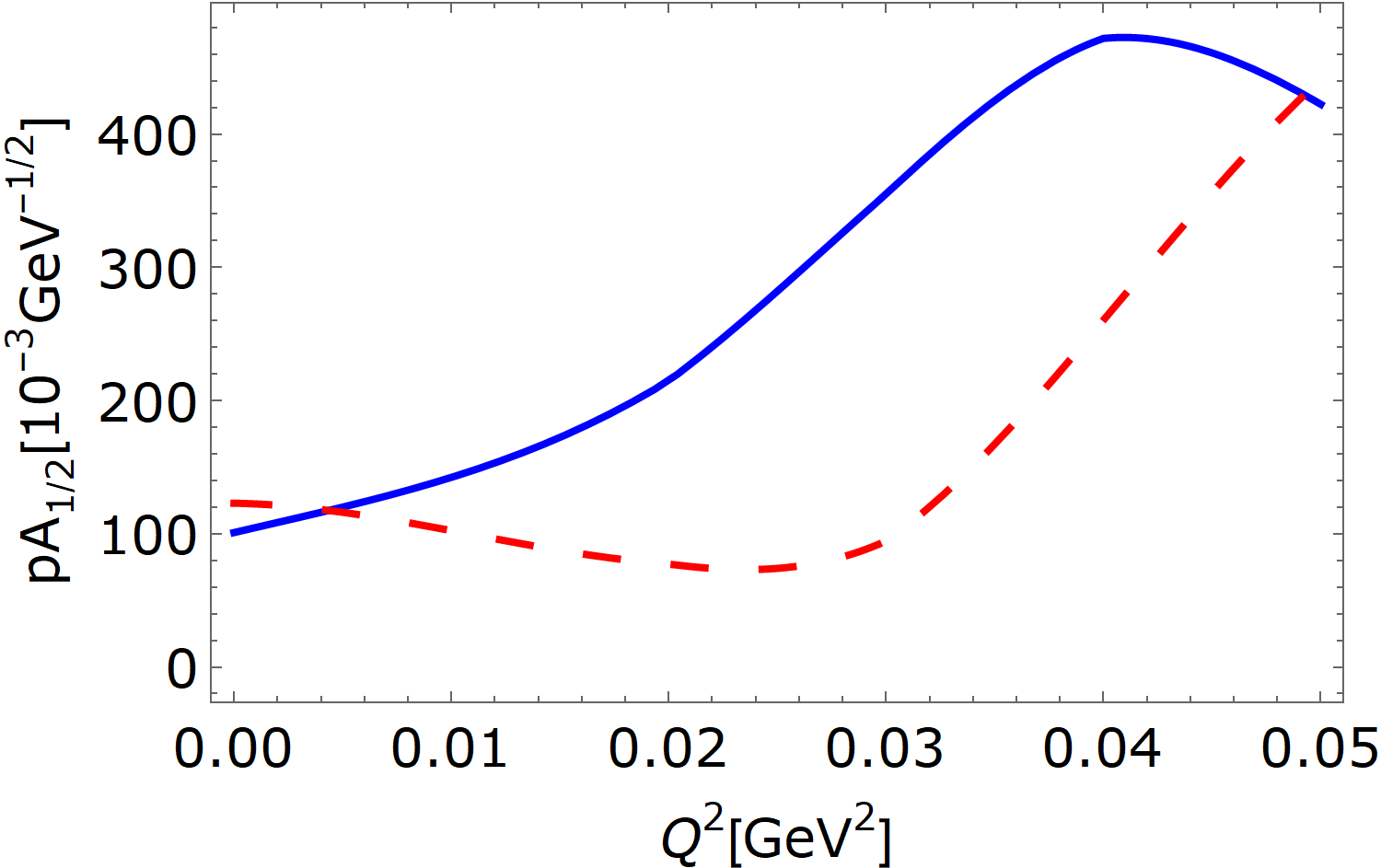}
    \vspace{0.02cm}
    \hspace{0.02cm}
    \end{minipage}
    \begin{minipage}[b]{0.48\linewidth}
    \includegraphics[width=7cm]{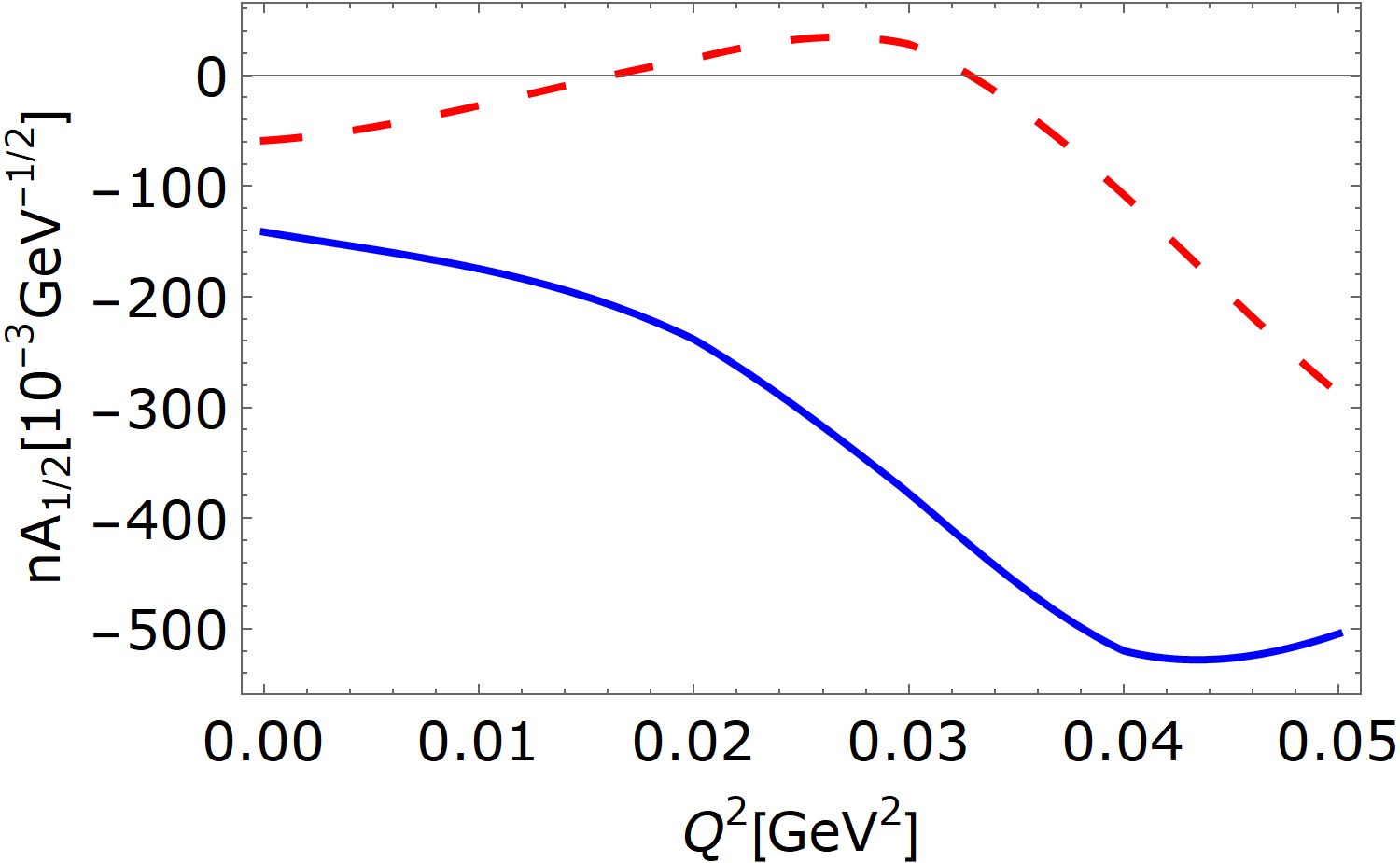}
    \vspace{0.02cm}
    \hspace{0.02cm}
    \end{minipage}
    
    \begin{minipage}[b]{0.48\linewidth}
    \includegraphics[width=7cm]{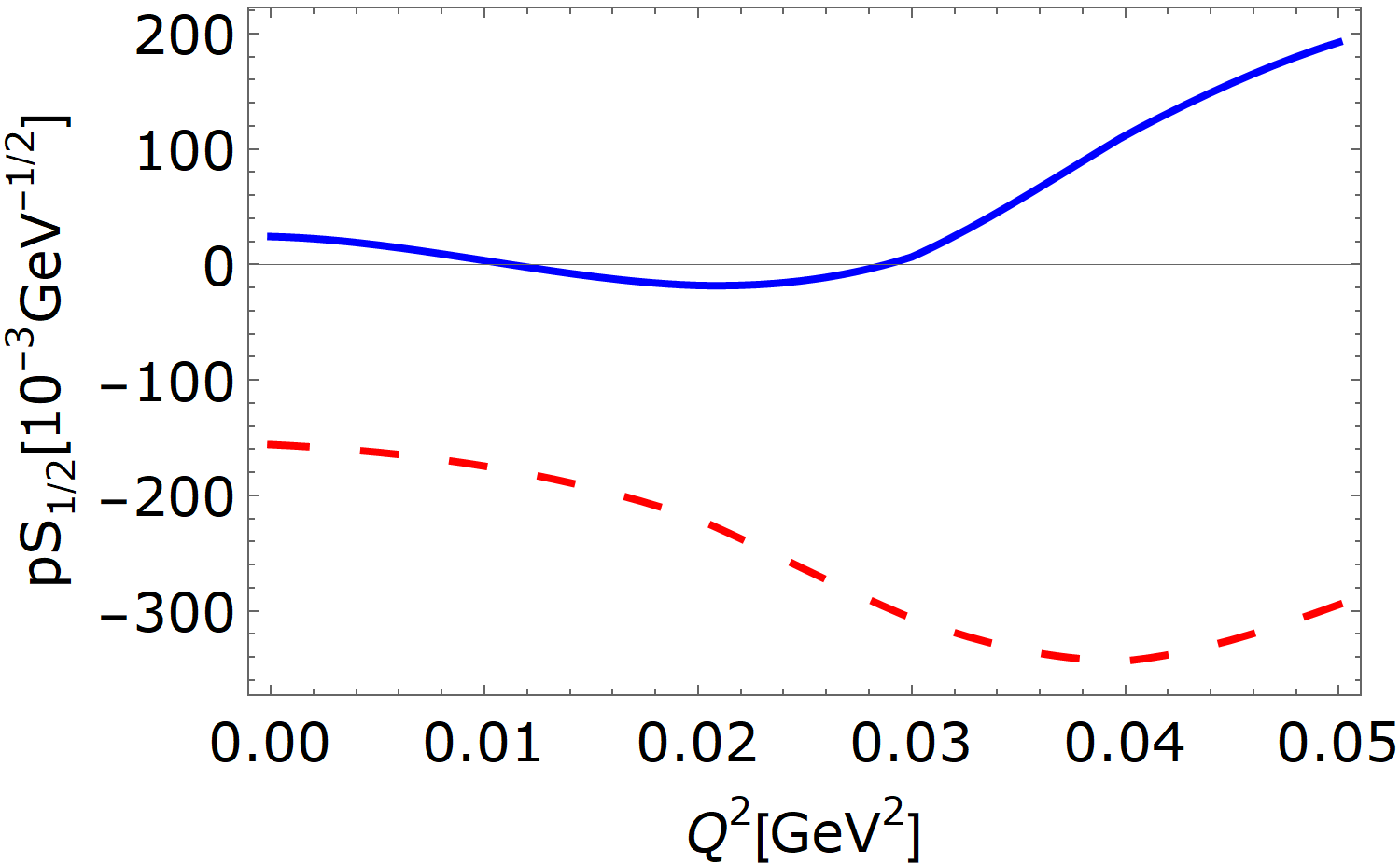}
    \vspace{0.02cm}
    \hspace{0.02cm}
    \end{minipage}
    \begin{minipage}[b]{0.48\linewidth}
    \includegraphics[width=7cm]{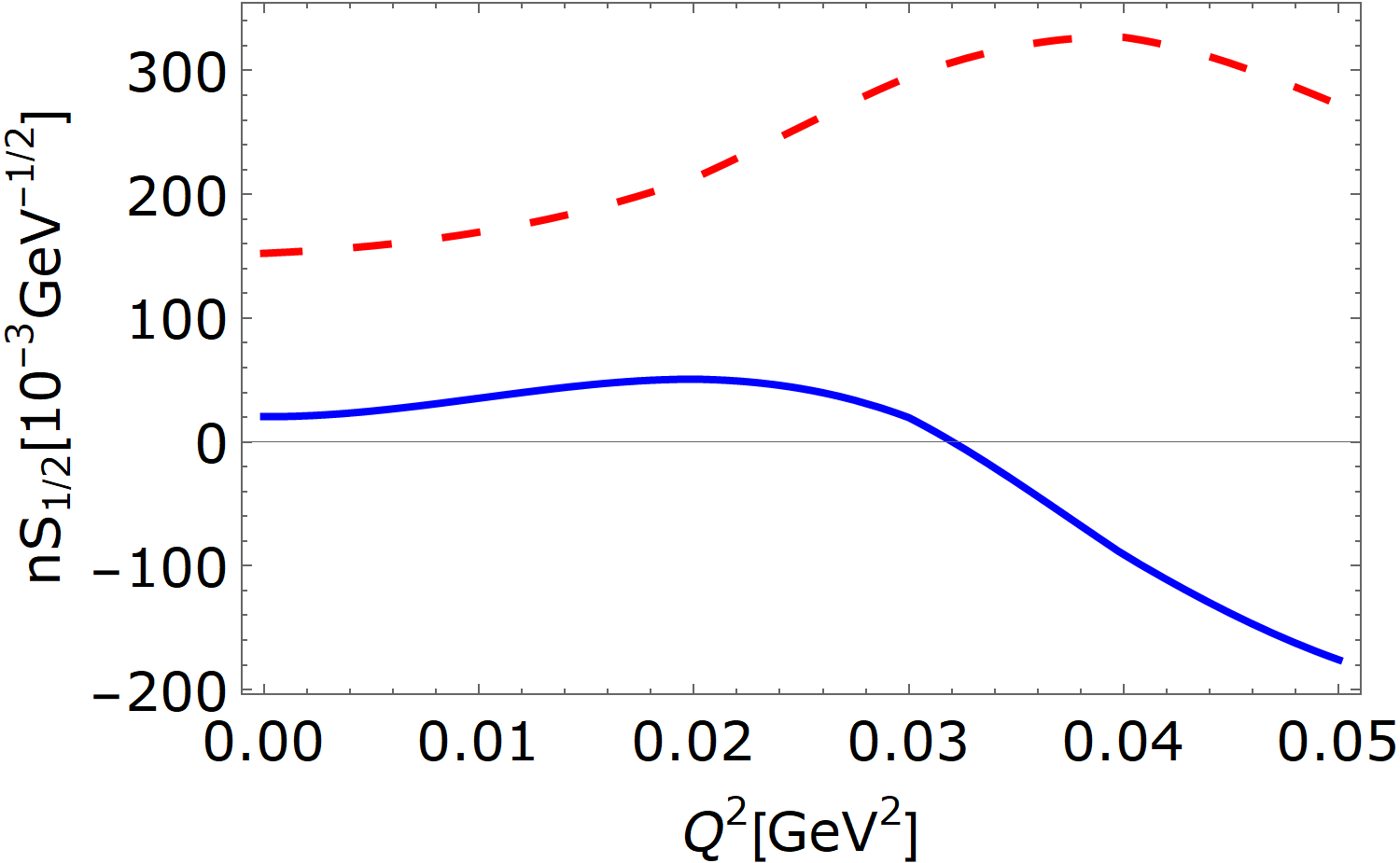}
    \vspace{0.02cm}
    \hspace{0.02cm}
    \end{minipage}  
    
    \end{minipage}
    }
\caption{The blue solid line and the red solid line represent real and imaginary parts of virtual-photon decay amplitudes at pole position, respectively.}\label{f:UIAS}
\end{figure}

The values of the decay amplitudes \(A_{1/2}\) and $S_{1/2}$ at the pole position are shown in Figs.~\ref{f:UIAS}.

\end{appendices}

\newpage

\bibliographystyle{hphysrev}
\bibliography{ephotoN}

\end{document}